\documentclass[11pt,notitlepage]{amsart}
\usepackage{mathpazo}
\usepackage[symbol,perpage]{footmisc}
\usepackage[nobottomtitles]{titlesec}
\usepackage{dcolumn,booktabs}

\pdfcatalog{/PageLayout/TwoPageRight }
\titleformat{\section}[block]{\normalfont\Large\filcenter\bf}{}{1em}{}
\titleformat{\subsection}[block]{\normalfont\filcenter\bf}{}{1em}{}

\vfuzz \hfuzz
\let\Person\textsc
\let\Emphasis\textbf
\let\Author\textsc

\let\Title\textit

\def\th{${}^\text{th}$\ }
\newcommand\Section[1]{\subsection{{#1}}}
\def\Crule{\begin{center}\rule[.5ex]{5em}{.2pt}\\[10pt]\end{center}}
\DeclareMathOperator{\cotan}{\textrm{cotan}}

\renewcommand{\rho}{\varrho}
\renewcommand{\phi}{\varphi}

\newcolumntype{.}[1]{D{,}{,}{#1}}  
\newcommand{\mc}[1]{\multicolumn{1}{c@{}}{#1}}

\begin{document}
\thispagestyle{empty}
\begin{center}
\textsc{\huge\bf An Essay on the Application of mathematical Analysis to the 
theories of Electricity and Magnetism.}\footnote{Originally 
published as book in Nottingham, 1828. Reprinted in three parts in Journal f\"ur die reine und angewandte Mathematik
Vol.~39, 1 (1850) p.~73--89, Vol.~44, 4 (1852) p.~356--74, and Vol.~47, 3 (1854)
p.~161--221.
From there transcribed by Ralf Stephan,
eMail: \texttt{mailto:ralf@ark.in-berlin.de}}\\[5pt]
By George Green\\[5pt]
{\small Fellow of Gonville- and Cains-Colleges at Cambridge.}\\
\rule[.5ex]{10em}{.2pt}\\[10pt]
\end{center}

\bigskip

MSC-Class: 31-3 01A50
\Crule

\section{Preface.}
\markboth{Preface.}{Preface.}
After I had composed the following Essay, I naturally felt anxious to
become acquainted with what had been effected by former writers on the same
subject, and, had it been practicable, I should have been glad to have given,
in this place, an historical sketch of its progress; my limited sources of information,
however, will by no means permit me to do so; but probably I may here be
allowed to make one or two observations on the few works which have fallen
in my way, more particularly as an opportunity will thus offer itself, of 
noticing an excellent paper, presented to the Royal Society by one of the most
illustrious members of that learned body, which appears to have attracted little
attention, but which, on examination, will be found not unworthy the man who
was able to lay the foundations of pneumatic chymistry, and to discover that
water, far from being according to the opinions then received, an elementary
substance, was a compound of two of the most important gasses in nature.

It is almost needless to say the author just alluded to is the celebrated
\Person{Cavendish}, who, having confined himself to such simple methods, as may
readily be understood by any one possessed of an elementary knowledge of
geometry and fluxions, has rendered his paper accessible to a great number
of readers; and although, from subsequent remarks, he appears dissatisfied
with an hypothesis which enabled him to draw some important conclusions,
it will readily be perceived, on an attentive perusal of his paper, that a trifling
alteration will suffice to render the whole perfectly legitimate\footnote{
In order to make this quite clear, let us select one of
\Person{Cavendish}'s propositions,
the twentieth for instance, and examine with some attention the method
there employed.
The object of this proposition is to show,
that when two similar conducting bodies communicate
by means of a long slender canal, and are charged with electricity,
the respective quantities of redundant fluid contained in them,
will be proportional to the $n-1$
power of their corresponding diameters:
supposing the electric repulsion to vary inversely
as the $n$ power of the distance.
This is proved by considering the canal as cylindrical,
and filled with incompressible fluid of uniform density:
then the quantities of electricity
in the interior of the two bodies are determined by a very simple
geometrical construction, so that the total action exerted
on the whole canal by one of them, shall exactly
balance that arising from the other; and from some remarks
in the 27\th proposition, it
appears the results thus obtained, agree very well with
experiments in which real canals
are employed, whether they are straight or crooked,
provided, as has since been shown
by \Person{Coulomb}, $n$ is equal to two.
The author however confesses he is by no means able
to demonstrate this, although, as we shall see immediately,
it may very easily be deduced
from the propositions contained in this paper.

For this purpose, let us conceive an incompressible
fluid of uniform density, whose
particles do not act on each other,
but which are subject to the same actions from all
the electricity in their vicinity, as real electric
fluid of like density would be; then supposing an
infinitely thin canal of this hypothetical fluid,
whose perpendicular sections are
all equal and similar, to pass from a point $a$
on the surface of one of the bodies, through
a portion of its mass, along the interior of the real canal,
and through a part of the
other body, so as to reach a point $A$ on its surface,
and then proceed from $A$ to $a$ in
a right line, forming thus a closed circuit,
it is evident from the principles of hydrostatics,
and may be proved from our author's $23^\text{d}$ proposition,
that the whole of the
hypothetical canal will be in equilibrium,
and as every particle of the portion contained
within the system is necessarily so,
the rectilinear portion $aA$ must therefore be in equilibrium.
This simple consideration serves to complete
\Person{Cavendish}'s demonstration,
whatever may be the form or thickness of the real canal,
provided the quantity of electricity
in it is very small compared with that contained in the bodies.
An analagous application
of it will render the demonstration of the $22^\text{d}$
proposition complete, when the two coatings
of the glass plate communicate with their respective conducting bodies,
by fine metallic wires of any form.}.

Little appears to have been effected in the mathematical theory of electricity,
except immediate deductions from known formulae, that first presented
themselves in researches on the figure of the earth, of which the principal
are, --- the determination of the law of the electric density on the surfaces of
conducting bodies differing little from a sphere, and on those of ellipsoids,
from 1771, the date of \Person{Cavendish}'s paper,
until about 1812, when M.~\Person{Poisson}
presented to the French Institute two memoirs of singular elegance, relative
to the distribution of electricity on the surfaces of conducting spheres, 
previously electrified and put in presence of each other. It would be quite 
impossible to give any idea of them here: to be duly appreciated they must he
read. It will therefore only be remarked, that they are in fact founded upon
the consideration of what have, in this Essay, been termed potential functions,
and by means of an equation in variable differences, which may immediately
be obtained from the one given in our tenth article, serving to express the
relation between the two potential functions arising from any spherical 
surface, the author deduces the values of these functions belonging to each of
the two spheres under consideration, and thence the general expression of the
electric density on the surface of either, together with their actions on any
exterior point.

I am not aware of any material accessions to the theory of electricity,
strictly so called, except those before noticed; but since the electric and
magnetic fluids are subject to one common law of action, and their theory,
considered in a mathematical point of view, consists merely in developing the
consequences which flow from this law, modified only by considerations arising
from the peculiar constitution of natural bodies with respect to these two kinds
of fluid, it is evident, the mathematical theory of the latter must be
very intimately connected with that of the former;
nevertheless, because it is here
necessary to consider bodies as formed of an immense number of insulated
particles, all acting upon each other mutually, it is easy to conceive that
superior difficulties must, on this account,
present themselves, and indeed, until
within the last four or five years, no successful attempt to overcome them had
been published. For this farther extension of the domain of analysis, we are
again indebted to M.~\Person{Poisson},
who has already furnished us with three memoirs on magnetism:
the two first contain the general equations on which the
magnetic state of a body depends, whatever may be its form, together with
their complete solution in case the body under consideration is a hollow 
spherical shell, of uniform thickness,
acted upon by any exterior forces, and also
when it is a solid ellipsoid subject to the influence of the earth's action. By
supposing magnetic changes to require time, although an exceedingly short one,
to complete them, it had been suggested that M.~\Person{Arago}'s
discovery relative
to the magnetic effects developed in copper, wood, glass, etc., by rotation,
might be explained. On this hypothesis M.~\Person{Poisson}
has founded his third memoir,
and thence deduced formulae applicable to magnetism in a state of motion.
Whether the preceding hypothesis will serve to explain the singular phenomena
observed by M.~\Person{Arago} or not,
it would ill become me to decide; but it is
probably quite adequate to account for those produced by the rapid rotation of
iron bodies.

We have just taken a cursory view of what has hitherto been written,
to the best of my knowledge, on subjects connected with the mathematical
theory of electricity; and although many of the artifices employed in the works
before mentioned are remarkable for their elegance, it is easy to see they are
adapted only to particular objects, and that some general method, capable of
being employed in every case, is still wanting. Indeed M.~\Person{Poisson},
in the commencement of his first memoir
(M\'em. de l'Institut~1811), has incidentally
given a method for determining the distribution of electricity on the surface
of a spheroid of any form, which would naturally present itself to a person
occupied in these researches, being in fact nothing more than the ordinary
one noticed in our introductory observations, as requiring the resolution of
the equation ($a$). Instead however of supposing, as we have done, that the
point $p$ must be upon the surface, in order that the equation may subsist,
M.~\Person{Poisson} availing himself of a general fact,
which was then supported by
experiment only, has conceived the equation to hold good wherever this point
may be situated, provided it is within the spheroid, but even with this 
extension the method is liable to the same objection as before.

Considering how desirable it was that a power of universal agency, like
electricity, should, as far as possible,
be submitted to calculation, and reflecting
on the advantages that arise in the solution of many difficult problems, from
dispensing altogether with a particular examination of each of the forces which
actuate the various bodies in any system, by confining the attention solely to
that peculiar function on whose differentials they all depend, I was induced to
try whether it would be possible to discover any general relations, existing
between this function and the quantities of electricity
in the bodies producing it.
The advantages \Person{Laplace} had derived in the
third book of the \Title{M\'ecanique Celeste},
from the use of a partial differential equation
of the second order, there given,
were too marked to escape the notice of any one engaged with the present
subject, and naturally served to suggest that this equation might be made 
subservient to the object I had in view. Recollecting, after some attempts to
accomplish it, that previous researches on partial differential equations, had
shown me the necessity of attending to what have, in this Essay, been 
denominated the singular values of functions, I found, by combining this 
consideration with the preceding, that the resulting method was capable of being
applied with great advantage to the electrical theory, and was thus, in a short
time, enabled to demonstrate the general formulae contained in the preliminary
part of the Essay. The remaining part ought to be regarded principally as 
furnishing particular examples of the use of
these general formulae; their number
might with great ease have been increased, but those which are given, it is
hoped, will suffice to point out to mathematicians, the mode of applying the
preliminary results to any case they may wish to investigate. The hypotheses
on which the received theory of magnetism is founded, are by no means so
certain as the facts on which the electrical theory rests; it is however not
the less necessary to have the means of submitting them to calculation, for
the only way that appears oppen to us in the investigation of these subjects,
which seem as it were desirous to conceal themselves from our view, is to
form the most probable hypotheses we can, to deduce rigorously the consequences
which flow from them, and to examine whether such consequences
agree numerically with accurate experiments.

The applications of analysis to the physical Sciences, have the double
advantage of manifesting the extraordinary powers of this wonderful instrument
of thought, and at the same time of serving to increase them; numberless
are the instances of the truth of this assertion. To select one we may remark,
that M.~\Person{Fourier}, by his investigations relative to heat,
has not only discovered the general equations
on which its motion depends, but has likewise
been led to new analytical formulae, by whose aid M.~M.~\Person{Cauchy}
and \Person{Poisson}
have been enabled to give the complete theory of the motion of the waves
in an indefinitely extended fluid. The same formulae have also put us in
possession of the solutions of many other interesting problems, too numerous
to be detailed here. --- It must certainly be regarded as a pleasing prospect
to analists, that at a time when astronomy, from the state of perfection to
which it has attained, leaves little room for farther applications of their art,
the rest of the physical sciences should show themselves daily more and more
willing to submit to it; and, amongst other things, probably the theory that
supposes light to depend on the undulations of a luminiferous fluid, and to
which the celebrated Dr.~T.~\Person{Young} has
given such plausibility, may furnish
a useful subject of research, by affording new opportunities of applying the
general theory of the motion of fluids. The number of these opportunities
can scarcely be too great, as it must be evident to those who have examined
the subject, that, although we have long been in possession of the general
equations on which this kind of motion depends, we are not yet well
acquainted with the various limitations it will be necessary to introduce, in
order to adapt them to the different physical circumstances which may occur.

Should the present Essay tend in any way to facilitate thc application
of analysis to one of the most interesting of the physical sciences, the author
will deem himself amply repaid for any labour he may have bestowed upon
it; and it is hoped the difficulty of the subject will incline mathematicians to
read this work with indulgence, more particularly when they are informed
that it was written by a young man, who has been obliged to obtain the little
knowledge he possesses, at such intervals and by such means, as other 
indispensable avocations which offer
but few opportunities of mental improvement,
afforded.
\Crule
\bigskip

\section{Introductory observations.}
\markboth{Introductory observations.}{Introductory observations.}
The object of this Essay is to submit to Mathematical Analysis the
phenomena of the equilibrium of the Electric and Magnetic Fluids, and to lay
down some general principles equally applicable to perfect and imperfect 
conductors; but, before entering upon the calculus, it may not be amiss to give
a general idea of the method that has enabled us to arrive at results, 
remarkable for their simplicity and generality,
which it would be very difficult if
not impossible to demonstrate in the ordinary way.

It is well known, that nearly all the attractive and repulsive forces
existing in nature are such, that if we consider any material point~$p$, the
effect, in a given direction, of all the forces acting upon that point, arising
from any system of bodies~$S$ under consideration, will be expressed by a
partial differential of a certain function of the co-ordinates which serve to
define the point's position in space. The consideration of this function is of
great importance in many inquiries, and probably there are none in which its
utility is more marked than in those about to engage our attention. In the
sequel we shall often have occasion to speak of this function, and will 
therefore, for abridgment, call it the potential
function arising from the system~$S$.
If $p$ be a particle of positive electricity
under the influence of forces arising
from any electrified body, the function in question, as is well known, will be
obtained by dividing the quantity of electricity in each element of the body,
by its distance from the particle~$p$,
and taking the total sum of these quotients
for the whole body, the quantities of electricity in those elements which are
negatively electrified, being regarded as negative.

It is by considering the relations existing between the density of the
electricity in any system, and the potential functions thence arising, that we
have been enabled to submit many electrical phenomena to calculation, which
had hitherto resisted the attempts of analysts; and the generality of the 
consideration here employed, ought necessarily, and does, in fact, introduce a
great generality into the results obtained from it. There is one consideration
peculiar to the analysis itself, the nature and utility of which will be best
illustrated by the following sketch:

Suppose it were required to determine the law of the distribution of
the electricity on a closed conducting surface $A$ without thickness,
when placed
under the influence of any electrical forces whatever: these forces, for greater
simplicity, being reduced to three, $X$, $Y$ and~$Z$,
in the direction of the rectangular co-ordinates,
and tending to increase them. Then $\rho$ representing the
density of the electricity on an element~$d\sigma$ of the surface,
and $r$ the distance
between $d\sigma$ and $p$, any other point of the surface,
the equation for determining~$\rho$
which would be employed in the ordinary method, when the problem
is reduced to its simplest form, is known to be
\[
\tag{$a$}
\text{cons}\quad=\quad a\quad=\quad
\int\frac{\rho\,d\sigma}{r}-\int(X\,dx+Y\,dy+Z\,dz);
\]
the first integral relative to $d\sigma$
extending over the whole surface~$A$, and the second
representing the function whose complete differential is
${X\,dx+Y\,dy+Z\,dz}$, $x$, $y$ and $z$ being the co-ordinates of~$p$.

This equation is supposed to subsist, whatever may be the position of~$p$,
provided it is situate upon~$A$. But we have no general theory of equations
of this description, and whenever we are enabled to resolve one of them, it
is because some consideration peculiar to the problem renders, in that 
particular case, the solution comparatively simple,
and must be looked upon as the
effect of chance, rather than of any regular and scientific procedure.

We will now take a cursory view of the method it is proposed to
substitute in the place of the one just mentioned.

Let us make $B=\int(X\,dx+Y\,dy+Z\,dz)$ whatever may be the position
of the point~$p$, $V=\int\frac{\rho\,d\sigma}{r}$
when $p$ is situate any where within the surface~$A$,
and $V'=\int\frac{\rho\,d\sigma}{r}$ when $p$ is
exterior to it; the two quantities $V$ and~$V'$,
although expressed by the same definite integral,
are essentially distinct functions
of $x$, $y$, and $z$, the rectangular co-ordinates of~$p$;
these functions, as is well
known, having the property of satisfying the partial differential equations
\[
\begin{aligned}
0 &= \frac{d^2V}{dx^2}+\frac{d^2V}{dy^2}+\frac{d^2V}{dy^2},\\
0 &= \frac{d^2V'}{dx^2}+\frac{d^2V'}{dy^2}+\frac{d^2V'}{dy^2}.
\end{aligned}
\]
If now we could obtain the values of $V$ and $V'$ from these equations, we
should have immediately, by differentiation, the required value
of~$\rho$, as will
be shown in the sequel.

In the first place, let us consider the function $V$, whose value at the
surface~$A$ is given by the equation~($a$), since this may be written
\[
a=\overline{V}-\overline{B}
\]
the horizontal line over a quantity indicating
that it belongs to the surface~$A$.
But, as the general integral of the partial
differential equation ought to contain
two arbitrary functions, some other condition is requisite for the complete
determination of~$V$. Now since $V=\int\frac{\rho\,d\sigma}{r}$,
it is evident that none of its
differential co-efficients can become infinite
when $p$ is situate any where within
the surface~$A$, and it is worthy of remark,
that this is precisely the condition
required: for, as will be afterwards shown,
when it is satisfied we shall have generally
\[
V=-\int(\rho)d\sigma \overline{V};
\]
the integral extending over the whole surface,
and $(\rho)$ being a quantity dependant
upon the respective positions of $p$ and~$d\sigma$.

All the difficulty therefore reduces itself to finding a function $V$, which
satisfies the partial differential equation, becomes equal to the known value
of~$V$ at the surface,
and is moreover such that none of its differential coefficients
shall be infinite when $p$ is within~$A$.

In like manner, in order to find $V'$, we shall obtain
$\overline{V'}$, its value at~$A$,
by means of the equation~($a$), since this evidently becomes
\[
a=\overline{V'}-\overline{B},\quad\text{i.~e.}\quad
\overline{V'}=\overline{V}.
\]
Moreover it is clear, that none of the differential co-efficients
of~${V'=\int\frac{\rho\,d\sigma}{r}}$
can be infinite when $p$ is exterior to the surface~$A$,
and when $p$ is at an
infinite distance from~$A$,
$V'$ is equal to zero. These two conditions combined
with the partial differential equation in~$V'$, are sufficient in conjunction
with its known value $\overline{V'}$
at the surface~$A$ for the complete determination of~$V'$,
since it will be proved hereafter, that when they are satisfied we shall have
\[
V'=-\int(\rho)d\sigma \overline{V'};
\]
the integral, as before, extending over the whole surface~$A$,
and $(\rho)$ being a
quantity dependant upon the respective position of $p$ and~$d\sigma$.

It only remains therefore to find a function $V'$ which satisfies the partial
differential equation, becomes equal to~$\overline{V'}$
when $p$ is upon the surface~$A$, vanishes
when $p$ is at an infinite distance from~$A$, and is besides such, that none of
its differential co-efficients shall be infinite,
when the point $p$ is exterior to~$A$.

All those to whom the practice of analysis is familiar, will readily perceive
that the problem just mentioned,
is far less difficult than the direct resolution
of the equation~($a$),
and therefore the solution of the question originally proposed
has been rendered much easier by what has preceded. The peculiar
consideration relative to the differential co-efficients
of $V$ and~$V'$, by restricting
the generality of the integral of the partial differential equation,
so that it can
in fact contain only one arbitrary function,
in the place of two which it ought
otherwise to have contained, and, which has thus enabled us to effect the
simplification in question, seems worthy of the attention of analysts, and may
be of use in other researches where equations of this nature are employed.

We will now give a brief account of what is contained in the following
Essay. The first seven articles are employed in demonstrating some
very general relations existing between the density of the electricity on 
surfaces and in solids, and the corresponding potential functions. These serve
as a foundation to the more particular applications which follow them. As it
would be difficult to give any idea of this part without employing analytical
symbols, we shall content ourselves with remarking, that it contains a number
of singular equations of great generality and simplicity, which seem capable of
being applied to many departments of the electrical theory besides those 
considered in the following pages.

In the eighth article we have determined the general values of the
densities of the electricity on the inner and outer surfaces of an insulated
electrical jar, when, for greater generality, these surfaces are supposed to be
connected with separate conductors charged in any way whatever; and have
proved, that for the same jar, they depend solely on the difference existing
between the two constant quantities, which express the values of the potential
functions within the respective conductors. Afterwards, from these general
values the following consequences have been deduced: ---

When in an insulated electrical jar we consider only the electricity
accumulated on the two surfaces of the glass itself, the total quantity on the
inner surface is precisely equal to that on the outer surface, and of a contrary
sign, notwithstanding the great accumulation of electricity on each of them:
so that if a communication were established between the two sides of the jar,
the sum of the quantities of electricity which would manifest themselves on
the two metallic coatings, after the discharge, is exactly equal to that which,
before it had taken place, would have been observed to have existed on the
surfaces of the coatings farthest from the glass,
the only portions then sensible
to the electrometer.

If an electrical jar communicates by means of a long slender wire
with a spherical conductor, and is charged in the ordinary way, the density
of the electricity at any point of the interior surface of the jar, is to the
density on the conductor itself, as the radius of the spherical conductor to the
thickness of the glass in that point.

The total quantity of electricity contained in the interior of any number
of equal and similar jars, when one of them communicates with the prime
conductor and the others are charged by cascade, is precisely equal to that,
which one only would receive, if placed in communication with the same
conductor, its exterior surface being connected with the common reservoir.
This method of charging batteries, therefore, must not be employed when any
great accumulation of electricity is required.

It has been shown by M.\Person{Poisson}, in his first Memoir on Magnetism
(M\'em. de l'Acad. de Sciences, 1821 et 1822), that when an electrified body
is placed in the interior of a hollow spherical conducting shell of uniform
thickness, it will not be acted upon in the slightest degree by any bodies
exterior to the shell, however intensely they may be electrified. In the ninth
article of the present Essay this is proved to be generally true, whatever may
be the form or thickness of the conducting shell.

In the tenth article there will be found some simple equations, by
means of which the density of the electricity induced on a spherical conducting
surface, placed under the influence of any electrical forces whatever, is 
immediately given; and thence the general value of the potential function for
any point either within or without this surface is determined from the arbitrary
value at the surface itself, by the aid of a definite integral. The proportion
in which the electricity will divide itself between two insulated conducting
spheres of different diameters, connected by a very fine wire, is afterwards
considered; and it is proved, that when the radius of one of them is small
compared with the distance between their surfaces, the product of the mean
density of the electricity on either sphere, by the radius of that sphere, and
again by the shortest distance of its surface
from the centre of the other sphere,
will be the same for both. Hence when their distance is very great, the
densities are in the inverse ratio of the radii of the spheres.

When any hollow conducting shell is charged with electricity, the whole
of the fluid is carried to the exterior surface,
without leaving any portion on
the interior one, as may be immediately shown
from the fourth and fifth articles.
In the experimental verification of this,
it is necessary to leave a small orifice
in the shell: it became therefore a problem of some interest to determine the
modification which this alteration would produce. We have, on this account,
terminated the present article,
by investigating the law of the distribution of
electricity on a thin spherical conducting shell,
having a small circular orifice,
and have found that its density is
very nearly constant on the exterior surface,
except in the immediate vicinity of the orifice; and the density at any
point $p$ of the inner surface, is to the constant density on the outer one, as
the product of the diameter of a circle
into the cube of the radius of the orifice,
is to the product of three times the circumference of that circle into the cube
of the distance of $p$ from the centre of the orifice; excepting as before those
points in its immediate vicinity. Hence, if the diameter of the sphere were
twelve inches, and that of the orifice one inch, the density at the point
on the inner surface opposite the centre of the orifice, would be less than
the hundred and thirty thousandth part of the constant density on the exterior
surface.

In the eleventh article some of the effects due to atmospherical electricity
are considered; the subject is not however insisted upon, as the great
variability of the cause which produces them, and the impossibility of measuring
it, gives a degree of vagueness to these determinations.

The form of a conducting body being given, it is in general a problem
of great difficulty,
to determine the law of the distribution of the electric fluid
on its surface: but it is possible to give different forms, of almost every 
imaginable variety of shape, to conducting bodies; such, that the values of the
density of the electricity on their surfaces may be rigorously assignable by
the most simple calulations:
the manner of doing this is explained in the twelfth
article, and two examples of its use are given. In the last, the resulting form
of the conducting body is an oblong spheroid, and the density of the 
electricity on its surface, here found,
agrees with the one long since deduced from
other methods.

Thus far perfect conductors only have been considered. In order to
give an example of the application of theory to bodies which are not so, we
have, in the thirteenth article, supposed the matter of which they are formed
to be endowed with a constant coercive force equal to~$\beta$, and analagous to
friction in its operation,
so that when the resultant of the electric forces acting
upon any one of their elements is less than~$\beta$,
the electrical state of this
element shall remain unchanged;
but, so soon as it begins to exceed~$\beta$, a
change shall ensue. Then imagining a solid of revolution to turn continually
about its axis, and to be subject to a constant electrical force $f$ acting in
parallel right lines, we determine the permanent electrical state at which the
body will ultimately arrive. The result of the analysis is, that in consequence
of the coercive force~$\beta$,
the solid will receive a new polarity, equal to that
which would be induced in it if it were a perfect conductor and acted upon
by the constant force~$\beta$,
directed in lines parallel to one in the body's equator,
making the angle~$90^\circ+\gamma$,
with a plane passing through its axis and parallel
to the direction of $f$: $f$ being supposed resolved into two forces, one in the
direction of the body's axis, the other $b$
directed along the intersection of its
equator with the plane just mentioned,
and $\gamma$ being determined by the equation
\[
\sin\gamma\quad=\quad\frac{\beta}{b}.
\]

In the latter part of the present article the same problem is considered
under a more general point of view, and treated by a different analysis: the
body's progress from the initial, towards that permanent state it was the object
of the former part to determine is exhibited, and the great rapidity of this
progress made evident by an example.

The phenomena which present themselves during the rotation of iron
bodies, subject to the influence of the earth's magnetism, having lately engaged
the attention of experimental philosophers, we have been induced to dwell a
little on the solution of the preceeding problem, since it may serve in some
measure to illustrate what takes place in these cases. Indeed, if there were
any substances in nature whose magnetic powers, like those of iron and nickel,
admit of considerable developement, and in which moreover the coercive force
was, as we have here supposed it, the same for all their elements, the results
of the preceding theory ought scarcely to differ from what would be observed
in bodies formed of such substances, provided no one of their dimensions was
very small, compared with the others. The hypothesis of a constant coercive
force was adopted in this article, in order to simplify the calculations: 
probably, however, this is not exactly the case of nature,
for a bar of the hardest
steel has been shown (I think by Mr.~\Person{Barlow})
to have a very considerable
degree of magnetism induced in it by the earth's action, which appears to
indicate, that although the coercive force
of some of its particles is very great,
there are others in which it is so small as not to be able to resist the feeble
action of the earth. Nevertheless, when iron bodies are turned slowly round
their axes, it would seem that our theory ought not to differ greatly from
observation;
and in particular, it is very probable the angle $\gamma$ might be rendered
sensible to experiment, by sufficiently reducing $b$
the component of the force~$f$.

The remaining articles treat of the theory of magnetism. This theory
is here founded on an hypothesis relative to the constitution of
magnetic bodies,
first proposed by \Person{Coulomb},
and afterwards generally received by philosophers,
in which they are considered as formed of an infinite number of conducting
elements, separated by intervals absolutely impervious to the magnetic fluid,
and by means of the general results contained in the former part of the Essay,
we readily obtain the necessary equations for determining the magnetic state
induced in a body of any form, by the action of exterior magnetic forces.
These equations accord with those M.~\Person{Poisson}
has found by a very different
method. (M\'em. de l'Acad. des Sciences, 1821 et 1822.)

If the body in question be a hollow spherical shell of constant
thickness, the analysis used by \Person{Laplace}
(\Title{M\'ec. Cel.} Liv.~3) is applicable, and
the problem capable of a complete solution, whatever may be the situation
of the centres of the magnetic forces acting upon it. After having given the
general solution, we have supposed the radius of the shell to become infinite,
its thickness remaining unchanged, and have thence deduced formula belonging
to an indefinitely extended plate of uniform thickness. From these it follows,
that when the point~$p$, and the centres of the magnetic forces are situate on
opposite sides of a soft iron plate of great extent,
the total action on~$p$ will
have the same direction as the resultant of all the forces, which would be
exerted on the points $p, p', p'', p'''$ etc.\ in
infinitum if no plate were interposed,
and will be equal to this resultant multiplied by a very small constant
quantity: the points $p, p', p'', p'''$ etc.\ being
all on a right line perpendicular
to the flat surfaces of the plate,
and receding from it so, that the distance between
any two consecutive points may be equal to twice the plate's thickness.

What has just been advanced will be sensibly correct, on the supposition
of the distances between the point~$p$ and the magnetic centres not being
very great, compared with the plate's thickness, for, when these distances
are exceedingly great, the interposition of the plate will make no sensible
alteration in the force with which $p$ is solicited.

When an elongated body, as a steel wire for instance, has, under the
influence of powerful magnets, received a greater degree of magnetism than
it can retain alone, and is afterwards left to itself,
it is said to be magnetized
to saturation. Now if in this state we consider any one of its conducting
elements, the force with which a particle $p$ of magnetism situate within the
element tends to move,
will evidently be precisely equal to its coercive force~$f$,
and in equilibrium with it. Supposing therefore this force to be the same for
every element, it is clear that the degree of magnetism retained by the wire
in a state of saturation, is, on account of its elongated form, exactly the same
as would be induced by the action of a constant force, equal to~$f$, directed
along lines parallel to its axis, if all the elements were perfect conductors;
and consequently, may readily be determined by the general theory. The number
and accuracy of \Person{Coulomb}'s experiments
on cylindric wires magnetized to saturation,
rendered an application of theory to this particular case very desirable,
in order to compare it with experience. We have therefore effected this in the
last article, and the result of the comparison is of the most satisfactory kind.
\bigskip
\Crule
\bigskip
\bigskip
\bigskip

\section{General preliminary results.}
\markboth{General preliminary results.}{1.}
\Section{1.}
The function which represents the sum of all the electric particles
acting on a given point divided by their respective distances from this point,
has the property of giving, in a very simple form, the forces by which it is
solicited, arising from the whole electrified mass. We shall, in what follows,
endeavour to discover some relations between this function, and the density
of the electricity in the mass or masses producing it, and apply the relations
thus obtained, to the theory of electricity.

Firstly, let us consider a body of any form whatever, through which
the electricity is distributed according to any given law, and fixed there, and
let $x',y',z'$, be the rectangular co-ordinates of a particle of this body,
$\rho'$
the density of the electricity in this particle,
so that $dx'dy'dz'$ being the volume of the particle,
$\rho'dx'dy'dz'$ shall be the quantity of electricity it contains:
moreover, let $r'$ be the distance between this particle and a point $p$
exterior to the body, and $V$ represent the sum of all the particles of
electricity divided by their respective distances from this point,
whose co-ordinates are supposed to be $,x,y,z$, then shall we have
\[
r'=\sqrt{\bigl((x'-x)^2+(y-y')^2+(z-z')^2\bigr)},
\]
and
\[
V=\int\frac{\rho'dx'dy'dz'}{r'};
\]
the integral comprehending every particle in the electrified mass under 
consideration.

\Person{Laplace} has shown, in his \Title{M\'ec. Celeste},
that the function $V$ has the
property of satisfying the equation
\[
0=\frac{d^2V}{dx^2}+\frac{d^2V}{dy^2}+\frac{d^2V}{dz^2}
\]
and as this equation will be incessantly recurring in what follows, we shall
write it in the abridged form~${0=\delta V}$;
the symbol $\delta$ being used in no other
sense throughout the whole of this Essay.

In order to prove that $0=\delta V$, we have only to remark, that by
differentiation we immediately obtain~$0=\delta\frac{1}{r'}$,
and consequently each element of~$V$ substituted for~$V$
in the above equation satisfies it; hence the
whole integral (being considered as the sum of all these elements) will also
satisfy it. This reasoning ceases to hold good when the point $p$ is within
the body, for then, the co-efficients of some of the elements which enter
into~$V$ becoming infinite,
it does not therefore necessarily follow that $V$ satisfies the equation
\[
0=\delta V,
\]
although each of its elements, considered separately, may do so.

In order to determine what $\delta V$ becomes for any point within the body,
conceive an exceedingly small sphere whose radius is~$a$ inclosing the point $p$
at the distance $b$ from its centre,
$a$ and $b$ being exceedingly small quantities.
Then, the value of $V$ may be considered as composed of two parts, one due
to the sphere itself, the other due to the whole mass exterior to it: but the
last part evidently becomes equal to zero
when substituted for $V$ in~$\delta V$,
we have therefore only to determine the value of $\delta V$
for the small sphere
itself, which value is known to be
\[
\delta(2\pi a^2\rho-\tfrac23\pi b^2\rho);
\]
$\rho$ being equal to the density within the sphere
and consequently to the value
of $\rho'$ at~$p$. If now $x_\prime, y_\prime, z_\prime,$
be the co-ordinates of the centre of the sphere, we have
\[
b^2=(x_\prime-x)^2+(y_\prime-y)^2+(z_\prime-z)^2,
\]
and consequently
\[
\delta(2\pi a^2\rho-\tfrac23\pi b^2\rho)=-4\pi\rho.
\]
Hence, throughout the interior of the mass
\[
0=\delta V+4\pi\rho;
\]
of which, the equation $0=\delta V$
for any point exterior to the body is a particular case,
seeing that, here~$\rho=0$.

Let now $q$ be any line terminating in the point~$p$, supposed without
the body, then $-(\frac{dV}{dq})=$ the force
tending to impel a particle of positive
electricity in the direction of~$q$,
and tending to increase it. This is evident,
because each of the elements of~$V$ substituted
for~$V$ in~$-(\frac{dV}{dq})$, will give
the force arising from this element
in the direction tending to increase~$q$, and
consequently, $-(\frac{dV}{dq})$ will give the sum
of all the forces due to every element of~$V$,
or the total force acting on~$p$ in the same direction. In order
to show that this will still hold good,
although the point $p$ be within the body;
conceive the value of $V$ to be divided into two parts as before, and moreover
let $p$ be at the surface of the small sphere
or~${b=a}$, then the force exerted
by this small sphere will be expressed by
\[
\tfrac43\pi a\rho \Bigl(\frac{da}{dq}\Bigr);
\]
$da$ being the increment of the radius~$a$,
corresponding to the increment~$dq$
of~$q$, which force evidently vanishes when~${a=0}:$
we need therefore have
regard only to the part due to the mass exterior to the sphere, and this is
evidently equal to
\[
V-\tfrac43\pi a^2\rho.
\]
But as the first differentials of this quantity are the same
as those of~$V$ when
$a$ is made to vanish, it is clear,
that whether the point $p$ be within or without
the mass, the force acting upon it
in the direction of~$q$ increasing, is always
given by~$-(\frac{dV}{dq})$.

Although in what precedes we have spoken of one body only, the
reasoning there employed is general, and will apply equally to a system of
any number of bodies whatever, in those cases even, where there is a finite
quantity of electricity spread over their surfaces, and it is evident that we
shall have for a point $p$ in the interior of any one of these bodies
\[
\tag{1.}
0=\delta V+4\pi\rho.
\]
Moreover, the force tending to increase a line $q$
ending in any point $p$ within
or without the bodies, will be likewise given
by~$-(\frac{dV}{dq})$; the function $V$
representing the sum of all the electric particles in the system divided by
their respective distances from~$p$. As this function, which gives in so simple
a form the values of the forces by which a particle $p$ of electricity, any how
situated, is impelled, will recur very frequently in what follows, we have
ventured to call it the potential function belonging to the system, and it will
evidently be a function of the co-ordinates of the particle $p$ under consideration.
\bigskip

\Section{2.}
\markboth{General preliminary results.}{2.}
It has been long known from experience, that whenever the electric
fluid is in a state of equilibrium in any system whatever of perfectly 
conducting bodies, the whole of the electric fluid
will be carried to the surface
of those bodies, without the smallest portion of electricity remaining in their
interior: but I do not know that this has ever been shown to be a necessary
consequence of the law of electric repulsion, which is found to take place in
nature. This however may be shown to be the case for every imaginable
system of conducting bodies, and is an immediate consequence of what has
preceded. For let $x,y,z,$ be the rectangular co-ordinates of any particle $p$
in the interior of one of the bodies; then will
$-(\frac{dV}{dx})$, be the force with
which $p$ is impelled in the direction of the co-ordinate~$x$,
and tending to
increase it. In the same way $-\frac{dV}{dy}$ and $-\frac{dV}{dz}$
will be the forces in~$y$ and~$z$,
and since the fluid is in equilibrium all these forces are equal to
\Emphasis{zero}: hence
\[
0=\frac{dV}{dx}dx+\frac{dV}{dy}dy+\frac{dV}{dz}dz=dV,
\]
which equation being integrated gives
\[
V \quad=\quad \text{const.}
\]
This value of $V$ being substituted in the equation~(1) of the preceding
number gives
\[
\rho = 0,
\]
and consequently shows, that the density of the electricity at any point in the
interior of any body in the system is equal to \Emphasis{zero}.

The same equation (1) will give the value of $\rho$ the density of the
electricity in the interior of any of the bodies, when there are not perfect
conductors, provided we can ascertain the value of the potential function $V$
in their interior.
\bigskip

\Section{3.}
\markboth{General preliminary results.}{3.}
Before proceeding to make known some relations which exist between
the density of the electric fluid at the surfaces of bodies, and the 
corresponding values of the potential functions
within and without those surfaces, the
electric fluid being confined to them alone, we shall in the first place, lay
down a general theorem which will afterwards be very useful to us. This
theorem may be thus enunciated:

Let $U$ and $V$ be two continuous functions of the rectangular co-ordinates
$x,y,z,$ whose differential co-efficients do not become infinite at any
point within a solid body of any form whatever; then will
\[
\int dx\,dy\,dz\,U\delta V
+\int d\sigma U\Bigl(\frac{dV}{dw}\Bigr)=
\int dx\,dy\,dz\,V\delta U
+\int d\sigma V\Bigl(\frac{dU}{dw}\Bigr);
\]
the triple integrals extending over the whole interior of the body, and those
relative to~$d\sigma$, over its surface,
of which $d\sigma$ represents an element: $dw$ being
an infinitely small line perpendicular to the surface, and measured from this
surface towards the interior of the body.

To prove this let us consider the triple integral
\[
\int dx\,dy\,dz\,
\Bigl\{
  \Bigl(\frac{dV}{dx}\Bigr)\Bigl(\frac{dU}{dx}\Bigr)
  +\Bigl(\frac{dV}{dy}\Bigr)\Bigl(\frac{dU}{dy}\Bigr)
  +\Bigl(\frac{dV}{dz}\Bigr)\Bigl(\frac{dU}{dz}\Bigr)
\Bigr\}.
\]
The method of integration by parts, reduces this to
\begin{multline*}
\int dy\,dz\,V''\frac{dU''}{dx}
-\int dy\,dz\,V'\frac{dU'}{dx}
+\int dx\,dz\,V''\frac{dU''}{dy}
-\int dx\,dz\,V'\frac{dU'}{dy}\\
+\int dx\,dy\,V''\frac{dU''}{dz}
-\int dx\,dy\,V'\frac{dU'}{dz}
-\int dx\,dy\,dz\,V
\Bigl\{\frac{d^2U}{dx^2}+\frac{d^2U}{dy^2}+\frac{d^2U}{dz^2}\Bigr\};
\end{multline*}
the accents over the quantities indicating, as usual, the values of those 
quantities at the limits of the integral,
which in the present case are on the surface
of the body, over whose interior the triple integrals are supposed to extend.

Let us now consider the part $\int dy\,dz\,V''\frac{dU''}{dx}$
due to the greater values
of~$x$. It is easy to see
since $dw$ is every where perpendicular to the surface
of the solid, that if $d\sigma''$ be the element
of this surface corresponding to~$dy\,dz$,
we shall have
\[
dy\,dz = -\frac{dx}{dw}\,d\sigma''
\]
and hence by substitution
\[
\int dy\,dz\,V''\frac{dU''}{dx}
=-\int d\sigma''\frac{dx}{dw}\,V''\frac{dU''}{dx}.
\]
In like manner it is seen, that in the part
$-\int dy\,dz\,V'\frac{dU'}{dx}$ due to the smaller
values of~$x$, we shall have $dy\,dz=+\frac{dx}{dw}\,d\sigma'$,
and consequently
\[
-\int dy\,dz\,V'\frac{dU'}{dx}=
\int d\sigma\frac{dx}{dw}V'\frac{dU'}{dx}.
\]
Then, since the sum of the elements represented by~$d\sigma'$,
together with those
represented by~$d\sigma''$,
constitute the whole surface of the body, we have by
adding these two parts
\[
\int dy\,dz\Bigl(V''\frac{dU''}{dx}-V'\frac{dU'}{dx}\Bigr)=
-\int d\sigma\frac{dx}{dw}V\frac{dU}{dx};
\]
where the integral relative to do is supposed to extend over the whole surface,
and $dx$ to be the increment of~$x$ corresponding to the increment~$dw$.

In precisely the same way we have
\[
\begin{aligned}
\int dx\,dz\Bigl(V''\frac{dU''}{dy}-V'\frac{dU'}{dy}\Bigr)&=
-\int d\sigma\frac{dy}{dw}V\frac{dU}{dy}, \text{~and}\\
\int dx\,dy\Bigl(V''\frac{dU''}{dz}-V'\frac{dU'}{dz}\Bigr)&=
-\int d\sigma\frac{dz}{dw}V\frac{dU}{dz};
\end{aligned}
\]
therefore, the sum of all the double integrals in the expression before given
will be obtained by adding together the three parts just found; we shall
thus have
\[
-\int d\sigma V\Bigl\{
  \frac{dU}{dx}\frac{dx}{dw}
  +\frac{dU}{dy}\frac{dy}{dw}
  +\frac{dU}{dz}\frac{dz}{dw}
\Bigr\}=-\int d\sigma V \frac{dU}{dw};
\]
where $V$ and $\frac{dU}{dw}$ represent the values
at the surface of the body. Hence, the integral
\[
\int dx\,dy\,dz\,
\Bigl\{
  \frac{dV}{dx}\frac{dU}{dx}
  +\frac{dV}{dy}\frac{dU}{dy}
  +\frac{dV}{dz}\frac{dU}{dz}
\Bigr\},
\]
by using the characteristic~$\delta$
in order to abridge the expression, becomes
\[
-\int d\sigma V \frac{dU}{dw}-\int dx\,dy\,dz\,V\,\delta U.
\]

Since the value of the integral just given remains unchanged when
we substitute $V$ in the place of~$U$ and reciprocally,
it is clear, that it will
also be expressed by
\[
-\int d\sigma U \frac{dV}{dw}-\int dx\,dy\,dz\,U\,\delta V.
\]
Hence, if we equate these two expressions of the same quantity, after having
changed their signs, we shall have
\[
\tag{2.}
\int d\sigma V \frac{dU}{dw}+\int dx\,dy\,dz\,V\,\delta U=
\int d\sigma U \frac{dV}{dw}+\int dx\,dy\,dz\,U\,\delta V.
\]
Thus the theorem appears to be completely established, whatever may be the
form of the functions $U$ and~$V$.

In our enunciation of the theorem, we have supposed the differentials
of $U$ and $V$ to be finite within the body under consideration, a condition,
the necessity of which does not appear explicitly in the demonstration, but,
which is understood in the method of integration by parts there employed.

In order to show more clearly the necessity of this condition, we will
now determine the modification which the formula must undergo, when one
of the functions, $U$ for example, becomes infinite within the body; and let us
suppose it to do so in one point $p'$ only: moreover, infinitely near this point
let $U$ be sensibly equal to $\frac1r$;
$r$ being the distance between the point $p'$ and
the element~$dx\,dy\,dz$. Then if we suppose an
infinitely small sphere whose
radius is~$a$ to be described round~$p'$,
it is clear that our theorem is applicable
to the whole of the body exterior to this sphere, and since,
$\delta U=\delta\frac1r=0$
within the sphere, it is evident, the triple integrals may still be supposed to
extend over the whole body, as the greatest error that this supposition can
induce, is a quantity of the order~$a^2$.
Moreover, the part of $\int d\sigma U\frac{dV}{dw}$, due
to the surface of the small sphere is only an infinitely small quantity of the
order~$a$; there only remains therefore to consider,
the part of $\int d\sigma V\frac{dU}{dw}$, due
to this same surface, which, since we have here
$\frac{dU}{dw}=\frac{dU}{dr}=\frac{d\frac1r}{dr}=\frac{-1}{r^2}=\frac{-1}{a^2}$,
becomes
\[
-4\pi V'
\]
when the radius $a$ is supposed to vanish. Thus, the equation~(2.) becomes
\[
\tag{3.}
\int dx\,dy\,dz\,U\,\delta V+\int d\sigma U \frac{dV}{dw}=
\int dx\,dy\,dz\,V\,\delta U+\int d\sigma V \frac{dU}{dw}-4\pi V';
\]
where, as in the former equation, the triple integrals extend over the whole
volume of the body, and those relative to~$d\sigma$, over its exterior surface:
$V'$ being the value of~$V$ at the point~$p'$.

In like manner, if the function $V$ be such, that it becomes infinite for
any point $p''$ within the body, and is moreover,
sensibly equal to~$\frac{1}{r'}$, infinitely
near this point, as $U$ is infinitely near to the point $p'$, it is evident
from what has preceded that we shall have
\[
\tag{3.'}
\begin{aligned}
&\int dx\,dy\,dz\,U\,\delta V+\int d\sigma U \frac{dV}{dw}-4\pi U''\\
=&\int dx\,dy\,dz\,V\,\delta U+\int d\sigma V \frac{dU}{dw}-4\pi V';
\end{aligned}
\]
the integrals being taken as before,
and $U''$ representing the value of~$U$, at
the point $p''$ where $V$ becomes infinite. The same process will evidently
apply, however great may be the number of similar points belonging to the
functions $U$ and~$V$.

For abridgment, we shall in what follows, call those singular values
of a given function, where its differential co-efficients become infinite, and
the condition originally imposed upon $U$ and~$V$ will be expressed by saying,
that neither of them has any singular values within the solid body under
consideration.
\bigskip

\Section{4.}
\markboth{General preliminary results.}{4.}
We will now proceed to determine some relations existing between the
density of the electric fluid at the surface of a body,
and the potential functions
thence arising, within and without this surface. For this,
let $\rho\,d\sigma$ be the
quantity of electricity on an element $d\sigma$
of the surface, and $V$, the value of
the potential function for any point $p$ within it, of which the co-ordinates
are~$x,y,z$. Then, if $V'$ be the value
of this function for any other point $p'$
exterior to this surface, we shall have
\[
V=\int\frac{\rho\,d\sigma}
{\sqrt{(\xi-x)^2+(\eta-y)^2+(\zeta-z)^2}};
\]
$\xi,\eta,\zeta$ being the co-ordinates of $d\sigma$, and
\[
V'=\int\frac{\rho\,d\sigma}
{\sqrt{(\xi-x')^2+(\eta-y')^2+(\zeta-z')^2}}:
\]
the integrals relative to $d\sigma$
extending over the whole surface of the body.

It might appear at first view, that to obtain the value of $V'$ from that
of~$V$, we should merely have to change
$x,y,z,$ into~$x',y',z':$ but, this is
by no means the case; for, the form of
the potential function changes suddenly,
in passing from the space within to that without the surface. Of this, we may
give a very simple example, by supposing the surface to be a sphere whose
radius is~$a$ and centre at the origin of the co-ordinates;
then, if the density $\rho$
be constant, we shall have
\[
V=4\pi\rho a
\quad\text{and}\quad
V'=\frac{4\pi a^2\rho}{\sqrt{x'^2+y'^2+z'^2}};
\]
which are essentially distinct functions.

With respect to the functions $V$ and $V'$ in the general case, it is
clear that each of them will satisfy \Person{Laplace}'s
equation, and consequently
\[
0=\delta V
\quad\text{and}\quad
0=\delta' V':
\]
moreover, neither of them will have singular values; for any point of the
spaces to which they respectively belong, and at the surface itself, we shall
have
\[
\overline{V}=\overline{V'},
\]
the horizontal lines over the quantities indicating that they belong to the 
surface. At an infinite distance from this surface, we shall likewise have
\[
V'=0.
\]

We will now show, that if any two functions whatever are taken,
satisfying these conditions, it will always be in our power to assign one, and
only one value of~$\rho$, which will produce them for corresponding potential
functions. For this we may remark, that the equation~(3) art.~3 being applied
to the space within the body, becomes, by making~$U=\frac1r$,
\[
\int \frac{d\sigma}{r}\biggl(\frac{\overline{dV}}{dw}\biggr)=
\int d\sigma\overline{V}\biggl(\frac{d\frac1r}{dw}\biggr)-4\pi V;
\]
since $U=\frac1r$, has but one singular point, viz.~$p$;
and, we have also $\delta V=0$
and~$\delta\frac1r=0:r$ being the distance
between the point $p$ to which $V$ belongs,
and the element~$d\sigma$.

If now, we conceive a surface inclosing the body at an infinite distance
from it, we shall have, by applying the formula~(2) of the same article
to the space between the surface of the body and this imaginary exterior
surface (seeing that here $\frac1r=U$ has no singular value)
\[
\int \frac{d\sigma}{r}\biggl(\frac{\overline{dV'}}{dw'}\biggr)=
\int d\sigma\overline{V'}\biggl(\frac{d\frac1r}{dw'}\biggr):
\]
since the part due to the infinite surface may be neglected, because $V'$ is
there equal to zero. In this last equation, it is evident that $dw'$ is measured
from the surface, into the exterior space, and hence
\[
\biggl(\frac{d\frac1r}{dw}\biggr)=-\biggl(\frac{d\frac1r}{dw'}\biggr)
\quad\text{i.~e.}\quad
0=\biggl(\frac{d\frac1r}{dw}\biggr)+\biggl(\frac{d\frac1r}{dw'}\biggr)
\]
which equation reduces the sum of the two just given to
\[
\int \frac{d\sigma}{r}\biggl\{
  \biggl(\frac{\overline{dV}}{dw}\biggr)
  +\biggl(\frac{\overline{dV'}}{dw'}\biggr)
\biggr\}=-4\pi V.
\]
In exactly the same way, for the point $p'$ exterior to the surface, we shall
obtain
\[
\int \frac{d\sigma}{r'}\biggl\{
  \biggl(\frac{\overline{dV}}{dw}\biggr)
  +\biggl(\frac{\overline{dV'}}{dw'}\biggr)
\biggr\}=-4\pi V'.
\]
Hence it appears, that there exists a value of~$\rho$,
viz. $\rho=\frac{-1}{4\pi}\{
  (\frac{\overline{dV}}{dw})+(\frac{\overline{dV'}}{dw'})\}$,
which will give $V$ and~$V'$, for the two potential functions,
within and without the surface.

Again, $-(\frac{\overline{dV}}{dw})=$force 
with which a particle of positive electricity $p$,
placed within the surface and infinitely near it,
is impelled in the direction $dw$
perpendicular to this surface,
and directed inwards; and $-(\frac{\overline{dV}}{dw})$ expresses
the force with which a similar particle $p'$ placed without this surface, on
the same normal with~$p$, and also infinitely near it, is impelled outwards in
the direction of this normal: but the sum of these two forces is equal to
double the force that an infinite plane would exert upon~$p$,
supposing it uniformly covered with electricity
of the same density as at the foot of the
normal on which $p$ is; and this last force is easily shown to be expressed
by~$2\pi\rho$, hence by equating
\[
\tag{4.}
4\pi\rho = -\biggl\{
  \frac{\overline{dV}}{dw}+\frac{\overline{dV'}}{dw'}
\biggr\},
\]
and consequently there is only one value of~$\rho$,
which can produce $V$ and $V'$
as corresponding potential functions.

Although in what precedes, we have considered the surface of one
body only, the same arguments apply, how great soever may be their number;
for the potential functions $V$ and $V'$ would still be given by the formulae
\[
V=\int\frac{\rho\,d\sigma}{r}
\quad\text{and}\quad
V'=\int\frac{\rho\,d\sigma}{r'};
\]
the only difference would be, that the integrations must now extend over the
surface of all the bodies, and, that the number of functions
represented by~$V$,
would be equal to the number of the bodies, one for each. In this case, if
there were given a value of $V$ for each body,
together with $V'$ belonging
to the exterior space; and moreover, if these functions satisfied to the above
mentioned conditions, it would always be possible to determine the density
on the surface of each body, so as to produce these values as potential
functions, and there would be but one density, viz.\ that given by
\[
\tag{4'.}
0=4\pi\rho+\frac{\overline{dV}}{dw}+\frac{\overline{dV'}}{dw'}
\]
which could do so:
$\rho$, $\frac{\overline{dV}}{dw}$ and $\frac{\overline{dV'}}{dw'}$
belonging to a point on the surface of
any of these bodies.
\bigskip

\Section{5.}
\markboth{General preliminary results.}{5.}
From what has been before established (art.~3), it is easy to prove,
that when the value of the potential function $V$ is given on any closed 
surface, there is but one function
which can satisfy at the same time the equation
\[
0=\delta V,
\]
and the condition, that $V$ shall have no singular values within this surface.
For the equation~(3) art.~3, becomes by supposing~$\delta U=0$,
\[
\int d\sigma\overline{U}\frac{\overline{dV}}{dw}
=\int d\sigma\overline{V}\frac{\overline{dU}}{dw}-4\pi V'.
\]
In this equation, $U$ is supposed to have only one singular value within the
surface, viz.\ at the point~$p'$, and,
infinitely near to this point, to be sensibly
equal to~$\frac1r$; $r$ being the distance from~$p'$.
If now we had a value of~$U$,
which, besides satisfying the above written conditions, was equal to zero at
the surface itself, we should have~$\overline{U}=0$,
and this equation would become
\[
\tag{5.}
0=\int d\sigma\overline{V}\frac{\overline{dU}}{dw}-4\pi V'.
\]
which shows, that $V'$ the value of $V$ at the point $p'$ is given,
when $\overline{V}$ its
value at the surface is known.

To convince ourselves, that there does exist such a function as we
have supposed $U$ to be; conceive the surface to be a perfect conductor put
in communication with the earth, and a unit of positive electricity to be 
concentrated in the point $p'$;
then the total potential function arising from $p'$ and
from the electricity it will induce upon the surface, will be the required value
of~$U$. For, in consequence of the communication established between the
conducting surface and the earth, the total potential function at this surface
must be constant, and equal to that of the earth itself,
i.~e. to \Emphasis{zero} (seeing
that in this state they form but one conducting body). Taking, therefore,
this total potential function for~$U$,
we have evidently~$0=\overline{U}$, $0=\delta U$, and
$U=\frac1r$ for those parts infinitely near to~$p'$.
As moreover, this function has
no other singular points within the surface, it evidently possesses all the 
properties assigned to~$U$ in the preceding proof.

Again, since we have evidently $U'=0$, for all the space exterior
to the surface, the equation~(4) art.~4 gives
\[
0=4\pi(\rho)+\frac{\overline{dU}}{dw};
\]
where $(\rho)$ is the density
of the electricity induced on the surface, by the
action of a unit of electricity concentrated in the point~$p'$.
Thus, the equation~(5) of this article becomes
\[
\tag{6.}
V'=-\int d\sigma(\rho)\overline{V}.
\]

This equation is remarkable on account of its simplicity and singularity,
seeing that it gives the value of the potential for any point~$p'$, within the
surface, when $\overline{V}$, its value
at the surface itself is known, together with $(\rho)$,
the density that a unit of electricity concentrated in~$p'$ would induce on this
surface, if it conducted electricity perfectly, and were put in communication
with the earth.

Having thus proved, that $V'$ the value of the potential function~$V$, at
any point $p'$ within the surface is given,
provided its value $\overline{V}$ is known at
this surface, we will now show,
that whatever the value of~$\overline{V}$ may be, the
general value of~$V$ deduced from it by the formula just given shall satisfy
the equation
\[
0 = \delta V.
\]
For, the value of $V$ at any point $p$ whose co-ordinates are~$x,y,z,$ deduced
from the assumed value of~$\overline{V}$, by the above written formula, is
\[
4\pi V=\int d\sigma\overline{V}
\Bigl(\frac{dU}{dw}\Bigr),
\]
$U$ being the total potential function within the surface,
arising from a unit of
electricity concentrated in the point~$p$,
and the electricity induced on the surface
itself by its action.
Then, since $\overline{V}$ is evidently independent of~$x,y,z,$ we
immediately deduce
\[
4\pi\delta V=\int d\sigma\overline{V}\delta
\biggl(\frac{\overline{dU}}{dw}\biggr).
\]

Now the general value of $U$ will depend upon the position of the
point $p$ producing it, and upon that of any other point $p'$ whose co-ordinates
are $x',y',z',$ to which it is referred, and will consequently be a function of
the six quantities~$x,y,z,x',y',z'$.
But we may conceive $U$ to be divided
into two parts, one $=\frac1r$ ($r$ being the distance~$pp'$)
arising from the electricity
in~$p$, the other, due to the electricity induced
on the surface by the action of~$p$,
and which we shall call~$U_\prime$.
Then since $U_\prime$ has no singular values within
the surface, we may deduce its general value from that at the surface, by a
formula similar to the one just given. Thus
\[
4\pi U_\prime=\int d\sigma\overline{U_\prime}
\biggl(\frac{\overline{dU'}}{dw}\biggr);
\]
where $U'$ is the total potential function, which would be produced by a unit
of electricity in~$p'$, and therefore,
$(\frac{\overline{dU'}}{dw})$ is independent of the co-ordinates
$x,y,z,$ of~$p$, to which $\delta$ refers. Hence
\[
4\pi\delta U_\prime=\int d\sigma
\biggl(\frac{\overline{dU'}}{dw}\biggr)\delta\overline{U_\prime}.
\]
We have before supposed
\[
U=\frac1r+U_\prime,
\]
and as $\delta\frac1r=0$, we immediately obtain
\[
\delta U=\delta U_\prime.
\]
Again, since we have at the surface itself
$0=\overline{U}=\frac{1}{\overline{r}}+\overline{U_\prime}$;
$\overline{r}$ being the
distance between~$p$ and the element~$d\sigma$, we hence deduce
\[
0=\delta\overline{U_\prime};
\]
this substituted in the general value of~$\delta U_\prime$,
before given, there arises~$\delta U_\prime=0$,
and consequently~$0=\delta U$.
The result just obtained being general, and applicable
to any point $p''$ within the surface, gives immediately
\[
0=\delta
\biggl(\frac{\overline{dU}}{dw}\biggr),
\]
and we have by substituting in the equation determining~$\delta V$,
\[
0=\delta V.
\]

In a preceding part of this article, we have obtained the equation
\[
0=4\pi(\rho)+\biggl(\frac{\overline{dU}}{dw}\biggr);
\]
which combined with $0=\delta(\frac{\overline{dU}}{dw})$, gives
\[
0=\delta(\rho)
\]
and therefore the density $(\rho)$ induced on
any element~$d\sigma$, which is evidently
a function of the co-ordinates~$x,y,z,$ of~$p$, is also such a function as will
satisfy the equation~$0=\delta(\rho):$
it is moreover evident, that $(\rho)$ can never become
infinite when $p$ is within the surface.

It now remains to prove, that the formula
\[
V=\frac{1}{4\pi}\int d\sigma\overline{V}
\biggl(\frac{\overline{dU}}{dw}\biggr)
=-\int d\sigma(\rho)\overline{V}.
\]
shall always give $V=\overline{V}$
for any point within the surface and infinitely near
it, whatever may be the assumed value of~$\overline{V}$.

For this, suppose the point $p$ to approach infinitely near the surface;
then it is clear that the value of~$(\rho)$,
the density of the electricity induced
by~$p$, will be insensible,
except for those parts infinitely near to~$p$, and in
these parts it is easy to see,
that the value of $(\rho)$ will be independent of the
form of the surface, and depend only on the distance $p,d\sigma$.
But, we shall
afterwards show (art.~10), that when this surface is a sphere of any radius
whatever, the value of $(\rho)$ is
\[
(\rho)=\frac{-\alpha}{2\pi\cdot f^3};
\]
$\alpha$ being the shortest distance between~$p$ and the surface,
and $f$ representing
the distance~$p,d\sigma$.
This expression will give an idea of the rapidity with
which $(\rho)$ decreases,
in passing from the infinitely small portion of the surface
in the immediate vicinity of~$p$,
to any other part situate at a finite distance
from it, and when substituted in the above written value of~$V$, gives, by
supposing $\alpha$ to vanish,
\[
V=\overline{V}.
\]
It is also evident, that the function $V$, determined by the above written 
formula, will have no singular values within the surface under consideration.

What was before proved, for the space within any closed surface,
may likewise be shown to hold good, for that exterior to a number of closed
surfaces, of any forms whatever, provided we introduce the condition, that $V'$
shall be equal to \Emphasis{zero}
at an infinite distance from these surfaces. For, 
conceive a surface at an infinite distance from those under consideration; then,
what we have before said, may be applied to the whole space within the
infinite surface and exterior to the others; consequently
\[
\tag{5'.}
4\pi V'=\int d\sigma\overline{V'}
\biggl(\frac{\overline{dU}}{dw}\biggr);
\]
where the sign of integration must extend over all the surfaces, (seeing that
the part due to the infinite surface is destroyed by the condition, that $V'$ is
there equal to \Emphasis{zero})
and $dw$ must evidently be measured from the surfaces,
into the exterior space to which~$V'$ now belongs.

The form of the equation (6) remains also unaltered, and
\[
\tag{6'.}
V'=-\int(\rho)d\sigma\overline{V'};
\]
the sign of integration extending over all the surfaces,
and $(\rho)$ being the density
of the electricity which would be induced
on each of the bodies, in presence
of each other, supposing they all communicated with the earth by means of
infinitely thin conducting wires.
\bigskip

\Section{6.}
\markboth{General preliminary results.}{6.}
Let now $A$ be any closed surface, conducting electricity perfectly,
and $p$ a point within it, in which a given quantity of
electricity $Q$ is concentrated,
and suppose this to induce an electrical state in~$A$; then will $V$,
the value of the potential function arising from the surface only, at any other
point~$p'$, also within it,
be such a function of the co-ordinates $p$ and~$p'$, that
we may change the co-ordinates of~$p$,
into those of~$p'$, and reciprocally, without
altering its value. Or, in other words,
the value of the potential function at~$p'$,
due to the surface alone,
when the inducing electricity $Q$ is concentrated in~$p$,
is equal to that which would have place at~$p$,
if the same electricity $Q$ were
concentrated in~$p'$.

For, in consequence of the equilibrium at the surface, we have evidently,
in the first case, when the inducing electricity is concentrated in~$p$,
\[
\frac{Q}{\overline{r}}+\overline{V}=\beta;
\]
$\overline{r}$ being the distance between $p$ and $d\sigma'$
an element of the surface~$A$, and $\beta$
a constant quantity dependant upon the quantity of electricity originally placed
on~$A$. Now the value of~$V$ at $p'$ is
\[
V=-\int(\rho')d\sigma'\overline{V},
\]
by what has been shown (art.~5);
$(\rho)$ being, as in that article, the density
of the electricity which would be induced
on the element $d\sigma'$ by a unit of
electricity in~$p'$, if the surface $A$
were put in communication with the earth.
This equation gives
\[
\delta V=-\int(\rho')d\sigma'\delta\overline{V}=0;
\]
since $\delta\overline{V}=-\delta\frac{Q}{\overline{r}}=0:$
the symbol $\delta$ referring to the co-ordinates $x,y,z,$
of~$p$. But we know that~$0=\delta'V$;
where $\delta'$ refers in a similar way to the
co-ordinates $x',y',z',$ of~$p'$ only. Hence we have simultaneously
\[
0=\delta V \quad\text{and}\quad 0=\delta'V;
\]
where it must be remarked,
that the function $V$ has no singular values, provided
the points $p$ and $p'$ are both situate within the surface~$A$. This being
the case the first equation evidently gives (art.~5)
\[
V=-\int(\rho)d\sigma\overline{V};
\]
$\overline{V}$ being what $V$ would become,
if the inducing point $p$ were carried to do,
$p'$ remaining fixed. Where $\overline{V}$ is
a function of~$x',y',z',$ and~$\xi,\eta,\zeta,$ the
co-ordinates of~$d\sigma$, whereas $(\rho)$
is a function of~$x,y,z,\xi,\eta,\zeta,$ independent
of~$x',y',z'$; hence by the second equation
\[
0=\delta'V=-\int(\rho)d\sigma\delta'\overline{V},
\]
which could not hold generally whatever might be the situation of~$p$, unless
we had
\[
0=\delta\overline{V};
\]
where we must be cautious, not to confound the present value
of~$\overline{V}$, with
that employed at the beginning of this article in proving
the equation~${0=\delta V}$,
which last, having performed its office, will be no longer employed.

The equation $0=\delta'V'$ gives in the same way
\[
V=-\int(\rho')d\sigma'\overline{\overline{V}};
\]
$\overline{\overline{V}}$ being what $\overline{V}$ becomes
by bringing the point~$p'$ to any other element $d\sigma'$
of the surface~$A$. This substituted for~$\overline{V}$
in the expression before given,
there arises
\[
V=+\iint(\rho)(\rho')d\sigma\,d\sigma'\overline{\overline{V}}:
\]
in which double integral, the signs of integration, relative to each of the 
independent elements $d\sigma$ and~$d\sigma'$,
must extend over the whole surface.

If now, we represent by $V'$, the value of the potential function at~$p$
arising from the surface~$A$,
when the electricity $Q$ is concentrated in~$p'$, we
shall evidently have
\[
V=+\int(\rho')(\rho)d\sigma'\,
d\sigma\overline{\stackrel\prime{\overline{V_\prime}}};
\]
where the order of integrations alone is changed,
the limits remaining unaltered:
$\overline{\stackrel\prime{\overline{V_\prime}}}$ being
what $V_\prime$ would become, by first bringing the electrical point~$p'$ to
the surface, and afterward the point $p$ to which $V_\prime$ belongs.
This being done,
it is clear that $\overline{\overline{V}}$ and
$\overline{\stackrel\prime{\overline{V_\prime}}}$
represent but one and the same quantity, seeing
that each of them serves to express the value of the potential function, at
any point of the surface~$A$,
arising from the surface itself, when the electricity
is induced upon it by the action of an electrified point, situate in any other
point of the same surface, and hence we have evidently
\[
V=V_\prime,
\]
as was asserted at the commencement of this article.

It is evident from art.~5, that our preceding arguments will be equally
applicable to the space exterior to the surfaces of any number of conducting
bodies, provided we introduce the condition, that the potential function~$V$,
belonging to this space, shall be equal to zero, when either $p$ or $p'$ shall
remove to an infinite distance from these bodies, which condition will 
evidently be satisfied,
provided all the bodies are originally in a natural state.
Supposing this therefore to be the case, we see that the potential function
belonging to any point $p'$ of the exterior space, arising from the electricity
induced on the surfaces of any number of conducting bodies, by an electrified
point in~$p$, is equal to that which would have place at~$p$, if the electrified
point were removed to~$p'$.

What has been just advanced, being perfectly independent of the number
and magnitude of the conducting bodies, may be applied to the case of an 
infinite number of particles, in each of which the fluid may move freely, but
which are so constituted that it cannot pass from one to another. This is
what is always supposed to take place in the theory of magnetism, and the
present article will be found of great use to us when in the sequel we come
to treat of that theory.
\bigskip

\Section{7.}
\markboth{General preliminary results.}{7.}
These things being established with respect to electrified surfaces; the
general theory of the relations between the density of the electric fluid and
the corresponding potential functions,
when the electricity is disseminated through
the interior of solid bodies as well as over their surfaces, will very readily
flow from what has been proved (art.~1).

For this let $V'$ represent the value of the potential function
at a point~$p'$,
within a solid body of any form, arising from the whole of the electric fluid
contained in it, and $\rho'$ be
the density of the electricity in its interior; $\rho'$ being
a function of the three rectangular coordinates~$x,y,z:$
then if $\rho$ be the density
at the surface of the body, we shall have
\[
V'=\int\frac{dx\,dy\,dz\,\rho'}{r'}+\int\frac{d\sigma\rho}{r};
\]
$r'$ being the distance between the point $p'$
whose co-ordinates are~$x',y',z',$
and that whose co-ordinates are $x,y,z,$
to which $\rho'$ belongs, also $r$ the
distance between $p'$ and~$d\sigma$,
an element of the surface of the body: $V'$ being
evidently a function of~$x',y',z'$. If now $V$ be
what $V'$ becomes by changing
$x',y',z',$ into~$x,y,z,$ it is clear from~(art.~1),
that $\rho'$ will be given by
\[
0=4\pi\rho'+\delta V.
\]
Substituting for $\rho'$,
the value which results from this equation, in that immediately
preceding we obtain
\[
V'=\int\frac{dx\,dy\,dz\,\delta V}{4\pi r'}+\int\frac{d\sigma\rho}{r},
\]
which, by means of the equation (3. art. 3), becomes
\[
\int\frac{d\sigma\rho}{r}=
\frac{1}{4\pi}\biggl\{
  \int d\sigma\overline{V}
  \biggl(\frac{d\frac1r}{dw}\biggr)
  -\int\frac{d\sigma}{r}
  \biggl(\frac{\overline{dV}}{dw}\biggr)
\biggr\};
\]
the horizontal lines over the quantities, indicating that they belong to the
surface itself.

Suppose $V_\prime$ to be the value of the potential function in the space 
exterior to the body, which, by~(art.~5), will depend
on the value of $V$ at the
surface only; and the equation~(2. art.~3), applied to this exterior space,
will give since ${\delta V=0}$ and ${\delta\frac1r=0}$,
\[
  \int d\sigma\overline{V}
  \biggl(\frac{d\frac1r}{dw'}\biggr)
  =\int d\sigma\overline{V}_\prime
  \biggl(\frac{d\frac1r}{dw}\biggr)
  =\int\frac{d\sigma}{r}
  \biggl(\frac{\overline{dV}_\prime}{dw'}\biggr);
\]
where $dw'$ is measured from the surface into the exterior space
to which $V_\prime$
belongs, as $dw$ is, into the interior space. Consequently ${dw=-dw'}$,
and therefore
\[
  \int d\sigma\overline{V}
  \biggl(\frac{d\frac1r}{dw}\biggr)
  =-\int d\sigma\overline{V}
  \biggl(\frac{d\frac1r}{dw'}\biggr)
  =-\int\frac{d\sigma}{r}
  \biggl(\frac{\overline{dV}_\prime}{dw'}\biggr).
\]
Hence the equation determining $\rho$ becomes,
by substituting for $\int d\sigma\overline{V}\frac{d\frac1r}{dw}$
its value just given,
\[
\int\frac{\rho d\sigma}{r}
=\frac{-1}{4\pi} \int\frac{d\sigma}{r}\biggl\{
  \biggl(\frac{\overline{dV}}{dw}\biggr)
  +\biggl(\frac{\overline{dV}_\prime}{dw'}\biggr)
\biggr\}.
\]
an equation which could not subsist generally, unless
\[
\tag{7.}
\rho=\frac{-1}{4\pi}\biggl\{
  \frac{\overline{dV}}{dw}+\frac{\overline{dV}_\prime}{dw'}
\biggr\}.
\]
Thus the whole difficulty is reduced to finding
the value $V_\prime$ of the potential
function exterior to the body.

Although we have considered only one body, it is clear that the same
theory is applicable to any number of bodies,
and that the values of $\rho$ and $\rho'$
will be given by precisely the same formulae, however great that number may
be; $V_\prime$ being the exterior potential function common to all the bodies.

In case the bodies under consideration are all perfect conductors, we
have seen (art.~1), that the whole of the electricity will be carried to their
surfaces, and therefore there is here no place for the application of the theory
contained in this article; but as there are probably no perfectly conducting
bodies in nature, this theory becomes indispensably necessary, if we would
investigate the electrical phenomena in all their generality.

Having in this, and the preceding articles, laid down the most general
principles of the electrical theory, we shall in what follows apply these 
principles to more special cases;
and the necessity of confining this Essay within
a moderate extent, will compel us to limit ourselves to a brief examination of
the more interesting phenomena.
\bigskip
\Crule

\section{Application of the preceding results to the theory of electricity.}
\Section{8.}
\markboth{Application to electricity.}{8.}
The first application we shall make of the foregoing principles, will
be to the theory of the Leyden phial. For this, we will call the inner surface
of the phial $A$, and suppose it to be of any form whatever, plane or curved,
then, $B$ being its outer surface, and $\theta$
the thickness of the glass measured
along a normal to~$A$; $\theta$ will be a
very small quantity, which, for greater
generality, we will suppose to vary in any way, in passing from one point
of the surface~$A$ to another. If now the inner coating of the phial be put
in communication with a conductor~$C$, charged with any quantity of electricity,
and the outer one be also made to communicate with another
conducting body~$C'$,
containing any other quantity of electricity,
it is evident, in consequence of the
communications here established, that the total potential function, arising from
the whole system, will be constant throughout the interior of the inner metallic
coating, and of the body~$C$.
We shall here represent this constant quantity by
\[
\beta.
\]
Moreover, the same potential function within the substance of the outer coating,
and in the interior of the conductor~$C'$,
will be equal to another constant quantity
\[
\beta'.
\]
Then designating by $V$, the value of this function, for the whole of the space
exterior to the conducting bodies of the system, and consequently for that
within the substance of the glass itself; we shall have (art.~4)
\[
\overline{V}=\beta
\quad\text{and}\quad
\overline{\overline{V}}=\beta'.
\]
One horizontal line over any quantity, indicating that it belongs to the inner
surface~$A$; and two showing that it belongs to the outer one~$B$.

At any point of the surface $A$, suppose a normal to it to be drawn,
and let this be the axes of~$\overline{w}$:
then $\overline{w}',\overline{w}''$,
being two other rectangular\footnote{orthogonal -- RS} axes,
which are necessarily in the plane tangent to~$A$ at this point; $V$ may be
considered as a function of~$\overline{w},\overline{w}',\overline{w}'',$
and we shall have by \Person{Taylor}s
theorem, since $\overline{w}'=0$ and
$\overline{w}''=0$ at the axis of~$\overline{w}$
along which $\theta$ is measured,
\[
\overline{\overline{V}}=\overline{V}
+\frac{d\overline{V}}{d\overline{w}}\cdot\frac\theta1
+\frac{d^2\overline{V}}{d\overline{w}^2}\cdot\frac{\theta^2}{1\cdot2}
+\text{etc.};
\]
where, on account of the smallness of~$\theta$,
the series converges very rapidly.
By writing in the above, for $\overline{V}$ and
$\overline{\overline{V}}$ their values just given, we obtain
\[
\beta'-\beta=
\frac{d\overline{V}}{d\overline{w}}\cdot\frac\theta1
+\frac{d^2\overline{V}}{d\overline{w}^2}\cdot\frac{\theta^2}{1\cdot2}
+\text{etc.};
\]
In the same way, if $\overline{\overline{w}}$ be a normal to~$B$,
directed towards~$A$, and $\theta_\prime$ be the
thickness of the glass measured along this normal, we shall have
\[
\beta'-\beta=
\frac{d\overline{\overline{V}}}{d\overline{\overline{w}}}
\cdot\frac{\theta_\prime}{1}
+\frac{d^2\overline{\overline{V}}}{d\overline{\overline{w}}^2}
\cdot\frac{\theta_\prime^2}{1\cdot2}
+\text{etc.}.
\]
But, if we neglect quantities of the order~$\theta$,
compared with those retained,
the following equation will evidently hold good,
\[
\frac{d^n\overline{\overline{V}}}{d\overline{\overline{w}}^n}
=(-1)^n\frac{d^n\overline{V}}{d\overline{w}^n};
\]
$n$ being any whole positive number,
the factor $(-1)^n$ being introduced because
$\overline{w}$ and $\overline{\overline{w}}$
are measured in opposite directions. Now by article~4
\[
-4\pi\overline{\rho}=
\frac{d\overline{V}}{d\overline{w}}
\quad\text{and}\quad
-4\pi\overline{\overline{\rho}}=
\frac{d\overline{\overline{V}}}{d\overline{\overline{w}}};
\]
$\overline\rho$ and $\overline{\overline\rho}$
being the densities of the electric fluid at the
surfaces $A$ and~$B$ respectively.
Permitting ourselves, in what follows, to neglect quantities of the
order~$\theta^2$ compared with those retained,
it is clear that we may write $\theta$ for~$\theta_\prime$,
and hence by substitution
\[
\begin{aligned}
\beta-\beta' &=
-4\pi\overline{\rho}\theta+
\biggl(\frac{d^2\overline{V}}{d\overline{w}^2}\biggr)
\frac{\theta^2}{1\cdot2}\\
\beta-\beta' &=
-4\pi\overline{\overline{\rho}}\theta+
\biggl(\frac{d^2\overline{\overline{V}}}{d\overline{\overline{w}}^2}\biggr)
\frac{\theta^2}{1\cdot2};
\end{aligned}
\]
where $V$ and $\rho$ are quantities of the order~$\frac1\theta$;
$\beta'$ and $\beta$ being the ordre~$\theta^0$
or unity. The only thing which now remains to be determined, is the value
of~$\frac{d^2\overline{V}}{d\overline{w}^2}$
for any point on the surface~$A$.

Throughout the substance of the glass, the potential function $V$ will
satisfy the equation ${0=\delta V}$,
and therefore at a point on the surface of~$A$,
where of necessity, $w$, $w'$, and $w''$,
are each equal to zero, we have
\[
0=\frac{d^2\overline{V}}{dw^2}
+\frac{d^2\overline{V}}{dw'^2}
+\frac{d^2\overline{V}}{dw''^2}
=\delta\overline{V};
\]
the horizontal mark over $w$, $w'$ and $w''$ being,
for simplicity, omitted. Then
since~$w'=0$,
\[
\frac{d^2\overline{V}}{dw'^2}
=(V_0-2V_{dw'}+V_{2dw'}):dw'^2,
\]
and as $V$ is constant and equal to $\beta$
at the surface~$A$, there hence arises
\[
V_0=\beta;\quad
V_{dw'}=\beta+\frac{d\overline{V}}{dw}\frac{dw'^2}{2R},\quad
V_{2dw'}=\beta+\frac{d\overline{V}}{dw}\frac{4dw'^2}{2R};
\]
$R$ being the radius of curvature of the surface~$A$,
in the plane~$(w, w')$. Substituting
these values in the expression immediately preceding, we get
\[
\frac{d^2\overline{V}}{dw'^2}
=\frac1R\frac{d\overline{V}}{dw}
=\frac{-4\pi\overline\rho}{R}.
\]
In precisely the same way we obtain, by writing $R'$ for the radius of 
curvature in the plane~$(w, w'')$,
\[
\frac{d^2\overline{V}}{dw''^2}
=\frac{-4\pi\overline\rho}{R'}:
\]
both rays being accounted positive on the side where $w$,
i.~e. $\overline{w}$ is negative.
These values substituted in $0=\delta V$, there results
\[
\frac{d^2\overline{V}}{dw^2}
=4\pi\overline\rho\Bigl(\frac1R+\frac{1}{R'}\Bigr)
\]
for the required value of $\frac{d^2V}{dw^2}$,
and thus the sum of the two equations into
which it enters, yields
\[
\overline\rho\,\Bigl\{1+\Bigl(\frac1R+\frac{1}{R'}\Bigr)\theta\Bigr\}
=-\overline{\overline\rho},
\]
and the difference of the same equations, gives
\[
\beta-\beta'=2\pi(\overline\rho-\overline{\overline\rho})\theta,
\]
therefore the required values of the densities
$\overline\rho$ and $\overline{\overline\rho}$ are
\[
\tag{8.}
\left\{
  \begin{aligned}
  \overline\rho &=
  \frac{\beta-\beta'}{4\pi\theta}
  \,\Bigl\{1+\tfrac12\theta\,\Bigl(\frac1R+\frac{1}{R'}\Bigr)\Bigr\}\\
  \overline{\overline\rho} &=
  \frac{\beta'-\beta}{4\pi\theta}
  \,\Bigl\{1-\tfrac12\theta\,\Bigl(\frac1R+\frac{1}{R'}\Bigr)\Bigr\};
  \end{aligned}
\right.
\]
which values are correct to quantities
of the order $\theta^2\overline\rho$ or, which is the same
thing, to quantities of the order~$\theta$;
these having been neglected in the latter
part of the preceding analysis, as unworthy of notice.

Suppose $d\sigma$ is an element of the surface $A$, the corresponding element
of~$B$, cut off by normals to~$A$, will be
$d\sigma\,\{1+\theta\,(\frac1R+\frac{1}{R'})\}$, and therefore
the quantity of fluid on this last element will be
$\overline{\overline\rho}d\sigma\,\{1+\theta\,(\frac1R+\frac{1}{R'})\}$
substituting for $\overline{\overline\rho}$ its value before found,
$\overline{\overline\rho}=
-\overline\rho\{1-\theta\,(\frac1R+\frac{1}{R'})\}$
and neglecting~$\theta^2\overline\rho$, we obtain
\[
-\overline\rho\,d\sigma.
\]
the same quantity as on the element $d\sigma$
of the first surface. If therefore, we
conceive any portion of the surface~$A$, bounded by a closed curve, and a
corresponding portion of the surface~$B$, which would be cut off by a normal
to~$A$, passing completely round this curve; the sum of the two quantities of
electric fluid, on these corresponding portions, will be equal to zero; and
consequently, in an electrical jar any how charged, the total quantity of
electricity in the jar may be found, by calculating the quantity, on the two
exterior surfaces of the metallic coatings farthest from the glass, as the 
portions of electricity,
on the two surfaces adjacent to the glass, exactly neutralise
each other. This results will appear singular, when we consider the immense
quantity of fluid collected on these last surfaces, and moreover, it would not
be difficult to verify it by experiment.

As a particular example of the use of this general theory: suppose
a spherical conductor whose radius~$a$, to communicate with the inside of an
electrical jar, by means of a long slender wire, the outside being in 
communication with the common reservoir; and let the whole be charged: then $P$
representing the density of the electricity on the surface of the conductor,
which will be very nearly constant, the value of the potential function within
the sphere, and, in consequence of the communication established, at the inner
coating~$A$ also, will be~$4\pi aP$ very nearly, since we may, without sensible
error, neglect the action of the wire and jar itself in calculating it. Hence
\[
\beta=4\pi aP \quad\text{and}\quad \beta'=0,
\]
and the equations (8), by neglecting quantities of the order $\theta$, give
\[
\overline\rho=\frac{\beta}{4\pi\theta}=\frac{a}{\theta}P
\quad\text{and}\quad
\overline{\overline\rho}=\frac{-\beta}{4\pi\theta}=-\frac{a}{\theta}P.
\]
We thus obtain, by the most simple calculation, the values of the densities,
at any point on either of the surfaces~$A$ and~$B$,
next the glass, when that
on the spherical conductor is known.

The theory of the condenser, electrophorous, etc.\ depends upon what
has been proved in this article; but these are details into which the limits of
this Essay will not permit me to enter; there is, however, one result, relative
to charging a number of jars by cascade, that appears worthy of notice,
and which flows so readily from the equations~(8), that I cannot refrain from
introducing it here.

Conceive any number of equal and similar insulated Leyden phials, of
uniform thickness, so disposed, that the exterior coating of the first, may 
communicate with the interior one of the second; the exterior one of the second,
with the interior one of the third; and so on throughout the whole series, to
the exterior surface of the last,
which we will suppose in communication with the
earth. Then, if the interior of the first phial, be made to communicate with
the prime conductor of an electrical machine,
in a state of action, all the phials
will receive a certain charge, and this mode of operating is called charging
\Emphasis{by cascade}. Permitting ourselves
to neglect the small quantities of free fluid on
the exterior surfaces of the metallic coatings, and other quantities of the same
order, we may readily determine
the electrical state of each phial in the series:
for thus, the equations~(8) become
\[
\overline\rho=\frac{\beta-\beta'}{4\pi\theta},\quad
\overline{\overline\rho}=\frac{\beta'-\beta}{4\pi\theta}.
\]
Designating now, by an index at the foot of any letter, the number of the
phial to which it belongs, so that,
$\overline\rho_1$ may belong to the first,
$\overline\rho_2$ to the second
phial, and so on; we shall have, by supposing their whole number to be~$n$,
since $\theta$ is the same for every one,
\begin{align*}
\overline\rho_1 &=\frac{\beta_1-\beta_1'}{4\pi\theta} &
\overline{\overline\rho}_1 &=\frac{\beta'_1-\beta_1}{4\pi\theta}\\
\overline\rho_2 &=\frac{\beta_2-\beta_2'}{4\pi\theta} &
\overline{\overline\rho}_2 &=\frac{\beta'_2-\beta_2}{4\pi\theta}\\
&\text{etc.} &&\text{etc.}\\
\overline\rho_n &=\frac{\beta_n-\beta_n'}{4\pi\theta} &
\overline{\overline\rho}_n &=\frac{\beta'_n-\beta_n}{4\pi\theta}
\end{align*}

Now $\beta$ represents the value of the total potential function, within the
prime conductor and interior coating of the first phial, and in consequence of
the communications established in this system, we have in regular succession,
beginning with the prime conductor, and ending with the exterior surface of
the last phial, which communicates with the earth,
\[
\beta=\beta_1;
\ \beta_1'=\beta_2;
\ \beta_2'=\beta_3;
\ \text{etc.\ \ldots}
\ \beta_{n-1}'=\beta_n;
\ \beta_n'=0
\]
\[
0=\overline{\overline\rho}_1+\overline\rho_2;
\quad 0=\overline{\overline\rho}_2+\overline\rho_3;
\quad\text{etc.\ \ldots}
\quad 0=\overline{\overline\rho}_{n-1}+\overline\rho_n.
\]
But the first system of equations gives
$0=\overline\rho_s+\overline{\overline\rho}_s$,
whatever whole number $s$
may be, and the second line of that just exhibited is expressed by
$0=\overline\rho_{s-1}+\overline{\overline\rho}_s$,
hence by comparing these two last equations
\[
\overline\rho_s=\overline\rho_{s-1},
\]
which shows that every phial of the system is equally charged. Moreover,
if we sum up vertically, each of the columns of the first system, there will
arise in virtue of the second
\[
\begin{aligned}
\overline\rho_1+\overline\rho_2+\overline\rho_3\cdots\cdots
+\overline\rho_n
&=\frac{\beta}{4\pi\theta}\\
\overline{\overline\rho}_1+\overline{\overline\rho}_2
+\overline{\overline\rho}_3\cdots\cdots
+\overline{\overline\rho}_n
&=\frac{\beta}{4\pi\theta}.
\end{aligned}
\]
We therefore see, that the total charge of all the phials is precisely
the same, as that which one only would receive, if placed in communication
with the same conductor, provided its exterior coating were connected with
the earth. Hence this mode of charging, although it may save time, will
never produce a greater accumulation of fluid, than would take place, if one
phial only were employed.
\bigskip

\Section{9.}
\markboth{Application to electricity.}{9.}
Conceive now, a hollow shell of perfectly conducting matter, of any
form and thickness whatever, to be acted upon by any electrified bodies,
situate without it; and suppose them to induce an electrical state in the shell;
then will this induced state be such, that the total action on an electrified
particle, placed any where within it, will be absolutely null.

For let $V$ represent the value of the total potential function, at any
point $p$ within the shell, then we shall have at its inner surface, which is a
closed one,
\[
\overline{V}=\beta;
\]
$\beta$ being the constant quantity, which expresses the value of the potential
function, within the substance of the shell, where the electricity is, by the
supposition, in equilibrium, in virtue of the actions of the exterior bodies,
combined with that arising from the electricity induced in the shell itself.
Moreover, $V$ evidently satisfies the equation $0=\delta V$,
and has no singular
value within the closed surface to which it belongs: it follows therefore, from
art.~5, that its general value is
\[
V=\beta,
\]
and as the forces acting upon $p$, are given by the differentials of~$V$, these
forces are evidently all equal to \Emphasis{zero}.

If, on the contrary, the electrified bodies are all within the shell, and
its exterior surface is put in communication with the earth, it is equally easy
to prove, that there will not be the slightest action on any electrified point
exterior to it; but, the action of the electricity induced on its inner surface,
by the electrified bodies within it, will exactly balance the direct action of
the bodies themselves. Or more generally:

Suppose we have a hollow, and perfectly conducting shell, bounded by any
two closed surfaces,
and a number of electrical bodies are placed, some within and
some without it, at will;
then, if the inner surface and interior bodies be called
the interior system; also, the outer surface and exterior bodies the exterior
system;
all the electrical phenomena of the interior system, relative to attractions,
repulsions, and densities, will be the same as would take place if there were
no exterior system, and the inner surface were a perfect conductor, put in
communication with the earth; and all those of the exterior system will be
the same, as if the interior one did not exist, and the outer surface were a
perfect conductor, containing a quantity of electricity, equal to the whole of
that originally contained in the shell itself, and in all the interior bodies.

This is so direct a consequence of what has been shown in articles
4 and~5, that a formal demonstration would be quite superfluous, as it is easy
to see, the only difference which could exist, relative to the interior system,
between the case where there is an exterior system, and where there is not
one, would be in the addition of a constant quantity, to the total potential
function within the exterior surface, which constant quantity must necessarily
disappear in the differentials of this function, and consequently, in the values
of the attractions, repulsions, and densities, which all depend on these 
differentials alone.
ln the exterior system there is not even this difference, but
the total potential function
exterior to the inner surface is precisely the same,
whether we suppose the interior system to exist or not.
\bigskip

\Section{10.}
\markboth{Application to electricity.}{10.}
The consideration of the electrical phenomena, which arise from spheres
variously arranged, is rather interesting, on account of the case with which
all the results obtained from theory, may be put to the test of experiment;
but, the complete solution of the simple case of two spheres only, previously
electrified, and put in presence of each other, requires the aid of a profound
analysis, and has been most ably treated by M.~\Person{Poisson}
(M\'em. de l'Institut.~1811).
Our object, in the present article, is merely to give one or two examples of
determinations,
relative to the distribution of electricity on spheres, which may
be expressed by very simple formulae.

Suppose a spherical surface whose radius is $a$, to be covered with
electric matter, and let its variable density be represented by~$\rho$;
then if, as
in the \Title{M\'ec. C\'eleste},
we expand the potential function $V$, belonging to a
point $p$ within the sphere, in the form
\[
V=U^{(0)}+U^{(1)}\frac{r}{a}
+U^{(2)}\frac{r^2}{a^2}
+U^{(3)}\frac{r^3}{a^3}
+\text{etc.};
\]
$r$ being the distance between $p$ and the centre of the sphere,
and $U^{(0)},U^{(1)}$
etc.\ functions of the two other polar co-ordinates of~$p$, it is clear, by
what has been shown in the admirable work just mentioned, that the potential
function~$V'$, arising from the same spherical surface, and belonging to a
point~$p'$, exterior to this surface,
at the distance $r'$ from its centre, and on
the radius $r$ produced, will be
\[
V'=U^{(0)}\frac{a}{r'}
+U^{(1)}\frac{a^2}{r'^2}
+U^{(2)}\frac{a^3}{r'^3}
+\text{etc.}
\]
If, therefore, we make $V=\phi(r)$, and $V'=\psi(r')$,
the two functions $\phi$ and~$\psi$
will satisfy the equation
\[
\psi(r)=\frac{a}{r}\phi\Bigl(\frac{a^2}{r}\Bigr)
\quad\text{or}\quad
\phi(r)=\frac{a}{r}\psi\Bigl(\frac{a^2}{r}\Bigr).
\]

But (art. 4)
\[
4\pi\varrho=
-\frac{d\overline{V}}{dw}-\frac{d\overline{V'}}{dw'}
=+\frac{d\overline{V}}{dr}-\frac{d\overline{V'}}{dr'}
=\phi'(a)-\psi'(a),
\]
and the equation between $\phi$ and $\psi$,
in its first form, gives, by differentiation,
\[
\psi'(r)=
-\frac{a}{r^2}\phi\Bigl(\frac{a^2}{r}\Bigr)
-\frac{a^3}{r^3}\phi'\Bigl(\frac{a^2}{r}\Bigr).
\]
Making now $r=a$ there arises
\[
\psi'(a)=-\frac{\phi(a)}{a}-\phi'(a);
\]
$\phi'$ and $\psi'$ being the characteristics of
the differential co-efficients of~$\phi$ and~$\psi$,
according to \Person{Lagrange}'s notation.

In the same way the equation in its second form yields
\[
\phi'(a)=-\frac{\psi(a)}{a}-\psi'(a);
\]
These substituted successively, in the equation
by which $\rho$ is determined, we
have the following
\[
\tag{9.}
\left\{\begin{aligned}
4\pi\rho&=2\phi'(a)+\frac{\phi(a)}{a}
=2\frac{d\overline{V}}{dr}+\frac{\overline{V}}{a}\\
4\pi\rho&=-2\psi'(a)-\frac{\psi(a)}{a}
=-2\frac{d\overline{V'}}{dr'}-\frac{\overline{V'}}{a}.
\end{aligned}\right.
\]
If, therefore, the value of the potential function be known, either for the
space within the surface, or, for that without it,
the value of the density $\rho$
will be immediately given, by one or other of these equations.

From what has preceded, we may readily determine how the electric
fluid will distribute itself, in a conducting sphere whose radius is~$a$, when
acted upon by any bodies situate without it;
the electrical state of these bodies
being given. In this case, we have immediately the value of the potential
function arising from them. Let this value, for any point $p$ within the sphere,
be represented by~$A$; $A$~being a function of the radius~$r$, and two other
polar co-ordinates. Then the whole of the electricity will be carried to the
surface~(art.~1), and if $V$ be the potential function
arising from this electrified
surface, for the same point~$p$, we shall have,
in virtue of the equilibrium
within the sphere,
\[
V+A=\beta
\quad\text{or}\quad
V=\beta-A
\]
$\beta$ being a constant quantity.
This value of $V$ being substituted in the first
of the equations (9), there results
\[
4\pi\rho=-2\frac{\overline{dA}}{dr}-\frac{\overline{A}}{a}+\frac{\beta}{a}:
\]
the horizontal lines indicating, as before,
that the quantities under them belong
to the surface itself.

In case the sphere communicates with the earth, $\beta$ is evidently equal
to \Emphasis{zero}, and $\rho$ is completely determined by the above:
but if the sphere is
insulated, and contains any quantity $Q$ of electricity,
the value of $\beta$ may be
ascertained as follows: Let $V'$ be the value
of the potential function without
the surface, corresponding to the value~${V=\beta-A}$ within it;
then, by what precedes
\[
V'=\frac{\beta}{r'}-A';
\]
$A'$ being determined from $A$ by the following equations:
\[
A=\phi_\prime(r),\quad
\psi_\prime(r)=\frac{a}{r}\phi_\prime\Bigl(\frac{a^2}{r}\Bigr),\quad
A'=\psi_\prime(r'),
\]
and $r'$, being the radius corresponding to the point~$p'$,
exterior to the sphere,
to which~$A'$ belongs. When $r'$ is infinite,
we have evidently~${V'=\frac{Q}{r'}}$.
Therefore by equating
\[
\frac{Q}{r'}=\frac{\beta}{r'}-A'
\quad\text{or}\quad
\beta=Q+r'A';
\]
$r'$ being made infinite. Having thus the value of~$\beta$,
the value of $\rho$ becomes known.

To give an example of the application of the second equation in~$\rho$;
let us suppose a spherical conducting surface, whose radius is~$a$, in 
communication with the earth,
to be acted upon by any bodies situate within it, and
$B'$ to be the value of the potential function arising from them,
for a point $p'$
exterior to it. The total potential function, arising from the interior bodies
and surface itself, will evidently be equal to zero at this surface, and 
consequently~(art.~5), at any point exterior to it.
Hence $V'+B'=0$; $V'$ being
due to the surface. Thus the second of the equations (9) becomes
\[
4\pi\rho=2\frac{\overline{dB'}}{dr'}+\frac{\overline{B'}}{a}.
\]
We are therefore able, by means of this very simple equation, to determine
the density of the electricity induced on the surface in question.

Suppose now, all the interior bodies to reduce themselves to a single
point~$P$, in which a unit of electricity is concentrated, and $f$ to be the
distance~$Pp'$: the potential function arising from~$P$
will be~$\frac1f$ and hence
\[
B'=\frac1f;
\]
$r'$ being, as before, the distance between $p'$ and the centre~$O$
of the shell.
Let now $b$ represent the distance~$OP$,
and $\theta$ the angle~$POp'$, then will
$f^2=b^2-2br'\cdot\cos\theta+r'^2$.
From which equation we deduce successively,
\[
\Bigl(\frac{df}{dr'}\Bigr)=\frac{r'-b\cos\theta}{f},
\quad\text{and}\quad
2\frac{dB'}{dr'}=-\frac{2}{f^2}\Bigl(\frac{df}{dr'}\Bigr)
=\frac{-2r'+2b\cdot\cos\theta}{f^3}
\]
Making $r'=a$ in this, and in the value of $B'$ before given, in order to obtain those which belong to the surface, there results
\[
2\frac{\overline{dB'}}{dr'}+\frac{\overline{B'}}{a}=
\frac{-2a^2+2ab\cdot\cos\theta+f^2}{f^3}=
\frac{b^2-a^2}{af^3}.
\]
This substituted in the general equation written above, there arises
\[
\rho=\frac{b^2-a^2}{4\pi af^3}.
\]
If $P$ is supposed to approach infinitely near to the surface,
so that~${b=a-\alpha}$;
$\alpha$~being an infinitely small quantity, this would become
\[
\rho=\frac{-\alpha}{2\pi f^3}.
\]

In the same way, by the aid of the equation between $A$ and~$p$, the
density of the electric fluid, induced on the surface of a sphere whose radius
is~$a$, when the electrified point $P$ is exterior to it, is found to be
\[
\rho=\frac{b^2-a^2}{4\pi af^3};
\]
supposing the sphere to communicate, by means of an infinitely fine wire,
with the earth, at so great a distance, that we might neglect the influence
of the electricity induced upon it by the action of~$P$. If the distance of~$P$
from the surface,
be equal to an infinitely small quantity~$\alpha$, we shall have in
this case, as in the foregoing,
\[
\rho=\frac{-\alpha}{2\pi f^3}.
\]

From what has preceded, we may readily deduce the general value
of~$V$, belonging to any point~$P$, within the sphere,
when $\overline{V}$ its value at the
surface is known. For $(\rho)$, the density induced upon
an element $d\sigma$ of the
surface, by a unit of electricity concentrated in~$P$, has just been shown to be
\[
\frac{b^2-a^2}{4\pi af^3};
\]
$f$ being the distance $P,d\sigma$.
This substituted in the general equation~(6),
art.~5, gives
\[
\tag{10.}
V=-\int d\sigma(\rho)\overline{V}
=\frac{a^2-b^2}{4\pi a}\int\frac{d\sigma}{f^3}\,\overline{V}.
\]
In the same way we shall have, when the point $P$ is exterior to the sphere,
\[
\tag{11.}
V=\frac{b^2-a^2}{4\pi a}\int\frac{d\sigma}{f^3}\,\overline{V}.
\]
The use of these two equations will appear almost immediately, when we
come to determine the distribution of the electric fluid, on a thin spherical
shell, perforated with a small circular orifice.

The results just given, may be readily obtained by means of \Person{Laplace}'s
much admired analysis
(\Title{M\'ec. C\'el.} Liv.~3, Ch.~2), and indeed, our general
equations~(9), flow very easily from the equation~(2) art.~10 of that chapter.
Want of room compels me to omit these confirmations of our analysis, and
this I do the more freely, as the manner of deducing them must immediately
occur, to any one who has read this part of the \Title{M\'ecanique C\'eleste}.

Conceive now, two spheres $S$ and $S'$, whose radii are $a$ and~$a'$, to
communicate with each other by means of an infinitely fine wire: it is required
to determine the ratio of the quantities of electric fluid on these spheres,
when in a state of equilibrium; supposing the distance of their centres to be
represented by~$b$.

The value of the potential function, arising from the electricity on the
surface of~$S$, at a point~$p$, placed in its centre, is
\[
\int\frac{\rho\,d\sigma}{a}=\frac1a\int\rho\,d\sigma=\frac{Q}{a};
\]
$d\sigma$ being an element of the surface of the sphere,
$\rho$ the density of the fluid
on this element, and $Q$ the total quantity on the sphere. If now, we
represent by $F'$, the value of the potential function for the same point~$p$,
arising from~$S'$, we shall have, by adding together both parts,
\[
F'+\frac{Q}{a};
\]
the value of the total potential function belonging to $p$, the centre of~$S$.
In
like manner, the value of this function at $p'$, the centre of~$S'$, will be
\[
F+\frac{Q'}{a'}:
\]
$F$ being the part arising from $S$,
and $Q'$ the total quantity of electricity on~$S'$.
But in consequence of the equilibrium of the system,
the total potential function
throughout its whole interior is a constant quantity. Hence
\[
F'+\frac{Q}{a}
=F+\frac{Q'}{a'}.
\]

Although it is difficult to assign the rigorous values of $F$ and~$F'$;
yet, when the distance between the surfaces of the two spheres is considerable,
compared with the radius of one of them, it is easy to see, that
$F$ and $F'$ will be very nearly the same, as if the electricity on each of the
spheres producing them, was concentrated in their respective centres, and
therefore, we have very nearly
\[
F=\frac{Q}{b}
\quad\text{and}\quad
F'=\frac{Q'}{b}.
\]
These substituted in the above, there arises
\[
\frac{Q}{b}+\frac{Q'}{a'}=\frac{Q'}{b}+\frac{Q}{a}
\quad\text{i.e.}\quad
Q\Bigl(\frac1a-\frac1b\Bigr)=Q'\Bigl(\frac1{a'}-\frac1b\Bigr).
\]
Thus the ratio of $Q$ to $Q'$ is given by a very simple equation, whatever
may be the form of the connecting wire, provided it be a very fine one.

If we wished to put this result of calculation to the test of experiment,
it would be more simple to write $P$ and $P'$ for the mean densities of the
fluid on the spheres, or those which would be observed when, after being
connected as above, they were separated to such a distance, as not to influence
each other sensibly. Then since
\[
Q=4\pi a^2P \quad\text{and}\quad Q'=4\pi a'^2P',
\]
we have by substitution, etc.
\[
\frac{P}{P'}=\frac{a(b-a)}{a'(b-a')}.
\]
We therefore see, that when the distance $b$ between the centres of the spheres
is very great, the mean densities will be inversely as the radii; and these
last remaining unchanged, the density on the smaller sphere will decrease,
and that on the larger increase in a very simple way, by making them approach
each other.

Lastly, let us endeavour to determine the law of the distribution of
the electric fluid, when in equilibrium on a very thin spherical shell, in which
there is a small circular orifice. Then, if we neglect quantities of the order
of the thickness of the shell, compared with its radius, we may consider it
as an infinitely thin spherical surface, of which the greater segment $S$ is a
perfect conductor, and the smaller one $s$ constitutes the circular orifice. In
virtue of the equilibrium,
the value of the potential function, on the conducting
segment, will be equal to a constant quantity, as~$F$, and if there were no
orifice, the corresponding value of the density would be
\[
\frac{F}{4\pi a};
\]
$a$ being the radius of the spherical surface, Moreover on this supposition,
the value of the potential function for any point $P$,
within the surface, would be
\[
F.
\]
Let therefore, $\frac{F}{4\pi a}+\rho$ represent
the general value of the density, at any point
on the surface of either segment of the sphere,
and $F+V$, that of the corresponding
potential function for the point~$P$. The value of the potential
function for any point on the surface of the sphere,
will be ${F+\overline{V}}$, which
equated to~$F$, its value on~$S$, gives for the whole of this segment
\[
0=\overline{V}.
\]
Thus the equation (10) of this article becomes
\[
V=\frac{a^2-b^2}{4\pi a}\int\frac{d\sigma}{f^3}\,\overline{V};
\]
the integral extending over the surface of the smaller segment $s$ only, which,
without sensible error, may be considered as a plane.

But, since it is evident, that $\rho$ is the density corresponding to the 
potential function~$V$, we shall have for any point on the segments,
treated as a plane,
\[
\rho=\frac{-1}{2\pi}\,\frac{d\overline{V}}{dw},
\]
as it is easy to see, from what has been before shown (art.~4); $dw$~being
perpendicular to the surface, and directed towards the centre of the sphere;
the horizontal line always serving to indicate quantities belonging to the 
surface. When the point $P$ is very near the plane~$s$,
and $z$ is a perpendicular
from $P$ upon~$s$, $z$~will be a very small quantity, of which the square and
higher powers may be neglected. Thus $b=a-z$, and by substitution
\[
V=\frac{z}{4\pi}\int\frac{d\sigma}{f^3}\,\overline{V};
\]
the integral extending over the surface of the small plane~$s$,
and $f$ being, as before,
the distance~$P,d\sigma$.
Now $\frac{d\overline{V}}{dw}=\frac{d\overline{V}}{dz}$
at the surface of~$s$, and $\frac{z}{f^3}=-\frac{d}{dz}\,\frac1f$; hence
\[
\rho=\frac{-1}{2\pi}\frac{d\overline{V}}{dw}
=\frac{-1}{2\pi}\frac{d\overline{V}}{dz}
=\frac{-1}{4\pi^2}\frac{d}{dz}\int\frac{z\,d\sigma}{f^3}\,\overline{V}
=\frac{1}{4\pi^2}\frac{d^2}{dz^2}\int\frac{d\sigma}{f}\,\overline{V};
\]
provided we suppose $z=0$ at the end of the calculus. Now the density
$\frac{F}{4\pi a}+\rho$,
upon the surface of the orifice~$s$, is equal to zero, and therefore,
we have for the whole of this surface~$\rho=-\frac{F}{4\pi a}$.
Hence by substitution
\[
\tag{12.}
\frac{-F\pi}{a}=
\frac{d^2}{dz^2}\int\frac{d\sigma}{f}\,\overline{V};
\]
the integral extending over the whole of the plane $s$,
of which $d\sigma$ is an
element, and $z$ being supposed equal to zero, after all the operations have
been effected.

It now only remains to determine the value of $V$ from this equation.
For this, let $\beta$ now represent the linear radius of~$s$,
and $y$, the distance
between its centre~$C$ and the foot of the perpendicular~$z$:
then if we conceive
an infinitely thin oblate spheroid, of uniform density, of which the circular
plane $s$ constitutes the equator, the value of the potential function at the
point~$P$, arising from this spheroid, will be
\[
\phi=k\int\frac{d\sigma}{f}\sqrt{\beta^2-\eta^2};
\]
$\eta$ being the distance $d\sigma,C$, and $k$ a constant quantity.
The attraction exerted
by this spheroid, in the direction of the perpendicular~$z$,
will be~$-\frac{d\phi}{dz}$, and
by the known formulae relative to the attractions of homogeneous spheroids,
we have
\[
-\frac{d\phi}{dz}=\frac{3Mz}{\beta^3}(\tan\theta-\theta);
\]
$M$ representing the mass of the spheroid,
and $\theta$ being determined by the
equations
\[
\alpha^2=\tfrac12(z^2+y^2-\beta^2)
+\tfrac12\sqrt{(z^2+y^2-\beta^2)^2+4\beta^2z^2}
\]
\[
\tan\theta=\frac\beta\alpha.
\]
Supposing now $z$ very small, since it is to vanish at the end of the calculus,
and~$y<\beta$, in order that the point $P$ may fall within the limits of~$s$,
we shall
have by neglecting quantities of the order~$z^2$ compared with those retained
\[
\theta=\tfrac12\pi-\frac{z}{\sqrt{\beta^2-y^2}};
\]
and consequently
\[
-\frac{d\phi}{dz}
=\frac{-d}{dz}\,k\int\frac{d\sigma}{f}\sqrt{\beta^2-\eta^2}
=\frac{3M\sqrt{\beta^2-y^2}}{\beta^3}-\frac{3M\pi}{2\beta^3}\,z.
\]
This expression, being differentiated again relative to $z$, gives
\[
\frac{d^2}{dz^2}\,k\int\frac{d\sigma}{f}\sqrt{\beta^2-\eta^2}
=\frac{3M\pi}{2\beta^3}.
\]
But the mass $M$ is given by
\[
M=k\int d\sigma\sqrt{\beta^2-\eta^2}
=2\pi k\int \eta\,d\eta\,\sqrt{\beta^2-\eta^2}
=\frac{2\pi k\beta^3}{3}.
\]
Hence by substitution
\[
\frac{d^2}{dz^2}\,k\int\frac{d\sigma}{f}\sqrt{\beta^2-\eta^2}
=\pi^2k:
\]
which expression is rigorously exact when $z=0$. Comparing this result
with the equation (12) of the present article, we see that
if~$\overline{V}=k\sqrt{\beta^2-\eta^2}$,
the constant quantity $k$ may be always determined, so as to satisfy~(12).
In fact, we have only to make
\[
\pi^2k=\frac{-F\pi}{a}
\quad\text{i.~e.}\quad
k=\frac{-F}{a\pi}.
\]
Having thus the value of $\overline{V}$,
the general value of $V$ is known, since
\begin{multline*}
V=\frac{a^2-b^2}{4\pi a}\int\frac{d\sigma}{f^3}\,\overline{V}
=-\frac{a^2-b^2}{4\pi az}\frac{d}{dz}\int\frac{d\sigma}{f}
\Bigl\{\overline{V}=k\sqrt{\beta^2-\eta^2}\Bigr\}\\
=\frac{a^2-b^2}{4\pi az}\times-\frac{d\phi}{dz}
=\frac{a^2-b^2}{4\pi az}\times\frac{3Mz}{\beta^3}(\tan\theta-\theta)
=-\frac{a^2-b^2}{2\pi a^2}F(\tan\theta-\theta).
\end{multline*}
The value of the potential function, for any point $P$ within the shell,
being~${F+V}$, and that in the interior of the conducting matter of the shell
being constant, in virtue of the equilibrium,
the value $\rho'$ of the density, at
any point on the inner surface of the shell, will be given immediately by the
general formula~(4) art.~4. Thus
\[
\rho'=\frac{-1}{4\pi}\,\frac{d\overline{V}}{dw}
=\frac{1}{4\pi}\,\frac{d\overline{V}}{db}
=\frac{+F}{4\pi^2a}(\tan\theta-\theta):
\]
in which equation, the point $P$ is supposed to be upon
the element $d\sigma'$ of the
interior surface, to which~$\rho'$ belongs.
If now, $R$ be the distance between~$C$,
the centre of the orifice, and~$d\sigma'$,
we shall have~$R^2=y^2+z^2$, and by neglecting
quantities of the order~$\frac{\beta^2}{R^2}$
compared with those retained, we have successively
\[
\alpha=R,\quad
\theta=\frac{\beta}{R}\quad\text{and}\quad
\tan\theta-\theta=\tfrac13\theta^3=\frac{\beta^3}{3R^3}.
\]
Thus the value of $\rho'$ becomes
\[
\rho'=\frac{F}{12\pi^2a}\,\frac{\beta^3}{R^3}.
\]

In the same way, it is easy to show from the equation (11) of this
article, that~$\rho''$,
the value of the density on an element~$d\sigma''$ of the exterior
surface of the shell, corresponding
to the element~$d\sigma'$ of the interior surface,
will be
\[
\rho''=\frac{F}{4\pi a}+\rho',
\]
which, on account of the smallness of $\rho'$
for every part of the surface, except
very near the orifice~$s$,
is sensibly constant and equal to~$\frac{F}{4\pi a}$, therefore
\[
\frac{\rho'}{\rho''}=\frac{\beta^3}{3\pi\cdot R^3}:
\]
which equation shows, how very small the density within the shell is, even
when the orifice is considerable.
\bigskip

\Section{11.}
\markboth{Application to electricity.}{11.}
The determination of the electrical phenomena, which result from long
metallic wires, insulated and suspended in the atmosphere, depends upon the
most simple calculations. As an example,
let us conceive two spheres $A$ and~$B$,
connected by a long slender conducting wire;
then $\rho\,dx\,dy\,dz$ representing the
quantity of electricity in an element~$dx\,dy\,dz$ of the exterior space,
(whether
it results from the ground in the vicinity of the wire having become slightly
electrical, or from a mist, or even a passing cloud,) and $r$ being the distance
of this element from~$A$'s centre;
also $r'$ its distance from~$B$'s, the value of
the potential function at $A$'s centre, arising from the whole exterior space,
will be
\[
\int\frac{\rho\,dx\,dy\,dz}{r},
\]
and the value of the same function at $B$'s centre, will be
\[
\int\frac{\rho\,dx\,dy\,dz}{r'},
\]
the integrals extending over all the space exterior to the conducting system
under consideration.

If now, $Q$ be the total quantity of electricity on $A$'s surface, and $Q'$
that on $B$'s, their radii being~$a$ and~$a'$;
it is clear, the value of the potential
function at $A$'s centre, arising from the system itself, will be
\[
\frac{Q}{a};
\]
seeing that, we may neglect the part due to the wire, on account of its
fineness, and that due to the other sphere, on account of its distance. In a
similar way, the value of the same function at $B$'s~centre, will be found to be
\[
\frac{Q'}{a'}.
\]
But (art.~1), the value of the total potential function must be
constant troughout
the whole interior of the conducting system, and therefore, its value at the
two centres must be equal; hence
\[
\frac{Q}{a}+
\int\frac{\rho\,dx\,dy\,dz}{r}
=\frac{Q'}{a'}+
\int\frac{\rho\,dx\,dy\,dz}{r'}.
\]

Although $\rho$, in the present case, is exceedingly small, the integrals
contained in this equation, may not only be considerable, but very great, since
they are of the second dimension relative to space. The spheres, when at
a great distance from each other, may therefore become highly electrical,
according to the observations of experimental philosophers, and the charge
they will receive in any proposed case may readily be calculated; the value
of $\rho$ being supposed given. When one of the spheres, $B$ for instance, is
connected with the ground, $Q'$~will be equal to \Emphasis{zero},
and consequently
$Q$~immediately given. If, on the contrary, the whole system were insulated and
retained its natural quantity of electricity, we should have, neglecting that
on the wire,
\[
0=Q+Q',
\]
and hence $Q$ and $Q'$ would be known.

If it were required, to determine the electrical state of the sphere~$A$,
when in communication with a wire, of which one extremity is elevated into
the atmosphere, and terminates in a fine point~$p$, we should only have to
make the radius of~$B$, and consequently, $Q'$, vanish in the expression before
given. Hence in this case
\[
\frac{Q}{a}=
\int\frac{\rho\,dx\,dy\,dz}{r'}
-\int\frac{\rho\,dx\,dy\,dz}{r};
\]
$r'$ being the distance between $p$ and the element $dx\,dy\,dz$.
Since the object
of the present article, is merely to indicate the cause of some phenomena of
atmospherical electricity, it is useless to extend it to a greater length, more
particularly, as the extreme difficulty of determining correctly the electrical
state of the atmosphere at any given time, precludes the possibility of putting
this part of the theory to the test of accurate experiment.
\bigskip

\Section{12.}
\markboth{Application to electricity.}{12.}
Supposing the form of a conducting body to be given, it is in general
impossible to assign, rigorously, the law of the density of the electric fluid
on its surface in a state of equilibrium, when not acted upon by any exterior
bodies, and, at present, there has not even been found any convenient mode
of approximation applicable to this problem. It is, however, extremely easy
to give such forms to conducting bodies, that this law shall be rigorously
assignable by the most simple means. The following method, depending upon
art.~4 and~5, seems to give to these forms the greatest degree of generality
of which they are susceptible, as, by a tentative process, any form whatever
might be approximated indefinitely.

Take any continuous function $V'$, of the rectangular
co-ordinates~$x',y',z'$,
of a point~$p'$,
which satisfies the partial differential equation~${0=\delta V'}$,
and vanishes when $p'$ is removed to an infinite distance from the origin of
the co-ordinates.

Choose a constant quantity $b$, such that $V'=b$ may be the equation
of a closed surface~$A$, and that $V'$ may have no singular values, so long as
$p'$ is exterior to this surface: then if we form a conducting body, whose
outer surface is~$A$, the density of the electric fluid in equilibrium upon it,
will be represented by
\[
\rho=\frac{-h}{4\pi}\,\frac{d\overline{V'}}{dw'}\,,
\]
and the potential function due to this fluid, for any point $p'$,
exterior to the body, will be
\[
hV';
\]
$h$ being a constant quantity dependant upon
the total quantity of electricity~$Q$,
communicated to the body. This is evident from what has been proved in
the articles cited.

Let $R$ represent the distance between $p'$, and any point within~$A$;
then the potential function arising from the electricity upon it, will be 
expressed by~$\frac{Q}{R}$, when $R$ is infinite. Hence the condition
\[
\frac QR=hV'\quad\text{($R$ \Emphasis{being infinite})}
\]
which will serve to determine $h$, when $Q$ is given.

In the application of this general method, we may assume for~$V'$,
either some analytical expression containing the co-ordinates of~$p'$, which is
known to satisfy the equation~${0=\delta V'}$,
and to vanish when $p'$ is removed
to an infinite distance from the origin of the co-ordinates; as, for instance,
some of those given by \Person{Laplace}
(\Title{M\'ec. C\'eleste}, Liv.~3, Ch.~2), or, the value
[of] a potential function,
which would arise from a quantity of electricity any how
distributed within a finite space,
at a point $p'$ without that space; since this
last will always satisfy the conditions to which $V'$ is subject.

It may be proper to give an example of each of these cases. In the
first place, let us take the general expression given by \Person{Laplace},
\[
V=\frac{U^{(0)}}{r}
+\frac{U^{(1)}}{r^2}
+\frac{U^{(2)}}{r^3}
+\text{etc.},
\]
then, by confining ourselves to the two first terms, the assumed value of $V'$
will be
\[
V'=\frac{U^{(0)}}{r}
+\frac{U^{(1)}}{r^2};
\]
$r$ being the distance of $p'$ from the origin of the co-ordinates,
and $U^{(0)}$, $U^{(1)}$, etc.\ functions
of the two other polar co-ordinates~$\theta$ and~$\varpi$. This expression
by changing the direction of the axes, may always be reduced to
the form
\[
V'=\frac{2a}{r}+\frac{h^2\cos\theta}{r^2};
\]
$a$ and $k$ being two constant quantities, which we will suppose positive. Then
if $b$ be a very small positive quantity, the form of the surface given by the
equation~${V'=b}$, will differ but little from a sphere,
whose radius is~$\frac{2a}{b}:$
by gradually increasing~$b$,
the difference becomes greater, until~$b=\frac{a^2}{k^2}$; and
afterwards, the form assigned by $V'=b$, becomes improper for our purpose.
Making therefore ${b=\frac{a^2}{k^2}}$,
in order to have a surface differing as much from
a sphere, as the assumed value of~$V'$ admits,
the equation of the surface~$A$ becomes
\[
V'=\frac{2a}{r}+\frac{h^2\cos\theta}{r^2}=\frac{a^2}{k^2}.
\]
From which we obtain
\[
r=\frac{k^2}{a}(1+\sqrt2\cos\tfrac12\theta).
\]
If now $\phi$ represents the angle formed by $dr$ and $dw'$, we have
\[
\frac{-dr}{r\,d\theta}=
\frac{\sqrt2\sin\frac12\theta}{2+2\sqrt2\cos\frac12\theta}
=\tan\phi,
\]
and as the electricity is in equilibrium upon $A$,
the force with which a particle~$p$,
infinitely near to it, would be repelled, must be directed along~$dw'$:
but the value of this force is~$-\frac{d\overline{V'}}{dw'}$,
and consequently its effect in the direction
of the radius~$r$, and tending to increase it,
will be~$-\frac{d\overline{V'}}{dw'}\cos\phi$. This last
quantity is equally represented by~$-\frac{d\overline{V'}}{dr}$, and therefore
\[
-\frac{d\overline{V'}}{dr}
=-\frac{d\overline{V'}}{dw'}\cos\phi;
\]
the horizontal lines over quantities, indicating, as before, that they belong
to the surface itself. The value of $-(\frac{d\overline{V'}}{dw'})$,
deduced from this equation, is
\[
\frac{d\overline{V'}}{dw'}
=\frac{1}{\cos\phi}\,\frac{d\overline{V'}}{dr}
=\frac{1}{\cos\phi}\biggl\{\frac{2a}{r^2}+\frac{2k^2\cos\theta}{r^3}\biggr\}
=\frac{2a\sqrt2\cos\frac12\theta}{r^2\cos\phi},
\]
this substituted in the general value of $\phi$, before given, there arises
\[
\rho=\frac{-h}{4\pi}\,\frac{d\overline{V'}}{dw'}
=\frac{ha\sqrt2\cos\frac12\theta}{2\pi r^2\cos\phi}.
\]
Supposing $Q$ is the quantity of electricity communicated to the surface, the
condition
\[
\frac QR=hV\quad\text{\Emphasis{(where $R$ is infinite)}}
\]
before given, becomes, since $r$ may here be substituted for~$R$, seeing that
it is measured from a point within the surface,
\[
\frac Qr=\frac{2ah}{r}
\quad\text{i.~e.}\quad
h=\frac{Q}{2a}.
\]
We have thus the rigorous value of $\rho$ for the surface $A$ whose equation is
$r=\frac{k^2}{a}(1+\sqrt2\cos\frac12\theta)$
when the quantity~$Q$ of electricity upon it is known,
and by substituting for $r$ and $h$ their values just given, there results
\[
\rho=\frac{Qa^2\sqrt2\cos\frac12\theta}
{4\pi k^4\cos\phi(1+\sqrt2\cos\frac12\theta)^2}.
\]
Moreover the value of the potential function for the point $p'$
whose polar co-ordinates are $r$, $\theta$, and~$\varpi$, is
\[
hV'=\frac Qr+\frac{Qk^2\cos\theta}{2ar^2}.
\]
From which we may immediately deduce the forces acting on any point $p'$
exterior to~$A$.

In tracing the surface $A$,
$\theta$ is supposed to extend from $\theta=0$ to~$\theta=\pi$,
and~$\varpi$, from~$\varpi=0$ to~$\varpi=2\pi:$
it is therefore evident, by constructing the
curve whose equation is
\[
r=\frac{k^2}{a}(1+\sqrt2\cos\frac12\theta),
\]
that the parts about $P$, where $\theta=\pi$,
approximate continually in form towards
a cone whose apex is~$P$,
and as the density of the electricity at~$P$ is null,
in the example before us, we may make this general inference: when any
body whatever, has a part of its surface in the form of a cone, directed
inwards; the density of the electricity in equilibrium upon it,
will be null at its
apex, precisely the reverse of what would take place, if it were directed
outwards, for then, the density at the apex would become
infinite.\footnote{Since this was written,
I have obtained formulae serving to express, generally,
the law of the distribution of the
electric fluid near the apex~$O$ of a cone, which forms
part of a conducting surface of revolution
having the same axis. From these formulae
it results that, when the apex of the cone is
directed inwards, the density of the electric
fluid at any point~$p$, near to it,
is proportional to~$r^{n-1}$; $r$ being the distance~$Op$, and
the exponent $n$ very nearly such as would satisfy
the simple equation~${(4n+2)\beta=3\pi:}$
where $2\beta$ is the angle at the summit of the cone.
If $2\beta$ exceeds~$\pi$, this summit is directed
outwards, and when the excess is not very considerable,
$n$ will be given as above:
but $2\beta$ still increasing,
until it becomes~$2\pi-2\gamma$;
the angle $2\gamma$ at the summit of the cone,
which is now directed outwards, being very small,
$n$ will be given by $2n\log\frac2\gamma=1$,
and in case the conducting body is a sphere whose radius is~$b$,
on which $P$ represents
the mean density of the electric fluid,
$\rho$, the value of the density near the apex~$O$, will
be determined by the formula
\[
\rho=\frac{2Pbn}{(a+b)\gamma}\biggl(\frac ra\biggr)^{n-1};
\]
$a$ being the length of the cone.}

As a second example, we will assume for $V'$, the value of the potential
function arising from the action of a line uniformly covered with electricity.
Let $2a$ be the length of the line, $y$ the perpendicular falling from any
point~$p'$ upon it, $x$ the distance of the foot of this perpendicular from the
middle of the line, and $x'$ that of the element~$dx'$ from the same point: then
taking the element~$dx'$, as the measure of the quantity of electricity it 
contains, the assumed value of~$V'$ will be
\[
V'=\int\frac{dx'}{\sqrt{y^2+(x-x')^2}}
=\log\frac{a-x+\sqrt{y^2+(a-x)^2}}{-a-x+\sqrt{y^2+(a+x)^2}};
\]
the integral being taken from $x'=-a$ to $x'=+a$. Making this equal to
a constant quantity~$\log b$, we shall have,
for the equation of the surface~$A$,
\[
\frac{a-x+\sqrt{y^2+(a-x)^2}}{-a-x+\sqrt{y^2+(a+x)^2}}=b,
\]
which by reduction becomes
\[
0=y^2(1-b^2)^2+x^2\cdot 4b(1-b)^2-a^2\cdot 4b(1+b)^2.
\]
We thus see that this surface is a spheroid produced by the revolution of an
ellipsis about its greatest diameter;
the semi-transverse axis being~${a\frac{1+b}{1-b}=\beta}$
and semi-conjugate~${a\frac{2\sqrt b}{1-b}=\gamma}$.

By differentiating the general value of $V'$, just given, and substituting
for $y$ its value at the surface~$A$, we obtain
\[
\frac{d\overline{V'}}{dx}
=\frac{-2x\frac{1-b}{1+b}}{(\frac{1+b}{1-b})^2a^2-(\frac{1-b}{1+b})^2x^2}
=\frac{-2a\beta x}{\beta^4-a^2x^2}.
\]
Now writing $\phi$ for the angle formed by $dw$ and~$dw'$, we have
\[
\frac{1}{\cos\phi}=\frac{ds}{-dy}
=\frac{1-b}{2x\sqrt b}\sqrt{\biggl(\frac{1+b}{1-b}\biggr)^4a^2-x^2}
=\frac{\sqrt{\beta^4-a^2x^2}}{\gamma x};
\]
$ds$ being an element of the generating ellipsis. Hence, as in the preceding
example, we shall have,
\[
\frac{d\overline{V'}}{dw'}=\frac{1}{\cos\phi}\cdot\frac{d\overline{V'}}{dx}
=\frac{-2a\beta}{\gamma\sqrt{\beta^4-a^2x^2}}.
\]
On the surface $A$ therefore, in this example, the general value of $\rho$ is
\[
\rho=\frac{-h}{4\pi}\,\frac{d\overline{V'}}{dw'}
=\frac{ah\beta}{2\pi\gamma\sqrt{\beta^4-a^2x^2}},
\]
and the potential function for any point $p'$, exterior to $A$, is
\[
hV'=h\log\frac{a-x+\sqrt{y^2+(a-x)^2}}{-a-x+\sqrt{y^2+(a+x)^2}}.
\]
Making now $x$ and $y$ both infinite, in order that $p'$ may be at an infinite
distance, there results
\[
hV'=\frac{2ah}{\sqrt{x^2+y^2}}\,,
\]
and thus the condition determining $h$, in $Q$, the quantity of electricity upon
the surface, is, since $R$ may be supposed equal to~$\sqrt{x^2+y^2}$,
\[
\frac QR=hV'=\frac{2ah}{\sqrt{x^2+y^2}}
\quad\text{i.~e.}\quad
h=\frac{Q}{2a}.
\]
These results of our analysis, agree with what has been long known concerning
the law of the distribution of electric fluid on the surface of a spheroid,
when in a state of equilibrium.
\bigskip

\Section{13.}
\markboth{Application to electricity.}{13.}
In what has preceded, we have confined ourselves to the consideration
of perfect conductors. We will now give an example of the application of
our general method, to a body that is supposed to conduct electricity 
imperfectly, and which will, moreover,
be interesting, as it serves to illustrate the
magnetic phenomena, produced by the rotation of bodies under the influence
of the earth's magnetism.

If any solid body whatever of revolution, turn about its axis, it is
required to determine what will take place, when the matter of this solid is
not perfectly conducting,
supposing it under the influence of a constant electrical
force, acting parallel to any given right line fixed in space, the body being
originally in a natural state.

Let $\beta$ designate the coercive force of the body, which we will suppose
analogous to friction in its operation,
so that as long as the total force acting
upon any particle within the body is less than~$\beta$,
its electrical state shall
remain unchanged, but when it begins to exceed~$\beta$, a change shall ensue.

In the first place, suppose the constant electrical force, which we will
designate by~$b$, to act in a direction parallel to a line passing through the
centre of the body, and perpendicular to its axis of revolution; and let us
consider this line as the axis of~$x$,
that of revolution being the axis of~$z$,
and $y$ the other rectangular co-ordinate of a point~$p$, within the body and
fixed in space. Thus, if $V$ be the value of the total potential function for
the same point~$p$, at any instant of time, arising from the electricity of the
body and the exterior force,
\[
bx+V
\]
will be the part due to the body itself at the same instant; since $-bx$ is
that due to the constant force~$b$, acting in the direction of~$x$, and tending
to increase it. If now we make
\[
z=r\cos\theta,\quad
x=r\sin\theta\cos\varpi,\quad
y=r\sin\theta\sin\varpi;
\]
the angle $\varpi$ being supposed to increase
in the direction of the body's revolution,
the part due to the body itself becomes
\[
br\sin\theta\cos\varpi+V.
\]

Were we to suppose the value of the potential function $V$ given at
any instant, we might find its value at the next instant, by conceiving, that
whilst the body moves forward through the infinitely small angle $d\omega$, the
electricity within it shall remain fixed, and then be permitted to move, until
it is in equilibrium with the coercive force.

Now the value of the potential function at $p$, arising from the body
itself, after having moved through the angle~$d\omega$
(the electricity being fixed),
will evidently be obtained by changing~$\varpi$
into~$\varpi-d\omega$ in the expression just
given, and is therefore
\[
br\sin\theta\cos\varpi+V+br\sin\theta\sin\varpi\,d\omega
-\frac{dV}{d\varpi}\,d\omega,
\]
adding now the part $-bx=-br\sin\theta\cos\varpi$
due to the exterior bodies, and
restoring $x$, $y$, etc.\ we have
since~$\frac{dV}{d\varpi}=-y\frac{dV}{dx}+x\frac{dV}{dy}$,
\[
V+d\omega\biggl\{
  by+y\frac{dV}{dx}-x\frac{dV}{dy}
\biggr\}
\]
for the value of the total potential function at the end of the next instant,
the electricity being still supposed fixed. We have now only to determine
what this will become, by allowing the electricity to move forward until the
total forces acting on points within the body, which may now exceed the
coercive force by an infinitely small quantity, are again reduced to an 
equilibrium with it. If this were done, we should, when the initial state of the
body was given, be able to determine, successively, its state for every one of
the following instants. But since it is evident from the nature of the problem,
that the body, by revolving, will quickly arrive at a permanent state, in
which the value of $V$ will afterwards remain unchanged and be independent
of its initial value,
we will here confine ourselves to the determination of this
permanent state. It is easy to see, by considering the forces arising from
the new total potential function, whose value has just been given, that in this
case the electricity will be in motion over the whole interior of the body, and
consequently
\[
\beta^2=
\biggl(\frac{dV}{dx}\biggr)^2
+\biggl(\frac{dV}{dy}\biggr)^2
+\biggl(\frac{dV}{dz}\biggr)^2,
\]
which equation expresses that the total force to move any particle $p$, within
the body, is just equal to~$\beta$, the coercive force. Now if we can assume any
value for~$V$, satisfying the above,
and such, that it shall reproduce itself after
the electricity belonging to the new total potential function (art.~7),
is allowed
to find its equilibrium with the coercive force, it is evident this will be the
required value, since the rest of the electricity is exactly in equilibrium with
the exterior force~$b$, and may therefore be here neglected. To be able to
do this the more easily, conceive two new axes~$X',Y'$, in advance of the
old ones~$X,Y$, and making the angle~$\gamma$ with them; then the value of the
new potential function, before given, becomes
\[
V+d\omega\biggl\{
  by'\cos\gamma+bx'\sin\gamma
  +y'\frac{dV}{dx'}-x'\frac{dV}{dy'}
\biggr\},
\]
which, by assuming $V=\beta\gamma'$, and determining $\gamma$ by the equation
\[
0=b\sin\gamma-\beta
\]
reduces itself to
\[
y'(\beta+b\cos\gamma\,d\omega).
\]
Considering now the symmetrical distribution of the electricity belonging to
this potential function, with regard to the plane whose equation is~${0=y'}$, it
will [be] evident that,
after the electricity has found its equilibrium, the value of~$V$
at this plane must be equal to \Emphasis{zero}:
a condition which, combined with the
partial differential equation before given, will serve to determine, completely,
the value of~$V$ at the next instant, and this value of $V$ will be
\[
V=\beta y'.
\]
We thus see that the assumed value of $V$ reproduces itself at the end of the
following instant, and is therefore the one required belonging to the permanent
state.

If the body had been a perfect conductor, the value of $V$ would
evidently have been equal to \Emphasis{zero},
seeing that it was supposed originally in
a natural state: that just found is therefore due to the rotation combined with
the coercive force, and we thus see that their effect is to polarise the body
in the direction of~$y'$ positive,
making the angle~${\frac12\pi+\gamma}$ with the direction
of the constant force~$b$; and the degree of polarity will be the same as would
be produced by a force equal to~$\beta$, acting in this direction on a perfectly
conducting body of the same dimensions.

We have hitherto supposed the constant force to act in a direction
parallel to the equatorial plane of the body, but whatever may be its direction,
we may conceive it decomposed into two; one equal to~$b$ as before, and
parallel to this plane, the other perpendicular to it, which last will evidently
produce no effect on the value of~$V$, as this is due to the coercive force,
and would still be equal to zero under the influence of the new force, if the
body conducted electricity perfectly.

Knowing the value of the potential function at the surface of the body,
due to the rotation, its value for all the exterior space may be considered as
determined (art.~5), and if the body be a solid sphere, may easily be expressed
analytically; for it is evident (art.~7), from the value of~$V$ just given, that
even in the present case all the electricity will be confined to the surface of
the solid; and it has been shown (art.~10), that when the value of the 
potential function for the point~$p$
within a spherical surface, whose radius is~$a$,
is represented by
\[
\phi(r),
\]
the value of the same function for a point $p'$, situate without this sphere, on
the prolongation of~$r$, and at the distance $r'$ from its centre, will be
\[
\frac{a}{r'}\phi\biggl(\frac{a^2}{r'}\biggr).
\]
But we have seen that the value of $V$ due to the rotation,
for the point~$p$, is
\[
V=\beta\gamma'=\beta r\cos\theta';
\]
$\theta'$ being the angle formed by the ray $r$ and the axis of~$y'$;
the corresponding
value for the point $p'$ will therefore be
\[
V'=\frac{\beta a^3\cos\theta'}{r'^2}.
\]
And hence, by differentiation, we immediately obtain the value of the forces
acting on any particle situate without the sphere,
which arise from its rotation;
but, if we would determine the total forces arising from the sphere, we must,
to the value of the potential function just found, add that part which would
be produced by the action of the constant force upon this sphere, when it is
supposed to conduct electricity perfectly, which will be given in precisely the
same way as the former. In fact, $f$ designating the constant force,
and $\theta''$
the angle formed by $r$ and a line parallel to the direction of~$f$,
the potential
function arising from it, for the point~$p$, will be
\[
-rf\cos\theta'',
\]
and consequently the part arising from the electricity, induced by its action,
must be
\[
+fr\cos\theta'',
\]
seeing that their sum ought to be equal to zero. The corresponding value
for the point~$p'$, exterior to the sphere, is therefore
\[
\frac{fa^3\cos\theta''}{r'^2}\,,
\]
this added to the value of $V'$, before found, will give the value of the total
potential function for the point~$p'$, arising from the sphere itself.

It will be seen when we come to treat of the theory of magnetism,
that the results of this theory, in general, agree very nearly with those which
would arise from supposing the magnetic fluid at liberty to move from one
part of a magnetized body to another; at least, for bodies whose magnetic
powers admit of considerable developement, as iron and nickel for example;
the errors of the latter supposition being of the order~$1-g$ only; $g$~being
a constant quantity dependant on the nature of the body, which in those just
mentioned, differs very little from unity. It is therefore evident that when
a solid of revolution, formed of iron, is caused to revolve slowly round its
axis, and placed under the influence of the earth's magnetic force~$f$, the act
of revolving, combined with the coercive force~$\beta$ of the body, will produce
a new polarity, whose direction and quantity will be very nearly the same
as those before determined. Now $f$ having been supposed resolved into two
forces, one equal to~$b$ in the plane of the body's equator, and another 
perpendicular to this plane;
if~$\beta$ be very small compared with~$b$, the angle $\gamma$
will be very small, and the direction of the new polarity will be very nearly
at right angles to the direction of~$b$, a result which has been confirmed by
many experiments: but by our analysis we moreover see that when $b$ is 
sufficiently reduced, the angle $\gamma$ may be rendered sensible,
and the direction of
the new polarity will then form with that of~$b$
the angle~$\tfrac12\pi+\gamma$; $\gamma$~being
determined by the equation
\[
\sin\gamma=\frac\beta b.
\]
This would be very easily put to the test of experiment by employing a
solid sphere of iron.

The values of the forces induced by the rotation of the body, which
would be observed in the space exterior to it, may be obtained by 
differentiating that of~$V'$ before given,
and will be found to agree with the observations
of Mr.~\Person{Barlow} (Phil. Tran. 1825),
on the supposition of $\beta$ being
very small.

As the experimental investigation of the magnetic phenomena developed
by the rotation of bodies,
has lately engaged the attention of several distinguished
philosophers, it may not be amiss to consider the subject in a more general
way, as we shall thus not only confirm the preceding analysis, but be able
to show with what rapidity the body approaches that permanent state, which
it has been the object of the preceding part of this article to determine.

Let us now, therefore, consider a body $A$ fixed in space, under the
influence of electric forces which vary according to any given law; then we
might propose to determine the electrical state of the body, after a certain
interval of time, from the knowledge of its initial state; supposing a constant
coercive force to exist within it. To resolve this in its most general form,
it would be necessary to distinguish between those parts of the body where
the fluid was at rest, from the forces acting there being less than the 
coercive force, and those where it would be in motion; moreover these parts
would vary at every instant, and the problem therefore become very intricate:
were we however to suppose the initial state so chosen, that the total force
to move any particle~$p$ within~$A$,
arising from its electric state and exterior
actions, was then just equal to the coercive force~$\beta$;
also, that the alteration
in the exterior forces should always be such, that if the electric fluid 
remained at rest during the next instant, this total force should no where be
less than~$\beta$;
the problem would become more easy, and still possess a great
degree of generality. For in this case, when the fluid is moveable, the whole
force tending to move any particle~$p$ within~$A$, will, at every instant, be
exactly equal to the coercive force. If therefore $x,y,z,$ represent the
co-ordinates of~$p$, and $V$ the value of the total
potential function at any instant
of time~$t$, arising from the electric state of the body and exterior forces, we
shall have the equation
\[
\tag{$a$.}
\beta^2=
\Bigl(\frac{dV}{dx}\Bigr)^2
+\Bigl(\frac{dV}{dy}\Bigr)^2
+\Bigl(\frac{dV}{dz}\Bigr)^2,
\]
whose general integral may be thus constructed:

Take the value of $V$ arbitrarily over any surface whatever~$S$, plane
or curved, and suppose three rectangular co-ordinates $w,w',w''$, whose
origin is at a point~$P$ on~$S$: the axis of~$w$ being a normal to~$S$,
and those of~$w',w''$, in its plane tangent.
Then the values of~$\frac{dV}{dw'}$ and~$\frac{dV}{dw''}$ are known
at the point~$P$,
and the value of $\frac{dV}{dw}$ will be determined by the equation
\[
\Bigl(\frac{dV}{dw}\Bigr)^2
+\Bigl(\frac{dV}{dw'}\Bigr)^2
+\Bigl(\frac{dV}{dw''}\Bigr)^2=\beta^2,
\]
which is merely a transformation of the above.

Take now another point $P_\prime$, whose co-ordinates referred to these
axes are $\frac{dV}{dw}$, $\frac{dV}{dw'}$ and~$\frac{dV}{dw''}$,
and draw a right line $L$ through the points~$P,P_\prime$,
then will the value of~$V$ at any point~$p$, on~$L$, be expressed by
\[
V_0+\beta\lambda;
\]
$\lambda$ being the distance $Pp$, measured along the line~$L$,
considered as increasing
in the direction $PP_\prime$, and~$V_0$,
the given value of~$V$ at~$P$. For it is very
easy to see that the value of~$V$ furnished by this construction, satisfies the
partial differential equation~($a$),
and is its general integral, moreover the system
of lines $L, L', L''$, etc.\ belonging to the
points $P, P', P''$, etc.\ on~$S$, are
evidently those along which the electric fluid tends to move, and will move
during the following instant.

Let now $V+DV$ represent what $V$ becomes at the end of the time~$t+dt$;
substituting this for~$V$ in~($a$) we obtain
\[
\tag{$b$.}
0=
\frac{dV}{dx}\cdot\frac{dDV}{dx}
+\frac{dV}{dy}\cdot\frac{dDV}{dy}
+\frac{dV}{dz}\cdot\frac{dDV}{dz}.
\]
Then, if we designate by $D'V$, the augmentation of the potential function,
arising from the change which takes place in the exterior forces during the
element of time~$dt$,
\[
DV-D'V
\]
will be the increment of the potential function, due to the corresponding
alterations $D\rho$ and~$D\rho'$
in the densities of the electric fluid at the surface of~$A$
and within it, which may be determined from~${DV-D'V}$ by art.~7. But,
by the known theory of partial differential equations, the most general value 
of~$DV$ satisfying~($b$),
will be constant along every one of the lines $L, L', L''$, etc.,
and may vary arbitrarily in passing from one of them to another: as it is
also along these lines the electric fluid moves during the instant~$dt$,
it is clear
the total quantity of fluid in any infinitely thin needle, formed by them, and
terminating in the opposite surfaces of~$A$, will undergo no alteration during
this instant. Hence therefore
\[
\tag{$c$.}
0=\int D\rho'dv+D\rho\,d\sigma+D\rho_\prime\,d\sigma_\prime\,;
\]
$dv$ being an element of the volume of the needle,
and $d\sigma,d\sigma_\prime$, the two elements
of $A$'s~surface by which it is terminated. This condition, combined
with the equation~($b$), will completely determine the value of~$DV$, and we
shall thus have the value of the potential function~$V+DV$, at the instant of
time~$t+dt$, when its value~$V$, at the time~$t$, is known.

As an application of this general solution; suppose the body $A$ is a
solid of revolution, whose axis is that of the co-ordinate~$z$, and let the two
other axes~$X,Y$ situate in its equator, be fixed in space. If now the exterior
electric forces are such that they may be reduced to two, one equal
to~$c$, acting parallel to~$z$,
the other equal to~$b$, directed parallel to a line in
the plane~$(xy)$, making the variable angle~$\phi$ with~$X$;
the value of the potential function arising
from the exterior forces, will be
\[
-zc-xb\cos\phi-yb\sin\phi;
\]
where $b$ and $c$ are constant quantities,
and $\phi$ varies with the time so as to
be constantly increasing. When the time is equal to~$t$, suppose the value
of~$V$ to be
\[
V=\beta(x\cos\varpi+y\sin\varpi):
\]
then the system of lines $L, L', L''$,
will make the angle $\varpi$ with the plane~$(xz)$,
and be perpendicular to another plane whose equation is
\[
0=x\cos\varpi+y\sin\varpi.
\]
If during the instant of time $dt$, $\phi$ becomes $\phi+D\phi$,
the augmentation of
the potential function due to the elementary change in the exterior forces,
will be
\[
D'V = (x\sin\phi-y\cos\phi)b\,D\phi;
\]
moreover the equation ($b$) becomes
\[
\tag{$b'$.}
0=\cos\varpi\cdot\frac{dDV}{dx}+\sin\varpi\cdot\frac{dDV}{dy},
\]
and therefore the general value of $DV$ is
\[
DV=DF\{y\cos\varpi-x\sin\varpi\;;\;z\};
\]
$DF$ being the characteristic of an infinitely small arbitrary function. But, it
has been before remarked that the value of $DV$ will be completely determined,
by satisfying the equation~($b$) and the condition~($c$). Let us then assume
\[
DF\{y\cos\varpi-x\sin\varpi\;;\;z\}=h\,D\phi(y\cos\varpi-x\sin\varpi);
\]
$h$ being a quantity independent of $x,y,z$, and see if it be possible to 
determine~$h$ so as to satisfy the condition~($c$). Now on this supposition
\begin{multline*}
DV-D'V=h\,D\phi(y\cos\varpi-x\sin\varpi)-(x\sin\phi-y\cos\phi)b\,d\phi\\
=D\phi\bigl[y(h\cos\varpi+b\cos\phi)-x(h\sin\varpi+b\cos\phi)\bigr].
\end{multline*}
The value of $D\rho'$ corresponding to this potential function
is (art.~7)
\[
D\rho'=0,
\]
and on account of the parallelism of the lines $L,L'$ etc.\ to each other, and
to $A$'s equator~$d\sigma=d\sigma_\prime$. The condition ($c$) thus becomes
\[
\tag{$c'$.}
0=D\rho+D\rho_1:
\]
$D\rho$ and $D\rho_1$,
being the elementary densities on $A$'s~surface at opposite ends
of any of the lines $L,L'$, etc.\ 
corresponding to the potential function~${DV-D'V}$.
But it is easy to see from the form of this function, that these elementary
densities at opposite ends of any line
perpendicular to a plane whose equation is
\[
0 = y(h\cos\varpi+b\cos\phi)-x(h\sin\varpi+b\cos\phi),
\]
are equal and of contrary signs, and therefore the condition ($c$) will be 
satisfied by making this plane coincide with that perpendicular
to $L,L'$, etc.,
whose equation, as before remarked, is
\[
0=x\cos\varpi+y\sin\varpi;
\]
that is the condition ($c$) will be satisfied,
if $h$ be determined by the equation
\[
\frac{h\cos\varpi+b\cos\phi}{\sin\varpi}
=-\frac{h\sin\varpi+b\sin\phi}{\cos\varpi},
\]
which by reduction becomes
\[
0=h+b\cos(\phi-\varpi),
\]
and consequently
\[
\begin{aligned}
V+DV=\beta(x\cos\varpi+y\sin\varpi) &+ h\,D\phi(y\cos\varpi-x\sin\varpi)\\
=\beta x\bigl(\cos\varpi+\frac b\beta\sin\varpi\cos(\phi-\varpi)D\phi\bigr)
&+\beta y\bigl(\sin\varpi-\frac b\beta\cos\varpi\cos(\phi-\varpi)D\phi\bigr)\\
=\beta x\cos\bigl(\varpi+\frac b\beta\cos(\phi-\varpi)D\phi\bigr)
&+\beta y\sin\bigl(\varpi-\frac b\beta\cos(\phi-\varpi)D\phi\bigr)
\end{aligned}
\]
When therefore $\phi$ is augmented
by the infinitely small angle~$D\phi$, $\varpi$~receives
the corresponding increment $-\frac b\beta\cos(\phi-\varpi)D\phi$,
and the form of~$V$ remains
unaltered; the preceding reasoning is consequently applicable to every instant,
and the general relation between $\phi$ and~$\varpi$ [is] expressed by
\[
0=D\varpi+\frac b\beta\cos(\phi-\varpi)D\phi:
\]
a common differential equation, which by integration gives
\[
H\cdot e^{\phi\cotan\gamma}=
\frac{\sin(\frac34\pi-\frac12\gamma+\frac12\varpi-\frac12\phi)}
{\sin(\frac14\pi+\frac12\gamma+\frac12\varpi-\frac12\phi)};
\]
$H$ being an arbitrary constant, and $\gamma$,
as in the former part of this article,
the smallest root of
\[
0=b\sin\gamma-\beta.
\]

Let $\varpi_0$ and $\phi_0$, be the initial values of $\varpi$ and $\phi$;
then the total potential function at the next instant,
if the electric fluid remained fixed, would be
\[
V_\prime=
\beta(x\cos\varpi_0+y\sin\varpi_0)+(x\sin\phi_0-y\cos\phi_0)b\,d\phi,
\]
and the whole force to move a particle $p$, whose co-ordinates are~$,x,y,z$,
\[
\sqrt{
  \Bigl(\frac{dV_\prime}{dx}\Bigr)^2
  +\Bigl(\frac{dV_\prime}{dy}\Bigr)^2
  +\Bigl(\frac{dV_\prime}{dz}\Bigr)^2
}=\beta+d\phi\cdot b\sin(\phi_0-\varpi_0),
\]
which, in order that our solution may be applicable, must not be less 
than~$\beta$,
and consequently the angle $\phi_0-\varpi_0$
must be between~$0$ and~$\pi$: when this is
the case, $\varpi$~is immediately determined from~$\phi$
by what has preceded. In
fact, by finding the value of~$H$ from the initial
values $\varpi_0$ and~$\phi_0$, and
making $\zeta=\frac14\pi+\frac12\gamma+\frac12\varpi-\frac12\phi$, we obtain
\[
\tan\zeta=
\frac{\tan\zeta_0}
{e^{(\phi-\phi_0)\cot\gamma}+\tan\gamma\tan\zeta_0
(e^{(\phi-\phi_0)\cot\gamma}-1)};
\]
$\zeta_0$ being the initial value of $\zeta$.

We have, in the latter part of this article, considered the body~$A$ at
rest, and the line~$X'$, parallel to the direction of~$b$,
as revolving round it:
but if, as in the former, we now suppose this line immovable and the body
to turn the contrary way, so that the relative motion of $X'$ to~$X$ may remain
unaltered, the electric state of the body referred to the axes~$X,Y,Z$, 
evidently depending on this relative motion only, will consequently remain the
same as before. In order to determine it on the supposition just made, let
$X'$ be the axis of~$x'$,
one of the co-ordinates of~$p$, referred to the rectangular
axes~$X',Y',Z'$, also $y',z$, the other two; the direction $X'Y'$, being
that in which $A$ revolves. Then, if $\varpi'$ be the angle the system of lines
$L,L'$, etc. forms with the plane~$(x',z)$, we shall have
\[
\varpi+\varpi'=\phi;
\]
$\phi$, as before stated, being the angle included by the axes~$X,X'$. Moreover
the general values of $V$ and $Q$ will be
\[
V=\beta(x'\cos\varpi'+y'\sin\varpi')
\quad\text{and}\quad
\zeta=\tfrac14\pi+\tfrac12\gamma-\tfrac12\varpi',
\]
and the initial condition, in order that our solution may be applicable, will
evidently become $\phi_0-\varpi_0=\varpi'_0=$\ a quantity betwixt $0$ and~$\pi$.

As an example, let $\tan\gamma=\frac{1}{10}$,
since we know by experiment that $\gamma$
is generally very small; then taking the most unfavorable case,
viz.\ where ${\varpi'_0=0}$,
and supposing the body to make one revolution only, the value of~$\zeta$,
determined from its initial one,
$\zeta_0=\frac14\pi+\frac12\gamma-\frac12\varpi'$,
will be found extremely
small and only equal to a unit in the 27th decimal place. We thus see with
what rapidity $\zeta$ decreases, and consequently, the body approaches to a 
permanent state, defined by the equation
\[
0=\zeta=\tfrac14\pi+\tfrac12\gamma-\tfrac12\varpi'.
\]
Hence, the polarity induced by the rotation is ultimately directed along a line,
making an angle equal to~$\frac12\pi+\gamma$
with the axis~$X'$, which agrees with what
was shown in the former part of this article.

The value of $V$ at the body's surface being thus known at any
instant whatever, that of the potential function at a point $p'$ exterior to the
body, together with the forces acting there, will be immediately determined
as before.
\Crule
\bigskip
\bigskip
\bigskip
\section{Application of the preceding results to the theory of magnetism.}
\Section{14.}
\markboth{Application to magnetism.}{14.}
The electric fluid appears to pass freely from one part of a continuous
conductor to another, but this is by no means the case with the magnetic
fluid, even with respect to those bodies which, from their instantly returning
to a natural state the moment the forces inducing a magnetic one are removed,
must be considered, in a certain sense, as perfect conductors of magnetism.
\Person{Coulomb}, I believe,
was the first who proposed to consider these as formed
of an infinite number of particles, each of which conducts the magnetic fluid
in its interior with perfect freedom, but which are so constituted that it is
impossible there shall be any communication of it from one particle to the
next. This hypotesis is now generally adopted by philosophers, and its 
consequences, as far as they have hitherto been developed, are found to agree
with observation; we will therefore admit it in what follows, and endeavour
thence to deduce, mathematically, the laws of the distribution of magnetism
in bodies of any shape whatever.

Firstly, let us endeavour to determine the value of the potential
function, arising from the magnetic state induced in a very small body~$A$, by
the action of constant forces directed parallel to a given right line; the body
being composed of an infinite number of particles, all perfect conductors of
magnetism and originally in a natural state. In order to deduce this more
immediately from art.~6, we will conceive these forces to arise from an 
infinite quantity $Q$ of magnetic fluid, concentrated
in a point~$p$ on this line, at
an infinite distance from~$A$. Then the origin~$O$ of the rectangular
co-ordinates being any where within~$A$, if $x,y,z$,
be those of the point~$p$, and~$x',y',z'$,
those of any other exterior point~$p'$, to which the potential
function~$V$ arising from~$A$ belongs, we shall have
(vide \Title{M\'ec. C\'el.} Liv.~3)
\[
V=\frac{U^{(0)}}{r'}
+\frac{U^{(1)}}{r'^2}
+\frac{U^{(2)}}{r'^3}
+\text{etc.};
\]
$r'=\sqrt{x'^2+y'^2+z'^2}$ being the distance $Op'$.

Moreover, since the total quantity of magnetic fluid in $A$ is equal to
\Emphasis{zero}, ${U^{(0)}=0}$.
Supposing now $r'$~very great compared with the dimensions of
the body, all the terms after~$\frac{U^{(0)}}{r'}$
in the expression just given will be exceedingly
small compared with this, by neglecting them, therefore, and substituting
for~$U^{(1)}$ its most general value, we obtain
\[
V=\frac{U^{(1)}}{r'^2}
=\frac{Ax'+By'+Cz'}{r'^3};
\]
$A,B,C$, being quantities independent of $x',y',z'$,
but which may contain~$x,y,z$.

Now (art.~6) the value of $V$ will remain unaltered, when we change
$x,y,z$, into~$x',y',z'$, and reciprocally. Therefore,
\[
V=\frac{Ax'+By'+Cz'}{r'^3}
=\frac{A'x+B'y+C'z}{r^3};
\]
$A',B',C'$, being the same functions of $x',y',z'$, as $A,B,C$, are of
$x,y,z$. Hence it is easy to see that $V$ must be of the form
\[
V=\frac{a''xx'+b''yy'+c''zz'+e''(xy'+x'y)+f''(xz'+x'z)+g''(yz'+y'z)}{r^3r'^3};
\]
$a'',b'',c'',e'',f'',g''$, being constant quantities.

If $X, Y, Z$, represent the forces arising from the magnetism concentrated
in~$p$, in the directions of $x,y,z$, positive, we shall have
\[
X=\frac{-Qx}{r^3};\quad
Y=\frac{-Qy}{r^3};\quad
Z=\frac{-Qz}{r^3};
\]
and therefore $V$ is of the form
\[
V=\frac{a'Xx'+b'Yy'+c'Zz'+e'(Xy'+Yx')+f'(Xz'+Zx')+g'(Yz'+Zy')}{r'^3}
\]
$a',b'$, etc.\ being other constant quantities.
But it will always be possible to
determine the situation of three rectangular axes, so that $e$, $f$,
and~$g$ may
each be equal to zero, and consequently $V$ be reduced to the following
simple form
\[
\tag{$a$.}
V=\frac{aXx'+bYy'+cZz'}{r'^3};
\]
$a$, $b$, and $c$ being three constant quantities.

When $A$ is a sphere, and its magnetic particles are either spherical,
or, like the integrant particles of non-crystalized bodies, arranged in a 
confused manner; it is evident the constant quantities $a',b',c'$, etc.\ in the
general value of~$V$, must be the same for every system of rectangular
co-ordinates, and consequently we must have $a'=b'=c'$, $e'=0$, $f'=0$,
and $g'=0$, therefore in this case
\[
\tag{$b$.}
V=\frac{a'(Xx'+Yy'+Zz')}{r'^3};
\]
$a'$ being a constant quantity dependant on the magnitude and nature of~$A$.

The formula~($a$) will give the value of the forces acting on any
point~$p'$, arising from a mass~$A$ of soft iron or other similar matter, whose
magnetic state is induced by the influence of the earth's action; supposing the
distance $Ap'$ to be great compared with the dimensions of~$A$, and if it be a
solid of revolution, one of the rectangular axes, say~$X$, must coincide with
the axis of revolution, and the value of~$V$ reduce itself to
\[
V=\frac{a'Xx'+b'(Yy'+Zz')}{r'^3};
\]
$a'$ and $b$' being two constant quantities dependant on the form and nature of
the body. Moreover the forces acting in the directions of~$x',y',z'$, positive,
are expressed by
\[
-\Bigl(\frac{dV}{dx'}\Bigr),\quad
-\Bigl(\frac{dV}{dy'}\Bigr),\quad
-\Bigl(\frac{dV}{dz'}\Bigr).
\]
We have thus the means of comparing theory with experiment, but these are
details into which our limits will not permit us to enter.

The formula~($b$), which is strictly correct for an infinitely small sphere,
on the supposition of its magnetic particles being arranged in a confused
manner, will, in fact, form the basis of our theory, and although the preceding
analysis seems sufficiently general and rigorous, it may not be amiss to
give a simpler proof of this particular case. Let, therefore, the origin $O$ of
the rectangular co-ordinates be placed at the centre of the infinitely small
sphere~$A$, and $OB$ be the direction of the parallel forces acting upon it;
then,
since the total quantity of magnetic fluid in~$A$ is equal to \Emphasis{zero},
the value of
the potential function~$V$, at the point~$p'$, arising from~$A$,
must evidently be
of the form
\[
V=\frac{k\cos\theta}{r'^2};
\]
$r'$ representing as before the distance $Op'$,
and $\theta$ the angle formed between
the line~$Op'$, and another line $OD$ fixed in~$A$. If now $f$ be the magnitude
of the force directed along~$OB$, the constant $k$ will evidently be of the
form~${k=a'f}$;
$a'$~being a constant quantity. The value of~$V$, just given, holds
good for any arrangement, regular or irregular, of the magnetic particles
composing~$A$, but on the latter supposition, the value of $V$ would evidently
remain unchanged, provided the sphere, and consequently the line~$OD$, revolved
round~$OB$ as an axis, which could not be the case unless $OB$ and~$OD$
coincided. Hence ${\theta=\text{angle $BOp'$}}$ and
\[
V=\frac{a'f\cos\theta}{r'^2}.
\]
Let now $\alpha,\beta,\gamma$, be the angles that the line $Op'=r'$
makes with the axes
of~$x,y,z$, and $\alpha',\beta',\gamma'$,
those which $OB$ makes with the same axes; then
substituting for $\cos\theta$ its value
$\cos\alpha\cos\alpha'+\cos\beta\cos\beta'+\cos\gamma\cos\gamma'$, we have,
since $f\cos\alpha=X$, $f\cos\beta=Y$, $f\cos\gamma=Z$,
\[
\tag{$b$.}
V=\frac{a'(X\cos\alpha+Y\cos\beta+Z\cos\gamma)}{r'^2}.
\]
Which agrees with the equation (b), seeing that $\cos\alpha=\frac{x'}{r'}$, 
$\cos\beta=\frac{y'}{r'}$, $\cos\gamma=\frac{z'}{r'}$.
\bigskip

\Section{15.}
\markboth{Application to magnetism.}{15.}
Conceive now, a body $A$, of any form, to have a magnetic state induced
in its particles by the influence of exterior forces, it is clear that if
$dv$ be an element of its volume, the value of the potential function arising
from this element, at any point~$p'$ whose co-ordinates are~$x',y',z'$, must,
since the total quantity of magnetic fluid in~$dv$ is equal to \Emphasis{zero},
be of the form
\[
\tag{$a$.}
\frac{dv\bigl[X(x'-x)+Y(y'-y)+Z(z'-z)\bigr]}{r^3};
\]
$x,y,z$, being the co-ordinates of $dv$, $r$ the distance $p',dv$ and~$X,Y,Z$,
three quantities dependant on the magnetic state induced in~$dv$, and serving
to define this state. If therefore $dv'$
be an infinitely small volume within the
body~$A$ and inclosing the point~$p'$,
the potential function arising from the whole~$A$
exterior to~$dv'$, will be expressed by
\[
\int dx\,dy\,dz\,
\frac{X(x'-x)+Y(y'-y)+Z(z'-z)}{r^3};
\]
the integral extending over the whole volume of~$A$ exterior to~$dv'$.

It is easy to show from this expression that, in general, although $dv'$
be infinitely small, the forces acting in its interior vary in magnitude and 
direction by passing from one part of it to another; but,
when $dv'$ is spherical,
these forces are sensibly constant in magnitude and direction, and consequently,
in this case, the value of the potential function
induced in~$dv'$ by their action,
may be immediately deduced from the preceding article.

Let $\psi'$ represent the value of the integral just given, when $dv'$ is an
infinitely small sphere. The force acting on~$p'$ arising from the mass exterior
to~$dv'$, tending to increase~$x'$, will be
\[
-\biggl(\frac{\overline{d\psi'}}{dx'}\biggl);
\]
the line above the differential co-efficient
indicating that it is to be obtained
by supposing the radius of~$dv'$ to vanish after differentiation,
and this may
differ from the one obtained by first making the radius vanish, and afterwards
differentiating the resulting function of~$x',y',z'$,
which last being represented
as usual by~$\frac{d\psi'}{~dx'}$, we have
\[
\begin{aligned}
\frac{\overline{d\psi'}}{dx'} &=
\frac{d}{dx'}\int dx\,dy\,dz\,\frac{X(x'-x)+Y(y'-y)+Z(z'-z)}{r^3}\\
\frac{{d\psi'}}{dx'} &=
\frac{d}{dx'}\int dx\,dy\,dz\,\frac{X(x'-x)+Y(y'-y)+Z(z'-z)}{r^3};
\end{aligned}
\]
the first integral being taken over the whole volume of $A$ exterior to~$dv'$,
and the second over the whole of~$A$ including~$dv'$. Hence
\[
\frac{{d\psi'}}{dx'}-\frac{\overline{d\psi'}}{dx'}=
=\frac{d}{dx'}\int dx\,dy\,dz\,\frac{X(x'-x)+Y(y'-y)+Z(z'-z)}{r^3};
\]
the last integral comprehending the volume of the spherical particle $dv'$ only,
whose radius $a$ is supposed to vanish after differentiation. In order to effect
the integration here indicated, we may remark that $X$, $Y$ and~$Z$ are sensibly
constant within~$dv'$, and may therefore be replaced by
$X_\prime$, $Y_\prime$ and~$Z_\prime$, their
values at the centre of the sphere~$dv'$, whose co-ordinates
are~$x_\prime,y_\prime,z_\prime$;
the required integral will thus become
\[
\int dx\,dy\,dz\,\frac{X_\prime(x'-x)+Y_\prime(y'-y)+Z_\prime(z'-z)}{r^3}.
\]
Making for a moment $E=X_\prime x+Y_\prime y+Z_\prime y$,
we shall have $X_\prime=\frac{dE}{dx},Y_\prime=\frac{dE}{dy},Z_\prime=\frac{dE}{dz}$,
and as also $\frac{x'-x}{r^3}=\frac{d\frac1r}{dx};
\frac{y'-y}{r^3}=\frac{d\frac1r}{dy};
\frac{z'-z}{r^3}=\frac{d\frac1r}{dz}$
this integral may be written
\[
\int dx\,dy\,dz\,\biggl\{
  \frac{dE}{dx}\cdot\frac{d\frac1r}{dx}
  +\frac{dE}{dy}\cdot\frac{d\frac1r}{dy}
  +\frac{dE}{dz}\cdot\frac{d\frac1r}{dz}
\biggr\},
\]
which since $\delta E=0$, and $\delta\frac1r=0$,
reduces itself by what is proved in art.~3, to
\[
-\int\frac{d\sigma}{r}\Bigl(\frac{dE}{dw}\Bigr)
=\text{(because $dw=-da$)}
\int\frac{d\sigma}{r}\frac{dE}{da};
\]
the integral extending over the whole surface of the sphere $dv'$,
of which $d\sigma$
is an element; $r$ being the distance $p',d\sigma$,
and $dw$ measured from the surface
towards the interior of~$dv'$.
Now $\int\frac{d\sigma}{r}\frac{dE}{da}$ expresses the value of the potential
function for a point~$p'$, within the sphere, supposing its surface every where
covered with electricity whose density is~$\frac{dE}{da}$,
and may very easily be obtained
by No.~13, Liv.~3, \Title{M\'ec. C\'eleste}.
In fact, using for a moment the notation
there employed, supposing the origin of the polar co-ordinates at the centre
of the sphere, we have
\[
E=E_\prime+a[X_\prime\cos\theta
+Y_\prime\sin\theta\cos\varpi
+Z_\prime\sin\theta\sin\varpi];
\]
$E_\prime$ being the value of $E$ at the centre of the sphere. Hence
\[
\frac{dE}{da}=
X_\prime\cos\theta
+Y_\prime\sin\theta\cos\varpi
+Z_\prime\sin\theta\sin\varpi,
\]
and as this is of the form $U^{(1)}$ (Vide \Title{M\'ec. C\'eleste} Liv.~3.),
we immediately obtain
\[
\int\frac{d\sigma}{r}\frac{dE}{da}=\tfrac43\pi r'\{
X_\prime\cos\theta'
+Y_\prime\sin\theta'\cos\varpi'
+Z_\prime\sin\theta'\sin\varpi'\},
\]
where $r',\theta',\varpi'$, are the polar co-ordinates of $p'$.
Or by restoring $x'$, $y'$, and~$z'$
\[
\int\frac{d\sigma}{r}\frac{dE}{da}=\tfrac43\pi \bigl\{
  X_\prime(x'-x_\prime)
  +Y_\prime(y'-y_\prime)
  +Z_\prime(z'-z_\prime)\bigl\}.
\]
Hence we deduce successively
\begin{multline*}
\frac{{d\psi'}}{dx'}-\frac{\overline{d\psi'}}{dx'}
=\frac{d}{dx'}\int dx\,dy\,dz\,\frac{X(x'-x)+Y(y'-y)+Z(z'-z)}{r^3}\\
=\frac{d}{dx'}\int\frac{d\sigma}{r}\frac{dE}{da}
=\tfrac43\pi \bigl\{
  X_\prime(x'-x_\prime)
  +Y_\prime(y'-y_\prime)
  +Z_\prime(z'-z_\prime)\bigl\}
=\tfrac43\pi X_\prime.
\end{multline*}
If now we make the radius $a$ vanish,
$X_\prime$ must become equal to~$X'$, the value
of~$X$ at the point~$p'$, and there will result
\[
\frac{{d\psi'}}{dx'}-\frac{\overline{d\psi'}}{dx'}=\tfrac43\pi X'
\quad\text{i.~e.}\quad
\frac{\overline{d\psi'}}{dx'}=\frac{{d\psi'}}{dx'}-\tfrac43\pi X'.
\]

But $\frac{\overline{d\psi'}}{dx'}$ expresses the value
of the force acting in the direction
of~$x$ positive, on a point~$p'$ within the infinitely small sphere~$dv'$, arising from
the whole of~$A$ exterior to~$dv'$;
substituting now for $\frac{\overline{d\psi'}}{dx'}$ its value just found,
the expression of this force becomes
\[
\tfrac43\pi X'-\frac{{d\psi'}}{dx'}.
\]
Supposing $V'$ to represent the value of the potential function at~$p'$, arising
from the exterior bodies which induce the magnetic state of~$A$, the force due
to them acting in the same direction, is
\[
-\frac{dV'}{dx'},
\]
and therefore the total force in the direction of $x'$ positive,
tending to induce
a magnetic state in the spherical element~$dv'$, is
\[
\tfrac43\pi X'-\frac{{d\psi'}}{dx'}-\frac{dV'}{dx'}
=\overline{X}.
\]
In the same way, the total forces in the directions of $y'$ and $z'$ positive,
acting upon~$dv'$, are shown to be
\[
\tfrac43\pi Y'-\frac{{d\psi'}}{dy'}-\frac{dV'}{dy'}=\overline{Y},
\quad\text{and,}\quad
\tfrac43\pi Z'-\frac{{d\psi'}}{dz'}-\frac{dV'}{dz'}=\overline{Z}.
\]

By the equation ($b'$) of the preceding article, we see that when $dv'$
is a perfect conductor of magnetism,
and its particles are not regularly arranged,
the value of the potential function at any point~$p''$,
arising from the magnetic
state induced in~$dv'$ by the action of the
forces~$\overline{X},\overline{Y},\overline{Z}$, is of the form
\[
\frac{a'(\overline{X}\cos\alpha+\overline{Y}\cos\beta+\overline{Z}\cos\gamma)}
{r'^2};
\]
$r'$ being the distance $p'',dv'$, and
$\alpha,\beta,\gamma$, the angles which $r'$ forms with
the axes of the rectangular co-ordinates. If then $x'',y'',z''$, be the
co-ordinates of~$p''$, this becomes, by observing that here~${a'=kdv'}$,
\[
\frac{kdv'\bigl[\,
\overline{X}(x''-x')+\overline{Y}(y''-y')+\overline{Z}(z''-z')
\bigr]}{r'^3},
\]
$k$ being a constant quantity dependant on the nature of the body. The same
potential function will evidently be obtained from the expression~($a$) of this
article, by changing~$dv,p'$, and their co-ordinates, into~$dv',p''$, and their
co-ordinates; thus we have
\[
\frac{dv'\bigl[
X'(x''-x')+Y'(y''-y')+Z'(z''-z')
\bigr]}{r'^3}.
\]
Equating these two forms of the same quantity, there results the three following
equations:
\[
\begin{aligned}
X'&=k\overline{X}=\tfrac43\pi k\,X'-k\frac{d\psi'}{dx'}-k\frac{dV'}{dx'}\\
Y'&=k\overline{Y}=\tfrac43\pi k\,Y'-k\frac{d\psi'}{dy'}-k\frac{dV'}{dy'}\\
Z'&=k\overline{Z}=\tfrac43\pi k\,Z'-k\frac{d\psi'}{dz'}-k\frac{dV'}{dz'},
\end{aligned}
\]
since the quantities $x'',y'',z''$, are perfectly arbitrary.
Multiplying the first
of these equations by~$dx'$, the second by~$dy'$,
the third by~$dz'$, and taking
their sum, we obtain
\[
0=(1-\tfrac43\pi k)(X'dx'+Y'dy'+Z'dz')+k\,d\psi'+k\,dV'.
\]
But $d\psi'$ and $dV'$ being perfect differentials,
$X'dx'+Y'dy'+Z'dz'$ must be
so likewise, making therefore
\[
d\phi'=X'dx'+Y'dy'+Z'dz'
\]
the above, by integration, becomes
\[
\text{const}=(1-\tfrac43\pi k)\phi'+k\psi'+kV'.
\]

Although the value of $k$ depends wholly on the nature of the body
under consideration, and is to be determined for each by experiment, we may
yet assign the limits between which it must fall. For we have, in this theory,
supposed the body composed of conducting particles, separated by intervals
absolutely impervious to the magnetic fluid; it is therefore clear the magnetic
state induced in the infinitely small sphere~$dv'$, cannot be greater than that
which would be induced, supposing it one continuous conducting mass, but
may be made less in any proportion, at will, by augmenting the non-conducting
intervals.

When $dv'$ is a continuous conductor, it is easy to see the value of
the potential function at the point~$p''$,
arising from the magnetic state induced
in it by the action of the forces~$\overline{X},\overline{Y},\overline{Z}$
will be
\[
\frac{3dv}{4\pi}\cdot
\frac{X(x''-x')+Y(y''-y')+Z(z''-z')}{r'^3},
\]
seeing that $\frac{3dv}{4\pi}=a^3$;
$a$ representing, as before, the radius of the sphere~$dv'$.
By comparing this expression with that before found, when $dv'$ was not a
continuous conductor, it is evident $k$ must be between
the limits $0$ and~$\frac34\pi$,
or, which is the same thing, ${k=\frac{3g}{4\pi}}$;
$g$~being any positive quantity less than~$1$.

The value of $k$, just found, being substituted in the equation serving
to determine~$\phi'$, there arises
\[
\text{const}=(1-g)\phi'+\frac{3g}{4\pi}(\psi'+V').
\]
Moreover
\begin{multline*}
\psi'=\int dx\,dy\,dz\,\frac{X(x'-x)+Y(y'-y)+Z(z'-z)}{r^3}\\
=\int dx\,dy\,dz\,\biggl\{
  \frac{d\phi}{dx}\cdot\frac{d\frac1r}{dx}
  +\frac{d\phi}{dy}\cdot\frac{d\frac1r}{dy}
  +\frac{d\phi}{dz}\cdot\frac{d\frac1r}{dz}
\biggr\}\\
=4\pi\phi'-\int d\sigma\,\overline\phi
\biggl(\frac{\overline{d\frac1r}}{dw}\biggr)
\quad\text{(art.~3)};
\end{multline*}
the triple integrals extending over the whole volume of $A$, and that relative
to~$d\sigma$ over its surface, of which $d\sigma$ is an element;
the quantities $\overline\phi$ and
$\frac{\overline{d\frac1r}}{dw}$~belonging to this element.
We have, therefore, by substitution
\[
\text{const}=(1+2g)\phi'+
\frac{3g}{4\pi}\Biggl(V'-
\int d\sigma\,\overline\phi\biggl(\frac{\overline{d\frac1r}}{dw}\biggr)\Biggr).
\]
Now $\delta'V'=0$, and
$\delta'\int d\sigma\,\overline\phi\bigl(\frac{\overline{d\frac1r}}{dw}\bigr)$
and consequently ${\delta'\phi'=0}$; the
symbol $\delta'$ referring to~$x',y',z'$,
the co-ordinates of~$p'$; or, since $x$', $y'$ and~$z'$
are arbitrary, by making them equal to $x$, $y$, and $z$,
respectively, there results
\[
0=\delta\phi,
\]
in virtue of which, the value of $\psi'$, by article~3, becomes
\[
\tag{$b$.}
\psi'=-\int\frac{d\sigma}{r}\biggl(\frac{\overline{d\phi}}{dw}\biggr);
\]
$r$ being the distance $p',d\sigma$,
and $(\frac{\overline{d\phi}}{dw})$  belonging to~$d\sigma$. The former equation
serving to determine~$\phi'$ gives, by changing~$x',y',z'$, into~$x,y,z$,
\[
\tag{$c$.}
\text{const}=(1-g)\phi+\frac{3g}{4\pi}(\psi+V);
\]
$\phi$, $\psi$ and $V$ belonging to a point $p$,
within the body, whose co-ordinates
are~$x,y,z$. It is moreover evident from what precedes that, the functions
$\phi$, $\psi$ and~$V$ satisfy the equations
$0=\delta\phi$, $0=\delta\psi$ and $0=\delta V$ and have
no singular values in the interior of~$A$.

The equations ($b$) and ($c$) serve to determine $\phi$ and~$\psi$, completely,
when the value of~$V$ arising from the exterior bodies is known, and therefore
they enable us to assign the magnetic state of every part of the body~$A$,
seeing that it depends on $X, Y, Z$, the differential co-efficients of~$\phi$.
It is
also evident that~$\psi'$, when calculated for any point~$p'$,
not contained within
the body~$A$, is the value of the potential function at this point arising from
the magnetic state induced in~$A$, and therefore this function is always given
by the equation~($b$).

The constant quantity $g$, which enters into our formulae, depends on
the nature of the body solely, and, in a subsequent article, its value is 
determined for a cylindric wire used by \Person{Coulomb}.
This value differs very little
from unity: supposing therefore~${g=1}$,
the equations~($b$) and~($c$) become
\[
\tag{$b'$.}
\psi'=-\int\frac{d\sigma}{r}\biggl(\frac{\overline{d\phi}}{dw}\biggr);
\]
\[
\tag{$c'$.}
\text{const}=\psi+V,
\]
evidently the same, in effect, as would be obtained by considering the magnetic
fluid at liberty to move from one part of the conducting body to another;
the density $\rho$ being here replaced
by~$\bigl(\frac{\overline{d\phi}}{dw}\bigr)$,
and since the value of the
potential function for any point exterior to the body is, on either supposition,
given by the formula~($b$), the exterior actions will be precisely the same in
both cases. Hence, when we employ iron, nickel, or similar bodies, in which
the value of $g$ is nearly equal to 1, the observed phenomena will differ little
from those produced on the latter hypothesis, except when one of their 
dimensions is very small compared with the others, in which case the results
of the two hypotheses differ widely, as will be seen in some of the 
applications which follow.

If the magnetic particles composing the body were not perfect conductors,
but indued with a coercive force, it is clear there might always be
equilibrium, provided the magnetic state of the element $dv'$
was such as would be
induced by the forces
$\frac{\overline{d\psi}}{dx'}+\frac{dV'}{dx'}+A'$,
$\frac{\overline{d\psi}}{dy'}+\frac{dV'}{dy'}+B'$ and
$\frac{\overline{d\psi}}{dz'}+\frac{dV'}{dz'}+C'$,
instead of
$\frac{\overline{d\psi}}{dx'}+\frac{dV'}{dx'}$,
$\frac{\overline{d\psi}}{dy'}+\frac{dV'}{dy'}$ and
$\frac{\overline{d\psi}}{dz'}+\frac{dV'}{dz'}$;
supposing the resultant of
the forces $A',B',C'$, no where exceeds a quantity~$\beta$, serving to measure
the coercive force. This is expressed by the condition
\[
A'^2+B'^2+C'^2<\beta^2
\]
The equation ($c$) would then be replaced by
\[
0=(1-g)d\phi+\frac{3g}{4\pi}(d\psi+dV+A\,dx+B\,dy+C\,dz);
\]
$A,B,C$, being any functions of $x,y,z$, as $A',B',C'$, are of~$x',y',z'$,
subject only to the condition just given.

It would be extremely easy so to modify the preceding theory, as to
adapt it to a body whose magnetic particles are regularly arranged, by using
the equation~($a$) in the place of the equation~($b$) of the preceding article;
but, as observation has not yet offered any thing which would indicate a
regular arrangement of magnetic particles, in any body hitherto examined, it
seems superfluous to introduce this degree of generality, more particularly as
the omission may be so easily supplied.
\bigskip

\Section{16.}
\markboth{Application to magnetism.}{16.}
As an application of the general theory contained in the preceding
article, suppose the body~$A$
to be a hollow spherical shell of uniform thickness,
the radius of whose inner surface is~$a$,
and that of its outer one~$a_\prime$; and let
the forces inducing a magnetic state in~$A$, arise from any bodies whatever,
situate at will, within or without the shell. Then since in the interior 
of~$A$'s
mass~${0=\delta\phi}$, and~${0=\delta V}$,
we shall have (\Title{M\'ec. C\'el.} Liv.~3)
\[
\phi=\sum\phi^{(i)}r^i+\sum\phi^{(i)}_\prime r^{-i-1}
\quad\text\quad
V=\sum U^{(i)}r^i+\sum U^{(i)}_\prime r^{-i-1};
\]
$r$ being the distance of the point $p$, to which $\phi$ and~$V$ belong,
from the
shell's centre, $\phi^{(0)}$, $\phi^{(1)}$, etc.\ --- $U^{(0)}$, $U^{(1)}$,
etc.\ functions of $\theta$ and~$\varpi$, the two
other polar co-ordinates of~$p$, whose nature has been fully explained by
\Person{Laplace} in the work just cited;
the finite integrals extending from $i=0$
to~$i=\infty$.

If now, to prevent ambiguity, we enclose the $r$ of equation~($b$) art.~15
in a parenthesis, it will become
\[
\psi=\int\frac{d\sigma}{(r)}\biggl(\frac{\overline{d\phi}}{dw}\biggr);
\]
$(r)$ representing the distance $p,d\sigma$,
and the integral extending over both
surfaces of the shell. At the inner surface we have
$\frac{\overline{d\phi}}{dw}=\frac{\overline{d\phi}}{dr}$
and~$r=a$:
hence the part of $\psi$ due to this surface is
\[
-\int\frac{d\sigma}{(r)}\frac{\overline{d\phi}}{dr}
=\int\frac{d\sigma}{(r)}\sum i\phi^{(i)}a^{i-1}
+\int\frac{d\sigma}{(r)}\sum (i+1)\phi_\prime^{(i)}a^{-i-2}
\]
the integrals extending over the whole of the inner surface, and $d\sigma$ being
one of its elements. Effecting the integrations
by the formulae of \Person{Laplace}
(\Title{M\'ec. C\'eleste}, Liv.~3),
we immediately obtain the part of~$\psi$, due to the inner
surface, viz.
\[
\frac{4\pi a^2}{r}\sum\frac{a^i}{(2i+1)r^i}
\bigl(-ia^{i-1}\phi^{(i)}+(i+1)\phi_\prime^{(i)}a^{-i-2}\bigr).
\]
In the same way the part of $\psi$ due to the outer surface, by observing that
for it $\frac{\overline{d\phi}}{dw}=-\frac{\overline{d\phi}}{dr}$
and~$r=a_\prime$, is found to be
\[
4\pi a_\prime \sum\frac{r^i}{(2i+1)a_\prime^i}
\bigl(ia_\prime^{i-1}\phi^{(i)}-(i+1)\phi_\prime^{(i)}a_\prime^{-i-2}\bigr).
\]
The sum of these two expressions is the complete value of $\psi$,
which, together
with the values of~$\phi$ and~$V$ before given,
being substituted in the equation~($c$)
art.~15, we obtain
\[
\begin{aligned}
\text{const}=
&(1-g)\sum\phi_\prime^{(i)}r^{-i-1}
+(1-g)\sum\phi^{(i)}r^{i}\\
&+\frac{3g}{4\pi}\sum U_\prime^{(i)}r^{-i-1}
+\frac{3g}{4\pi}\sum U^{(i)}r^{i}\\
&+\frac{3ga^2}{r}\sum\frac{a^i}{(2i+1)r^i}
\bigl(-ia^{i-1}\phi^{(i)}+(i+1)\phi_\prime^{(i)}a^{-i-2}\bigr)\\
&+3ga_\prime \sum\frac{r^i}{(2i+1)a_\prime^i}
\bigl(ia_\prime^{i-1}\phi^{(i)}-(i+1)\phi_\prime^{(i)}a_\prime^{-i-2}\bigr).
\end{aligned}
\]
Equating the co-efficients of like powers of the variable $r$,
we have generally,
whatever $i$ may be,
\[
\begin{aligned}
0 &= (1-g)\phi_\prime^{(i)}+\frac{3ga^{i+2}}{2i+1}
\bigl(-ia^{i-1}\phi^{(i)}+(i+1)\phi_\prime^{(i)}a^{-i-2}\bigr)
+\frac{3g}{4\pi}U_\prime^{(i)}\\
0 &= (1-g)\phi^{(i)}+\frac{3g}{(2i+1)a_\prime^{i-1}}
\bigl(ia_\prime^{i-1}\phi^{(i)}-(i+1)\phi_\prime^{(i)}a_\prime^{-i-2}\bigr)
+\frac{3g}{4\pi}U^{(i)};
\end{aligned}
\]
neglecting the constant on the right side of the equation in $r$ as superfluous,
since it may always be made to enter into~$\phi^{(0)}$. If now, for abridgment,
we make
\[
D=(2i+1)^2(1+g)+(i-1)(i+2)g^2-9g^2i(i+1)\Bigl(\frac{a}{a_\prime}\Bigr)^{2i+1},
\]
we shall obtain by elimination
\[
\begin{aligned}
\phi^{(i)} &=
-\frac{3g}{4\pi}U^{(i)}\frac{(2i+1)\bigl(2i+1+(i+2)g\bigr)}{D}
-\frac{3g}{4\pi}U_\prime^{(i)}\frac{3g(i+1)(2i+1)a_\prime^{-2i-1}}{D}\\
\phi_\prime^{(i)} &=
-\frac{3g}{4\pi}U^{(i)}\frac{3gi(2i+1)a^{2i+1}}{D}
-\frac{3g}{4\pi}U_\prime^{(i)}\frac{(2i+1)\bigl(2i+1+(i-1)g\bigr)}{D}.
\end{aligned}
\]
These values substituted in the expression
\[
\phi=\sum\phi^{(i)}r^i+\sum\phi_\prime^{(i)}r^{-i-1},
\]
give the general value of $\phi$ in a series of the powers of $r$,
when the potential
function due to the bodies inducing a magnetic state in the shell is known,
and thence we may determine the value of the potential function or arising
from the shell itself, for any point whatever, either within or without it.

When all the bodies are situate in the space exterior to the shell, we
may obtain the total actions exerted on a magnetic particle in its interior, by
the following simple method, applicable to hollow shells of any shape and
thickness.

The equation ($c$) art.~15 becomes, by neglecting the superfluous constant,
\[
0=(1-g)\phi+\frac{3g}{4\pi}(\psi+V).
\]
If now $(\phi)$ represent the value of the potential function,
corresponding to~$\overline\phi$
the value of~$\phi$ at the inner surface of the shell,
each of the functions $(\phi)$,
$\psi$ and~$V$, will satisfy the equations
$0=\delta(\phi)$, $0=\delta\psi$ and~$0=\delta V$, and
moreover, have no singular values in the space within the shell; the same
may therefore be said of the function
\[
(1-g)(\phi)+\frac{3g}{4\pi}(\psi+V),
\]
and as this function is equal to \Emphasis{zero} at the inner surface,
it follows (art.~5)
that it is so for any point~$p$ of the interior space. Hence
\[
0=(1-g)(\phi)+\frac{3g}{4\pi}(\psi+V).
\]
But $\psi+V$ is the value of the total potential function at the point~$p$,
arising
from the exterior bodies and shell itself: this function will therefore be 
expressed by
\[
-\frac{4\pi(1-g)}{3g}(\phi).
\]
In precisely the same way, the value of the total potential function at any
point~$p'$, exterior to the shell,
when the inducing bodies are all within it, is
shown to be
\[
-\frac{4\pi(1-g)}{3g}(\phi');
\]
$(\phi')$ being the potential function
corresponding to the value of $\phi$ at the exterior
surface of the shell. Having thus the total potential functions, the total
action exerted on a magnetic particle in any direction, is immediately given
by differentiation.

To apply this general solution to our spherical shell, the inducing bodies
being all exterior to it, we must first determine~$\overline\phi$,
the value of~$\phi$ at its
inner surface, making $0=\sum U_\prime^{(i)}r^{-i-1}$
since there are no interior bodies, and
thence deduce the value of~$(\phi)$.
Substituting for~$\phi^{(i)}$ and~$\phi_\prime^{(i)}$ their values
before given, making~$U_\prime^{(i)}=0$ and~$r=a$, we obtain
\[
\overline\phi=
\frac{-3g}{4\pi}(1+2g)\sum U^{(i)}\frac{(2i+1)^2a^i}{D},
\]
and the corresponding value of $(\phi)$ is (\Title{M\'ec. C\'el.} Liv.~3)
\[
(\phi)=
\frac{-3g}{4\pi}(1+2g)\sum U^{(i)}\frac{(2i+1)^2r^i}{D}.
\]
The value of the total potential function at any point $p$ within the shell, whose
polar coordinates are~$r,\theta,\varpi$ is
\[
-\frac{3g}{4\pi}(1-g)(\phi)=(1-g)(1+2g)\sum U^{(i)}\frac{(2i+1)^2r^i}{D}.
\]
In a similar way, the value of the same function at a point $p'$ exterior to
the shell, all the inducing bodies being within it, is found to be
\[
(1-g)(1-2g)\sum U_\prime^{(i)}\frac{(2i+1)^2}{D\cdot r^{i+1}};
\]
$r$, $\theta$ and $\varpi$ in this expression
representing the polar co-ordinates of~$p'$.

To give a very simple example of the use of the first of these formulae,
suppose it were required to determine the total action exerted in the
interior of a hollow spherical shell, by the magnetic influence of the earth;
then making the axis of~$x$ to coincide
with the direction of the dipping needle,
and designating by~$f$,
the constant force tending to impel a particle of positive
fluid in the direction of~$x$ positive,
the potential function~$V$, due to the exterior
bodies, will here become
\[
V=-f\cdot x=-f\cos\theta\cdot r=U^{(1)}\cdot r.
\]
The finite integrals expressing the value of $V$ reduce themselves therefore,
in this case, to a single term, in which~$i=1$, and the corresponding value
of~$D$ being~$9(1+g-2g^2\frac{a^3}{a_\prime^3})$,
the total potential function within the shell is
\[
-(1-g)(1+2g)U^{(1)}\frac{r}{1+g-2g^2\frac{a^3}{a_\prime^3}}
=\frac{1+g-2g^2}{1+g-2g^2\frac{a^3}{a_\prime^3}}f\cdot x.
\]
We therefore see that the effect produced by the intervening shell, is to reduce
the directive force which would act on a very small magnetic needle,
\[
\text{from~}\mathbold{f},\qquad
\text{to~}\frac{1+g-2g^2}{1+g-2g^2\frac{a^3}{a_\prime^3}}
\mathbold{f}.
\]
In iron and other similar bodies, $g$ is very nearly equal to $1$,
and therefore the
directive force in the interior of a hollow spherical shell
is greatly diminished,
except when its thickness is very small compared with its radius, in which
case, as is evident from the formula,
it approaches towards the original value~$f$,
and becomes equal to it when this thickness is infinitely small.

To give an example of the use of the second formula, let it be proposed
to determine the total action upon a point~$p$,
situate on one side of an infinitely
extended plate of uniform thickness, when another point~$P$, containing a unit
of positive fluid, is placed on the other side of the same plate
considering it as
a perfect conductor of magnetism. For this,
let fall the perpendicular $PQ$ upon
the side of the plate next~$P$, on $PQ$ prolonged,
demit the perpendicular~$pq$,
and make $PQ=b$, $Pq=u$, $pq=v$, and $t=$ the thickness of the plate;
then, since its action is evidently equal to that of an infinite sphere of the
same thickness, whose centre is upon the line~$QP$ at an infinite distance
from~$P$, we shall have the required value
of the total potential function at~$p$
by supposing $a_\prime=a+t$, $a$ infinite,
and the line $PQ$ prolonged to be the
axis from which the angle $\theta$ is measured. Now in the present case
\[
V=\frac{1}{Pp}
=\frac{1}{\sqrt{r^2-2r(a-b)\cos\theta+(a-b)^2}}
=\sum U_\prime^{(i)}r^{-i-1},
\]
and the value of the potential function, as before determined, is
\[
(1-g)(1-2g)\sum\frac{(2i+1)^2}{D}\,U_\prime^{(i)}r^{-i-1}.
\]

From the first expression we see that the
general term $U_\prime^{(i)}r^{-i-1}$ is
a quantity of the order~$(a-b)^ir^{-i-1}$.
Moreover, by substituting for $r$ its
value in~$u$
\[
(a-b)^ir^{-i-1}=(a-b)^i(a-b+u)^{-i-1}=\frac1ae^{-\frac{iu}{a}};
\]
neglecting such quantities as are
of the order $\frac1a$ compared with those retained.
The general term~$U_\prime^{(i)}r^{-i-1}$, and consequently~$U_\prime^{(i)}$,
ought therefore to be
considered as functions of~${\frac ia=\gamma}$.
In the finite integrals just given, the
increment of~$i$ is~$1$, and the corresponding increment
of~$\gamma$ is $\frac1a=d\gamma$ (because
$a$ is infinite), the finite integrals thus
change themselves into ordinary integrals
or fluents. In fact (\Title{M\'ec. C\'el.} Liv.~3),
$U_\prime^{(i)}$ always satisfies the equation
\[
\frac{d^2U_\prime^{(i)}}{d\theta^2}
+\frac{\cos\theta}{\sin\theta}\frac{dU_\prime^{(i)}}{d\theta}
+i(i+1)U_\prime^{(i)}=0,
\]
and as $\theta$ is infinitely small whenever $V$
has a sensible value, we may eliminate
it from the above by means of the equation~${a\theta=v}$,
and we obtain by
neglecting infinitesimals of higher orders than those retained,
since~${\frac ia=\gamma}$,
\[
0=\frac{d^2U_\prime^{(i)}}{dv^2}
+\frac{dU_\prime^{(i)}}{v\,dv}
+\gamma^2U_\prime^{(i)}.
\]
Hence the value $U_\prime^{(i)}$ is of the form
\[
U_\prime^{(i)}=
A\int_0^1\frac{d\beta}{\sqrt{1-\beta^2}}\,\cos(\beta\gamma v);
\]
seeing that the remaining part of the general integral becomes infinite when $v$
vanishes, and ought therefore to be rejected. It now only remains to determine
the value of the arbitrary constant~$A$. Making, for this purpose, $\theta=0$,
i.~e. $v=0$, we have
\[
U_\prime^{(i)}=(a-b)^i
\quad\text{and}\quad
\int_0^1\frac{d\beta}{\sqrt{1-\beta^2}}=\tfrac12\pi:
\quad\text{hence}\quad
(a-b)^i=\tfrac12(A\pi)
\]
\[
\text{i.~e.}\quad
A=\frac2\pi(a-b)^i.
\]
By substituting for $A$ and $r$ their values, there results
\[
\begin{aligned}
U_\prime^{(i)}r^{-i-1} &=
\frac2\pi(a-b)^i(a-b+u)^{-i-1}
\int_0^1\frac{d\beta}{\sqrt{1-\beta^2}}\,\cos(\beta\gamma v)\\
&=\frac{2d\gamma}{\pi}e^{-\gamma u}
\int_0^1\frac{d\beta}{\sqrt{1-\beta^2}}\,\cos(\beta\gamma v);
\end{aligned}
\]
because $\frac ia=\gamma$ and $\frac1a=d\gamma$.
Writing now in the place of~$i$ its value~$a\gamma$,
and neglecting infinitesimal quantities, we have
\[
\frac{(2i+1)^2}{D}=\frac{4}{4+4g+g^2-9g^2e^{-2\gamma t}}.
\]
Hence the value of the total potential function becomes
\[
\frac8\pi(1-g)(1+2g)
\int_0^\infty\frac{d\gamma\cdot e^{-\gamma u}}{4+4g+g^2-9g^2e^{-2\gamma t}}
\int_0^1\frac{d\beta}{\sqrt{1-\beta^2}}\,\cos(\beta\gamma v);
\]
where the integral relative to $\gamma$ is taken
from~$\gamma=0$ to~$\gamma=\infty$, to correspond
with the limits $0$ and~$\infty$ of~$i$, seeing that~${i=a\gamma}$.

The preceding solution is immediately applicable to the imaginary case
only, in which the inducing bodies reduce themselves to a single point~$P$,
but by the following simple artifice we may give it a much greater degree
of generality:

Conceive another point~$P'$, on the line $PQ$, at an arbitrary distance~$c$
from~$P$, and suppose the unit of positive fluid concentrated in $P'$
instead of~$P$;
then if we make~$r'=Pp$,
and~$\theta=\angle pPQ$,
we shall have~$u=r'\cos\theta'$, $v=r'\sin\theta'$,
and the value of the potential function arising from~$P'$ will be
\[
\frac{1}{P'p}=
\frac{1}{\sqrt{r'^2-2r'c\cos\theta'+c^2}}
=Q^{(0)}\frac{1}{r'}
+Q^{(1)}\frac{c}{r'^2}
+Q^{(2)}\frac{c^2}{r'^3}
+\text{etc.}
\]
Moreover, the value of the total potential function at $p$ due to this, arising
from~$P'$ and the plate itself, will evidently be obtained by changing $u$ into
$u-c$ in that before given, and is therefore
\[
\frac8\pi(1-g)(1+2g)
\int_0^\infty\frac{e^{\gamma c}d\gamma e^{-\gamma u}}
{(2+g)^2-9g^2e^{-2\gamma t}}
\int_0^1\frac{d\beta}{\sqrt{1-\beta^2}}\,\cos(\beta\gamma v).
\]
Expanding this function in an ascending series of the powers of~$c$, the term
multiplied by~$c^i$ is
\[
\frac8\pi(1-g)(1+2g)
\int_0^\infty\frac
{\frac{\gamma^ie^i}{1\cdot2\cdot3\cdots n}\,d\gamma e^{-\gamma u}}
{(2+g)^2-9g^2e^{-2\gamma t}}
\int_0^1\frac{d\beta}{\sqrt{1-\beta^2}}\,\cos(\beta\gamma v),
\]
which, as $c$ is perfectly arbitrary,
must be the part due to the term~$Q^{(i)}\frac{c^i}{r^{i+1}}$
in the potential function arising from the inducing bodies.
If then this function
had been
\[
Q^{(0)}\frac{k_0}{r'}
+Q^{(1)}\frac{k_1}{r'^2}
+Q^{(2)}\frac{k_2}{r'^3}
+Q^{(3)}\frac{k_3}{r'^4}
+\text{etc.};
\]
where the successive powers $c^0,c^1,c^2$ etc.\ of $c$
are replaced by the arbitrary
constant quantities $k_0,k_1,k_2$, etc.,
the corresponding value of the total potential
function will be given by making a like change in that due to~$P'$.
Hence if,
for abridgment, we make
\[
\phi(\gamma)=
k_0+\frac{k_1}{1}\gamma+\frac{k_2}{1\cdot2}\gamma^2
+\frac{k_3}{1\cdot2\cdot3}\gamma^3+
\text{etc.},
\]
the value of this function at the point $p$ will be
\[
\frac8\pi(1-g)(1+2g)
\int_0^\infty\frac{\phi(\gamma)\,d\gamma e^{-\gamma u}}
{(2+g)^2-9g^2e^{-2\gamma t}}
\int_0^1\frac{d\beta}{\sqrt{1-\beta^2}}\,\cos(\beta\gamma v).
\]
Now, if the original one due to the point $P$ be called $F$, it is clear the
expression just given may be written
\[
\phi\Bigl(\frac{-d}{du}\Bigr)\cdot F;
\]
where the symbols of operation are separated from those of quantity, according
to \Person{Arbogast}'s method;
thus all the difficulty is reduced to the determination of~$F$.

Resuming therefore the original supposition of the plate's magnetic state
being induced by a particle of positive fluid concentrated in~$P$, the value of
the total potential function at~$p$ will be
\[
F=\frac8\pi(1-g)(1+2g)
\int_0^\infty\frac{d\gamma e^{-\gamma u}}
{(2+g)^2-9g^2e^{-2\gamma t}}
\int_0^1\frac{d\beta}{\sqrt{1-\beta^2}}\,\cos(\beta\gamma v),
\]
as was before shown.

Let $\frac{3g}{2+g}=m:$  we shall have
\[
\begin{aligned}
F &= \frac2\pi(1-m^2)\int_0^1\frac{d\beta}{(1-\beta^2)^{\frac12}}
\int_0^\infty\frac{d\gamma e^{-u\gamma}}
{1-m^2e^{-2\gamma t}}\,\cos(\beta\gamma v)\\
&= \frac2\pi(1-m^2)\int_0^1\frac{d\beta}{(1-\beta^2)^{\frac12}}
\int_0^\infty d\gamma e^{-u\gamma}
(1+m^2e^{-2\gamma t}+m^4e^{-4\gamma t}+\text{etc.})\,\cos(\beta\gamma v)\\
&= \frac2\pi(1-m^2)\int_0^1\frac{d\beta}{(1-\beta^2)^{\frac12}}
\Bigl\{ \frac{u}{u^2+\beta^2v^2}+\frac{m^2(u+2t)}{(u+2t)^2+\beta^2v^2}
+\frac{m^4(u+4t)}{(u+4t)^2+\beta^2v^2}+\text{etc.} \Bigr\}\\
&= \frac2\pi(1-m^2)\sum\int_0^{\frac12\pi}
\frac{m^{2i}u_i\,d\theta}{u_i^2+v^2\sin^2\theta},
\quad\text{where $u_i=u+2it$}\\
&= (1-m^2)\sum\frac{m^{2i}}{u_i^2+v^2)^{\frac12}}.
\end{aligned}
\]

Writing now $e^{-v\beta\gamma\sqrt{-1}}$
in the place of~$\cos(\beta\gamma v)$, we obtain
\[
F=\frac8\pi(1-g)(1+2g)
\int_0^1\frac{d\beta}{\sqrt{1-\beta^2}}
\int_0^\infty\frac{d\gamma e^{-\gamma(u+\beta v\sqrt{-1})}}
{(2+g)^2-9g^2e^{-2\gamma t}},
\]
provided we reject the imaginary quantities which may arise. In order to
transform this double integral let~$z=\frac{3g}{2+g}e^{-\gamma t}$,
and we shall have
\[
F=\frac{8(1-g)(1+2g)}{9\pi g^2t}\Bigl(\frac{2+g}{3g}\Bigr)^{\frac ut-2}
\int_0^1\frac{d\beta}{\sqrt{1-\beta^2}}
\Bigl( \frac{2+g}{3g} \Bigr)^{\frac{\beta v\sqrt{-1}}{t}}
\int\frac{dz\cdot z^{\frac ut-1+\frac{\beta v\sqrt{-1}}{t}}}{1-z^2};
\]
the integral relative to $z$ being taken from $z=0$ to~$z=\frac{3g}{2+g}$.

The value of $1-g$, for iron and other similar bodies, is very small,
neglecting therefore quantities which are of the order~$(1-g)$
compared with those
retained, there results
\[
\tag{$a$.}
F=\frac{8(1-g)}{3\pi t}
\int_0^1\frac{d\beta}{\sqrt{1-\beta^2}}
\int_0^1\frac{dz}{1-z^2}\,z^{\frac ut-1+\frac{\beta v}{t}\sqrt{-1}};
\]
where $u$ and $v$ may have any values whatever provided they are not very
great and of the order~$\frac{t}{1-g}$.
If $F_1$~represents what $F$ becomes by changing
$u$ into~$u+2t$, we have
\[
F_1=\frac{8(1-g)}{3\pi t}
\int_0^1\frac{d\beta}{\sqrt{1-\beta^2}}
\int_0^1\frac{z^2dz}{1-z^2}\,z^{\frac ut-1+\frac{\beta v}{t}\sqrt{-1}};
\]
and consequently
\[
F-F_1=\frac{8(1-g)}{3\pi t}
\int_0^1\frac{d\beta}{\sqrt{1-\beta^2}}
\int_0^1dz\cdot z^{\frac ut-1+\frac{\beta v}{t}\sqrt{-1}}
\]
which, by effecting the integrations and rejecting the imaginary quantities,
becomes
\[
F-F_1=\frac{4(1-g)}{3\sqrt{u^2+v^2}}=\frac{4(1-g)}{3r'}.
\]
Suppose now $pO$ is a perpendicular falling from the point $p$ upon the surface
of the plate, and on this line, indefinitely extended in the direction~$Op$,
take
the points $p_1,p_2,p_3$, etc., at the distances $2t,4t,6t$,
etc.\ from~$p$; then
$F_1,F_2,F_3$, etc.\ being the values of~$F$,
calculated for the points $p_1,p_2,p_3$, etc.\ by
the formula~($a$) of this article,
and $r'_1,r'_2,r'_3$, etc.\ the corresponding
values of~$r'$, we shall equally have
\[
F_1-F_2=\frac{4(1-g)}{3r'_1},\quad
F_2-F_3=\frac{4(1-g)}{3r'_2},\quad\text{etc.};
\]
and consequently
\[
F=\tfrac43(1-g)
\biggl\{ \frac{1}{r'}+\frac{1}{r'_1}+\frac{1}{r'_2}
+\text{etc. in infinitum} \biggr\};
\quad\text{seeing that $F_\infty=0$.}
\]

From this value of $F$, it is evident the total action exerted upon the
point~$p$, in any given direction~$pn$, is equal to the sum of the actions which
would be exerted without the interposition of the plate, on each of the points
$p,p_1,p_2$, etc.\ in infinitum,
in the directions $pn,p_1n_1,p_2n_2$, etc.\ multiplied
by the constant factor~$\frac43(1-g):$
the lines $pn,p_1n_1,p_2n_2$, etc.\ being all
parallel. Moreover, as this is the case wherever the inducing point $P$ may
be situate, the same will hold good when, instead of~$P$, we substitute a body
of any figure whatever magnetized at will. The only condition to be observed,
is, that the distance between~$p$ and every part of the inducing body be not
a very great quantity of the order~$\frac{1}{1-g}$.

On the contrary, when the distance between $p$ and the inducing body
is great enough to render $\frac{(1-g)r'}{t}$
a very considerable quantity, it will be
easy to show, by expanding~$F$ in a descending series of the powers of~$r'$,
that the actions exerted upon~$p$ are very nearly the same as if no plate
were interposed.

We have before remarked (art.~15), that when the dimensions of a
body are all quantities of the same order, the results of the true theory differ
little from those, which would be obtained by supposing the magnetic like the
electric fluid,
at liberty to move from one part of a conducting body to another;
but when, as in the present example, one of the dimensions is very small
compared with the others, the case is widely different; for if we make $g$
rigorously equal to~$1$ in the preceding formulae,
they will belong to the latter
supposition (art.~15), and as $F$ will then vanish, the interposing plate will
exactly neutralize the action of any magnetic bodies however they may be
situate, provided they are on the side opposite the attracted point.
This differs
completely from what has been deduced above by employing the correct
theory. A like difference between the results of the two suppositions takes
place, when we consider the action exerted by the earth on a magnetic
particle, placed in the interior of a hollow spherical shell, provided its
thickness is very small compared with its radius, as will be evident by making
$g=1$ in the formulae belonging to this case, which are given in a preceding
part of the present article.
\bigskip

\Section{17.}
\markboth{Application to magnetism.}{17.}
Since \Person{Coulomb}'s experiments on cylindric wires magnetized to saturation
are numerous and very accurate, it was thought this little work could not be
better terminated, than by directly deducing from theory such consequences
as would admit of an immediate comparison with them, and in order to effect
this, we will, in the first place, suppose a cylindric wire whose radius is~$a$
and length~$2\lambda$,
is exposed to the action of a constant force, equal to~$f$, and
directed parallel to the axis of the wire, and then endeavour to determine
the magnetic state which will thus be induced in it. For this, let $r$ be a
perpendicular falling from a point $p$ within the wire upon its axis, and~$x$,
the distance of the foot of this perpendicular from the middle of the axis;
then $f$ being directed along or positive, we shall have for the value of the
potential function due to the exterior forces
\[
V=-fx,
\]
and the equations ($b$), ($c$) (art.~15) become,
by omitting the superfluous constant,
\[
\tag{$b$}
\psi'=-\int\frac{d\sigma}{(r)}\biggl(\frac{\overline{d\phi}}{dw}\biggr),
\]
\[
\tag{$c$}
0=(1-g)\phi+\frac{3g}{4\pi}\psi-\frac{3gf}{4\pi}x:
\]
$(r)$, the distance $p',d\sigma$ being inclosed
in a parenthesis to prevent ambiguity,
and $p'$ being the point to which $\psi'$ belongs. By the same article we have
$0=\delta\phi$ and~$0=\delta\psi$,
and as $\phi$ and $\psi$ evidently depend on $x$ and~$r$ only,
these equations being written at length are
\[
\begin{aligned}
0 &= r^2\frac{d^2\phi}{dx^2}+\frac{rd}{dr}\Bigl(\frac{r\,d\phi}{dr}\Bigr)\\
0 &= r^2\frac{d^2\psi}{dx^2}+\frac{rd}{dr}\Bigl(\frac{r\,d\psi}{dr}\Bigr).
\end{aligned}
\]
Since $r$ is always very small compared with the length of the wire, we may
expand~$\phi$ in an ascending series of the powers of~$r$, and thus
\[
\phi=X+X_1r+X_2r^2+\text{etc.};
\]
$X,X_1,X_2$ etc.\ being functions of $x$ only. By substituting this value in
the equation just given, and comparing the co-efficients of like powers of~$r$,
we obtain
\[
\phi=X-\frac{d^2X}{dx^2}\frac{r^2}{2^2}
+\frac{d^4X}{dx^4}\frac{r^4}{2^2\cdot4^2}
+\text{etc.}
\]
In precisely the same way the value of $\psi$ is found to be
\[
\psi=Y-\frac{d^2Y}{dx^2}\frac{r^2}{2^2}
+\frac{d^4Y}{dx^4}\frac{r^4}{2^2\cdot4^2}
-\text{etc.}
\]
It now only remains to find the values of $X$ and $Y$ in functions of~$x$. By
supposing $p'$ placed on the axis of the wire, the equation~($c$) becomes
\[
Y=-\int\frac{d\sigma}{(r)}\biggl(\frac{\overline{d\phi}}{dw}\biggr);
\]
the integral being extended over the whole surface ofthe wire: $Y'$ belonging
to the point~$p'$, whose co-ordinates will be marked with an accent.

The part of $Y'$ due to the circular plane at the end of the cylinder,
where~$x=\lambda$, is
\[
-\frac{dX''}{dx}\int_0^a\frac{2\pi r\,dr}{(r)}
=-2\pi\frac{dX''}{dx}
\biggl\{\sqrt{(\lambda+x')^2+a^2}-\lambda-x'\biggr\},
\]
since here $d\sigma=2\pi r\,dr$
and $\frac{\overline{d\phi}}{dw}=\frac{dX''}{dx}$,
by neglecting quantities of the order
$a^2$ on account of their smallness;
$X''$~representing the value of~$X$ when~${x=-\lambda}$.

At the other end where $x=+\lambda$ we have
$d\sigma=2\pi r\,dr$, $\frac{{d\phi}}{dw}=\frac{dX'''}{dx}$
and consequently the part due to it is
\[
\frac{dX'''}{dx}\int_0^a\frac{2\pi r\,dr}{(r)}
=2\pi\frac{dX'''}{dx}
\biggl\{\sqrt{(\lambda-x')^2+a^2}-\lambda+x'\biggr\},
\]
$X'''$ designating the value of $X$ when $x=+\lambda$.

At the curve surface of the cylinder
\[
d\sigma=2\pi a\,dx
\quad\text{and}\quad
\frac{\overline{d\phi}}{dw}
=-\frac{\overline{d\phi}}{dr}
=\tfrac12a\frac{d^2X}{dx^2}
\]
provided we omit quantities of the order $a^2$ compared with those retained.
Hence the remaining part due to this surface is
\[
-\pi a^2\int\frac{dx}{(r)}\frac{d^2X}{dx^2};
\]
the integral being taken from $x=-\lambda$ to $x=+\lambda$.
The total value of $Y'$
is therefore
\begin{multline*}
Y'=
2\pi\frac{dX'''}{dx}
\biggl\{\sqrt{(\lambda-x')^2+a^2}-\lambda+x'\biggr\}\\
-2\pi\frac{dX''}{dx}
\biggl\{\sqrt{(\lambda+x')^2+a^2}-\lambda-x'\biggr\}\\
-\pi a^2\int\frac{dx}{(r)}\frac{d^2X}{dx^2};
\end{multline*}
the limits of the integral being the same as before. If now we substitute for
$(r)$ its value $\sqrt{(x-x')^2+a^2}$ we shall have
\[
-\pi a^2\int\frac{dx}{(r)}\frac{d^2X}{dx^2}
=-\pi a^2\int\frac{dx}{\sqrt{(x-x')^2+a^2}}\frac{d^2X}{dx^2};
\]
both integrals extending from $x=-\lambda$ to $x=+\lambda$.

On account of the smallness of $a$, the elements of the last integral
where $x$ is nearly equal to~$x'$ are very great compared with the others, and
therefore the approximate value of the expression just given, will be
\[
-\pi a^2A\frac{d^2X'}{dx'^2}
\quad\text{where}\quad
A=\int\frac{dx}{\sqrt{(x-x')^2+a^2}}=2\log\frac{2\mu}{a}
\text{\ very nearly;}
\]
the two limits of the integral being $-\mu$ and $+\mu$ and $\mu$ so chosen that
when $p'$ is situate any where on the wire's axis, except in the immediate
vicinity of either end, the approximate shall differ very little from the true
value, which may in every case be done without difficulty. Having thus, by
substitution, a value of~$Y'$ free from the sign of integration,
the value of~$Y$
is given by merely changing $x'$ into~$x$ and $X'$ into~$X$; in this way
\begin{multline*}
Y=
2\pi\frac{dX'''}{dx}
\biggl\{\sqrt{(\lambda-x)^2+a^2}-\lambda+x\biggr\}\\
-2\pi\frac{dX''}{dx}
\biggl\{\sqrt{(\lambda+x)^2+a^2}-\lambda-x\biggr\}
-\pi a^2A\frac{d^2X}{dx^2};
\end{multline*}
The equation ($c$), by making $r=0$, becomes
\[
0=(1-g)X+\frac{3g}{4\pi}Y-\frac{3gf}{4\pi}x,
\]
or by substituting for $Y$
\begin{multline*}
0=(1-g)X-\tfrac34(ga^2A)\frac{d^2X}{dx^2}-\frac{3gf}{4\pi}x\\
+\tfrac32g\frac{dX'''}{dx}
\biggl\{\sqrt{(\lambda-x)^2+a^2}-\lambda+x\biggr\}\\
-\tfrac32g\frac{dX''}{dx}
\biggl\{\sqrt{(\lambda+x)^2+a^2}-\lambda-x\biggr\};
\end{multline*}
an equation which ought to hold good, for every value of $x$, from 
$x=-\lambda$ to $x=+\lambda$.

In those cases to which our theory will be applied, $1-g$ is a small
quantity of the same order as~$a^2A$, and thus the three terms of the first line
of our equation will be of the order~$a^2AX$;
making now $x=+\lambda$, $\tfrac32g\frac{dX'''}{dx}a$
is shown to be of the order~$a^2AX'''$,
and therefore $\frac{dX'''}{dx}\div X'''$ is a small
quantity of the order~$aA$; but for any other value of~$x$ the function 
multiplying $\frac{dX'''}{dx}$ becomes of the order~$a^2$,
and therefore we may without sensible
error neglect the term containing it, and likewise suppose
\[
\frac{dX'''}{dx}\div X'''=0.
\]
In the same way by making $x=-\lambda$, it may be shown that the term 
containing $\frac{dX''}{dx}$ is negligible, and
\[
\frac{dX''}{dx}\div X''=0.
\]
Thus our equation reduces itself to
\[
0=(1-g)X-\tfrac34(ga^2A)\frac{d^2X}{dx^2}-\frac{3gf}{4\pi}x,
\]
of which the general integral is
\[
X=\frac{3gfx}{4\pi(1-g)}+Be^{-\beta x}+Ce^{+\beta x};
\]
where $\beta^2=\frac{4(1-g)}{3ga^2A}:$ $B$ and $C$ being
two arbitrary constants. Determining
these by the conditions $0=\frac{dX'''}{dx}\div X'''$
and $0=\frac{dX''}{dx}\div X''$, we ultimately obtain
\[
X=\frac{3gfx}{4\pi(1-g)}
\biggl\{
  x-\frac{e^{\beta x}-e^{-\beta x}}{\beta(e^{\beta\lambda}+e^{-\beta\lambda})}
\biggr\}.
\]
But the density of the fluid at the surface of the wire, which would produce
the same effect as the magnetized wire itself, is
\[
-\frac{\overline{d\phi}}{dw}
=\frac{\overline{d\phi}}{dr}
=-\tfrac12a\frac{d^2X}{dx^2}
\text{\ very nearly,}
\]
and therefore the total quantity in an infinitely thin section whose breadth
is~$dx$, will be
\[
-\pi a^2\frac{d^2X}{dx^2}\,dx=
\frac{3gf\beta a^2}{4(1-g)}\cdot
\frac{e^{\beta x}-e^{-\beta x}}{e^{\beta\lambda}+e^{-\beta\lambda}}\,dx.
\]

As the constant quantity $f$ may represent the coercive force of steel
or other similar matter, provided we are allowed to suppose this force the
same for every particle of the mass, it is clear that when a wire is magnetized
to saturation, the effort it makes to return to a natural state must, in every
part, be just equal to~$f$, and therefore, on account of its elongated form, the
degree of magnetism retained by it will be equal to that which would be
induced in a conducting wire of the same form by the force~$f$, directed along
lines parallel to its axis. Hence the preceding formulae are applicable to
magnetized steel wires. But it has been shown
by M.~\Person{Biot} (\Title{Trait\'e de Phy.}
Tome~3, Chap.~6), from \Person{Coulomb}'s experiments,
that the apparent quantity of
free fluid in any infinitely thin section is represented by
\[
A'(\mu'^{-x}-\mu'^{+x})\,dx.
\]
This expression agrees precisely with the one before deduced from theory,
and gives, for the determination of the constants $A'$ and~$\mu'$,
the equations
\[
\beta=-\log\mu';\quad
A'=\frac{3gf\beta a^2}{4(1-g)(e^{\beta\lambda}+e^{-\beta\lambda})}.
\]

The chapter in which these experiments are related, contains also a
number of results, relative to the forces with which magnetized wires tend
to turn towards the meridian, when retained at a given angle from it, and
it is easy to prove that this force for a fine wire,
whose variable section is~$s$,
will be proportional to the quantity
\[
\int s\,dx\,\frac{d\phi}{dx}
\]
where the wire is magnetized in any way either to saturation or otherwise,
the integral extending over its whole length. But in a cylindric wire 
magnetized to saturation, we have, by neglecting quantities of the order~$a^2$,
\[
\frac{d\phi}{dx}=\frac{dX}{dx}=
\frac{3gf}{4\pi(1-g)}\biggl\{1-
\frac{e^{\beta x}-e^{-\beta x}}{e^{\beta\lambda}+e^{-\beta\lambda}}\biggr\}
\quad\text{and}\quad
s=\pi a^2,
\]
and therefore for this wire the force in question is proportional to
\[
\frac{3gfa^2}{4(1-g)}\biggl\{2\lambda-
\frac{2(e^{\beta\lambda}-e^{-\beta\lambda})}
{\beta(e^{\beta\lambda}+e^{-\beta\lambda})}\biggr\}.
\]
The value of $g$, dependant on the nature of the substance of which
the needles are formed, being supposed given as it ought to he, we have only
to determine~$\beta$ in order to compare this result with observation.
But $\beta$
depends upon~$A=2\log\frac{\mu}{a}$,
and on account of the smallness of~$a$, $A$ undergoes
but little alteration for very considerable variations in~$\mu$, so that we
shall be able in every case to judge with sufficient accuracy what value 
of~$\mu$
ought to be employed: nevertheless, as it is always desirable to avoid every
thing at all vague, it will be better to determine~$A$ by the condition, that
the sum of the squares of the errors committed by employing, as we have
done, $A\frac{d^2X'}{dx'^2}$ for the approximate
value of $\int_{-\lambda}^{+\lambda}\frac{dx}{\sqrt{(x-x')^2+a^2}}$
shall be a minimum for the whole length of the wire.
In this way I find when $\lambda$ is so
great that quantities of the order $\frac{1}{\beta\lambda}$ may be neglected:
\[
A = ,231863-2\log a\beta+2a\beta;
\]
where $,231863$ etc.\ $= 2\log2-2(A)$; $(A)$ being the quantity represented
by~$A$ in \Person{Lacroix}: \Title{Trait\'e du Cal. Diff.}
Tome~3, p.~521. Substituting the value
of~$A$ just found in the equation $\beta^2=\frac{4(1-g)}{3ga^2A}$
before given, we obtain
\[
\frac{4(1-g)}{3g\cdot a^2\beta^2}=,231863-2\log a\beta+2a\beta.
\]
We hence see that when the nature of the substance of which the wires are
formed remains unchanged, the quantity $a\beta$ is constant,
and therefore $\beta$ varies
in the inverse ratio of~$a$. This agrees with
what M.~\Person{Biot} has found by experiment
in the chapter before cited, as will be evident by recollecting
that~$\beta=-\log\mu'$.

From an experiment made with extreme care by \Person{Coulomb}, on a
magnetized wire whose radius was $\frac{1}{12}$ inch,
M.~\Person{Biot} has found the value
of $\mu'$ to be~$,517948$
(\Title{Trait\'e de Phy.} Tome~3, p.~78). Hence we have in
this case
\[
a\beta=\frac{-1}{12}\log\mu'=,054823,
\]
which, according to a remark just made, ought to serve for all steel wires.
Substituting this value in the equation ($a$) of the present article,
we obtain
\[
g = ,986636.
\]
With this value of $g$ we may calculate the forces with which different lengths
of a steel wire whose radius is $\frac{1}{12}$ inch,
tend to turn towards the meridian,
in order to compare the results with the table of \Person{Coulomb}'s
observations,
given by M.~\Person{Biot} (\Title{Trait\'e de Phy.} Tome~3, p.~84).
Now we have before
proved that this force for any wire may be represented by
\[
K\biggl(\beta\lambda-
\frac{e^{\beta\lambda}-e^{-\beta\lambda}}{e^{\beta\lambda}+e^{-\beta\lambda}}\biggr)
=K\biggl(\beta\lambda-
\frac{1-e^{-2\beta\lambda}}{1+e^{-2\beta\lambda}}\biggr);
\]
where, for abridgment, we have supposed
\[
K=\frac{3gfa^2}{2\beta(1-g)}.
\]
It has also been shown that for any steel wire:
\[
a\beta = ,0548235,
\]
the French inch being the unit of space, and as in the present
case~$a=\frac{1}{12}$,
there results~$\beta=,657882$.
It only remains therefore to determine~$K$ from
one observation, the first for example, from which we
obtain~$K=58^\circ,5$ very
nearly; the forces being measured by their equivalent torsions. With this
value of~$K$ we have calculated the last column of the following table:
\bigskip
\begin{center}
\begin{tabular}{@{}*{3}{.{1}}l@{}}\toprule
\mc{\textit{Length $2\lambda$.}} &
\mc{\textit{Observed}} &
\mc{\textit{Calculated}}\\
\mc{\textit{(inch)}} &
\mc{\textit{Torsion (${}^\circ$).}} &
\mc{\textit{Torsion (${}^\circ$).}}\\ \midrule
18 & 288 & 287,9 \\
12 & 172 & 172,1 \\
9 & 115 & 115,3 \\
6 & 59  & 59,3 \\
4,5 & 34 & 33,9 \\
3 & 13 & 13,5 \\ \bottomrule
\end{tabular}
\end{center}
\bigskip
The three last observations have been purposely omitted, because the 
approximate equation~($a$) does not bold good for very short wires.

The very small difference existing between the observed and calculated
results will appear the more remarkable, if we reflect that the value of~$\beta$
was determined from an experiment of quite a different kind to any of the
present series, and that only one of these has been employed for the 
determination of the constant quantity~$K$,
which depends on~$f$, the measure of the
coercive force.

The table page 87 of the volume just cited, contains another set of
observed torsions, for different lengths of a much finer wire whose radius
$a=\frac{1}{12}\sqrt{\frac{38}{865}}$: hence we find
the corresponding value of~$\beta=3,13880$, and
the first observation in the table gives~$K={}^\circ,6448$.
With these values the
last column of the following table has been calculated as before:
\bigskip
\begin{center}
\begin{tabular}{@{}*{3}{.{3}}l@{}}\toprule
\mc{\textit{Length $2\lambda$.}} &
\mc{\textit{Observed}} &
\mc{\textit{Calculated}}\\
\mc{\textit{(inch)}} &
\mc{\textit{Torsion (${}^\circ$).}} &
\mc{\textit{Torsion (${}^\circ$).}}\\ \midrule
12 & 11,50 & 11,50 \\
9 & 8,50 & 8,46 \\
6 & 5,30 & 5,43 \\
3 & 2,30 & 2,39 \\
2 & 1,30 & 1,38 \\
1 & ,35 & ,42 \\
,5 & ,07 & ,084 \\
,25 & ,02 & ,012 \\\bottomrule
\end{tabular}
\end{center}
\bigskip

Here also the differences between the observed and calculated values
are extremely small, and as the wire is a very fine one, our formula is
applicable to much shorter pieces than in the former case. In general, when
the length of the wire exceeds 10 or 15 times its diameter, we may employ
it without hesitation.
\Crule

\end{document}